\documentclass[prd, twocolumn, nofootinbib, floatfix]{revtex4-1}
\usepackage{dcolumn}    

\newcolumntype{d}[1]{D{.}{.}{#1}}
\usepackage{lmodern} 
\usepackage[T1]{fontenc}
\usepackage{textcomp}
\usepackage[scr=boondoxo,frak=boondox,bb=boondox]{mathalfa}
\pdfoutput=1
\usepackage{makecell, multirow}
\setcellgapes{3pt}
\usepackage{hyperref}
\usepackage{amsmath}
\usepackage{amssymb}
\usepackage{tabularx}
\usepackage{tikz}
\usepackage[vlines]{tabularht}
\usepackage{adjustbox}
\usepackage{tikz}
\usepackage[]{draftfigure}
\usepackage{amssymb}
\usepackage{graphicx}
\usepackage{float} 
\usepackage{color}
\usepackage{color,soul}
\usepackage{ulem}
\usepackage{multirow}
\usepackage{epsfig}
\usepackage{epstopdf}
\usepackage{rotating}
\usepackage{afterpage}
\usepackage{soul}
\usepackage{fancyvrb}
\usepackage{lineno}
\usepackage{amsmath}
\usepackage{xspace}
\usepackage{hyperref}
\usepackage{url}
\usepackage{amssymb,latexsym,mathrsfs}
\usepackage{multirow,bigdelim, booktabs}
\usepackage{bigints}
\usepackage{soul}
\usepackage{graphicx}
\usepackage{subcaption}
\usepackage{dcolumn}
\usepackage{bm}
\usepackage{epsfig}
\usepackage{array}
\usepackage{caption}
\usepackage{color}
\usepackage{multirow}
\usepackage{makecell}
\usepackage{textcomp}
\usepackage[title]{appendix}
\usepackage{tabularx}
\usepackage{natbib}
\usepackage{graphicx}
\setcitestyle{authoryear,open={(},close={)}} 

\hypersetup{
    colorlinks=true,
    linkcolor=blue,
    citecolor=blue,
}

\definecolor{mediumpurple}{rgb}{0.58, 0.44, 0.86}

\begin{document} 
\title{  A Fully Photometric Approach to  Type Ia Supernova Cosmology in the LSST Era: Host Galaxy  Redshifts and  Supernova  Classification
}


\newcommand{\stddmuccprior}{1.03} 
\newcommand{\stdmubiasccprior}{0.15}
\newcommand{\Dzbias}{\overline{\Dz}}
\newcommand{\URLSNANA}{\url{https://github.com/RickKessler/SNANA}}
\newcommand{\URLLSST}{\url{www.lsst.org}}
\newcommand{\URLDESC}{\url{https://lsstdesc.org}}
\newcommand{\URLOPSIM}{\url{http://opsim.lsst.org/runs/minion_1016/data/minion_1016_sqlite.db.gz}}
\newcommand{\HOSTLIB}{{\tt HOSTLIB}}
\newcommand{\SALTII}{{\sc SALT-II}}
\newcommand{\DESSN}{DES-SN}
\newcommand{\lowz}{low-$z$~}
\newcommand{\mosfit}{{\tt MOSFiT}\xspace}
\newcommand{\bands}{$ugrizY$}
\newcommand{\PLASTICC}{Photometric LSST Astronomical Time Series Classification Challenge}
\newcommand{\acro}{{\tt ELAsTiCC}}
\newcommand{\SNANA}{{\tt SNANA}}
\newcommand{\pippin}{{\tt Pippin}}
\newcommand{\LSST}{Large Synoptic Survey Telescope}
\newcommand{\DES}{Dark Energy Survey}
\newcommand{\OPSIM}{\verb|OpSim|}
\newcommand{\Spec}{Spectroscopic}
\newcommand{\spec}{spectroscopic}
\newcommand{\specy}{spectroscopically}
\newcommand{\ZPHOT}{{\bf\tt ZPHOT}}
\newcommand{\zSpec}{z_{\rm spec}}
\newcommand{\LCDM}{\Lambda{\rm CDM}}
\newcommand{\OL}{\Omega_{\Lambda}}
\newcommand{\OM}{\Omega_{\rm M}}
\newcommand{\zcmb}{z_{\rm cmb,true}}
\newcommand{\RateUnit}{{\rm yr}^{-1}{\rm Mpc}^{-3}}
\newcommand{\Ftrue}{F_{\rm true}}
\newcommand{\sigF}{\sigma_{F}}
\newcommand{\sigFi}{\sigma_{F,i}}
\newcommand{\sigFtrue}{\sigma_{\rm Ftrue}}
\newcommand{\zSN}{z_{\rm SN}}
\newcommand{\zHOST}{z_{\rm HOST}}
\newcommand{\Ngen}{$N_{\rm gen}$} %

\newcommand{\FiData}{F_i^{\rm data}}
\newcommand{\FiModel}{F_i^{\rm model}}
\newcommand{\xvecSALT}{\vec{x}_5}
\newcommand{\be}{\begin{equation}}
\newcommand{\ee}{\end{equation}}
\newcommand{\bea}{\begin{eqnarray}}
\newcommand{\eea}{\end{eqnarray}}
\newcommand{\iu}{{i\mkern1mu}} 
\newcommand{\sigFiTilde}{\Tilde{\sigma}_{F,i}}
\newcommand{\logsigmodel}{2\ln({\sigFi/\sigFiTilde})}
\newcommand{\sigzhost}{\sigma_{z,{\rm host}}}
\newcommand{\alphaTrue}{\alpha^{\rm true}}
\newcommand{\betaIntTrue}{\beta_{\rm int}^{\rm true}}
\newcommand{\DESVYR}{DES-SN5YR}
\newcommand{\PPLUS}{Pantheon+}
\newcommand{\chisqflux}{\chi^2_{F_i}}
\newcommand{\chisqhost}{\chi^{2}_{\rm host}}
\newcommand{\chisqsyst}{\chi^{2}_{\rm syst}}
\newcommand{\probzhost}{\mathcal{P}_{host}(\zphot)}
\newcommand{\rcc}{R_{CC}}
\newcommand{\NSAMPLE}{25} 
\newcommand{\NBIASCORTOT}{$1 \times 10^8$}
\newcommand{\NBIASCORHIZ}{$9 \times 10^7$}
\newcommand{\NBIASCORLOZ}{$4 \times 10^6$}
\newcommand{\NSYST}{7}
\newcommand{\dz}{\delta z}
\newcommand{\fout}{f_{\rm out}}
\newcommand{\sigIQR}{\sigma_{\rm IQR}}
\newcommand{\ztrue}{z_{\rm true}}
\newcommand{\zphot}{z_{\rm phot}}
\newcommand{\zspec}{z_{\rm spec}}
\newcommand{\zcheat}{z_{\rm cheat}}
\newcommand{\zhost}{z_{\rm host}}
\newcommand{\Dz}{\Delta z_{(1+z)}}
\newcommand{\Pfit}{P_{\rm fit}}
\newcommand{\ayan}[1]{\textcolor{red}{#1}} 
\newcommand{\rk}[1]{\textcolor{brown}{#1}}
\newcommand{\rkout}[1]{\textcolor{red}{\sout{\protect#1}}}
\newcommand{\rkred}[1]{\textcolor{red}{#1}}
\newcommand{\IR}[1]{\textcolor{cyan}{#1}}
\newcommand{\G}{G_{\rm host}}
\newcommand{\mubias}{\Delta\mu_{\rm bias}}
\newcommand{\mutrue}{\mu_{\rm true}}
\newcommand{\pkref}{p_k^{\rm ref}}
\newcommand{\pk}{p_k}
\newcommand{\ddlr}{d_{DLR}}
\newcommand{\sigkref}{\sigma_k^{\rm ref}}
\newcommand{\Pia}{P_{\text{Ia}}}
\newcommand{\dzsyst}{\Delta z_{\rm syst-z}}
\newcommand{\dmusyst}{\Delta\mu_{\rm syst-z}}
\newcommand{\meanlamb}{\langle\lambda_{\rm cen}\rangle}
\newcommand{\alphaTrueSym}{\alpha_{\rm true}}
\newcommand{\betaTrueSym}{\beta_{\rm true}}
\newcommand{\alphaTrueVal}{0.15}
\newcommand{\betaTrueVal}{3.1}
\newcommand{\nrebinx}{$4$}
\newcommand{\nrebinc}{$8$}
\newcommand{\HDSIZEREBIN}{448} 
\newcommand{\hdsizerebin}{300} 
\newcommand{\sigmu}{\sigma_{\mu}}
\newcommand{\sigmubar}{\overline{\sigma_{\mu}}}
\newcommand{\URLDIFFSKY}{\url{https://diffsky.readthedocs.io/en/latest}}
\newcommand{\sigz}{\sigma_{z}}
\newcommand{\sigR}{\sigma_{R}}
\newcommand{\sigint}{\sigma_{\rm int}}

\newcommand{\COVsyst}{{\rm COV}_{\rm syst}}
\newcommand{\COVsysti}{{\rm COV}_{{\rm syst},i}}
\newcommand{\COVstat}{{\rm COV}_{\rm stat}}

\newcommand{\wCDM}{$w$CDM}
\newcommand{\wwCDM}{$w_0w_a$CDM}
\newcommand{\ww}{$w_0$-$w_a$}
\newcommand{\URLDDF}{\url{https://www.lsst.org/scientists/survey-design/ddf}}
\newcommand{\URLELASTICC}{\url{https://portal.nersc.gov/cfs/lsst/DESC_TD_PUBLIC/ELASTICC}}
\newcommand{\NZBIN}{\textcolor{red}{14}}

\newcommand{\AVGwbias}{\langle w$-bias$\rangle}
\newcommand{\AVGwsigbiaszspecsyst}{$0.025$}
\newcommand{\AVGwsigbiaszphotsyst}{$0.025$}
\newcommand{\AVGwsig}{\langle\sigma_w\rangle}
\newcommand{\STDw}{{\rm STD}_w}

\newcommand{\AVGwwbias}{\langle {w_0}$-bias$\rangle}
\newcommand{\AVGwwsig}{\langle\sigma_{w_0}\rangle}
\newcommand{\STDww}{{\rm STD}_{w_0}}

\newcommand{\desfv}{DES-SN5YR}
\newcommand{\AVGwabias}{\langle {w_a}$-bias$\rangle}
\newcommand{\AVGwasig}{\langle\sigma_{w_a}\rangle}
\newcommand{\STDwa}{{\rm STD}_{w_a}}

\newcommand{\AVGFoM}{{\langle}{\rm FoM}{\rangle}}

\newcommand{\RatioFoM}{{\cal R}_{{\rm FoM},i}}
\newcommand{\FoM}{{\rm FoM}}
\newcommand{\FoMStat}{{\rm FoM}_{\rm stat}}
\newcommand{\FoMSysti}{{\rm FoM}_{{\rm syst},i}}

\newcommand{\AVGFOMzspecstat}{$114$}
\newcommand{\AVGFOMzspecsyst}{$92$}
\newcommand{\AVGFOMzspecstatU}{$115$}
\newcommand{\AVGFOMzspecsystU}{$106$}
\newcommand{\URLDIA}{\url{https://github.com/LSSTDESC/dia_pipe}}
\newcommand{\URLMINION}{\url{http://ls.st/Collection-4604}}

\newcommand{\AVGFOMzphotstat}{$193$}
\newcommand{\AVGFOMzphotsyst}{$131$}
\newcommand{\AVGFOMzphotstatU}{$182$}
\newcommand{\AVGFOMzphotsystU}{$156$}
\newcommand{\AVGFOMzphotsystR}{$159$}

\newcommand{\AVGwbiaszphotsyst}{$0.0091$}
\newcommand{\AVGwabiaszphotsyst}{$0.0363$}
\newcommand{\AVGwbiaszspecsyst}{$0.0140$}
\newcommand{\AVGwabiaszspecsyst}{$0.0662$}

\author{
Ayan Mitra${}^{1,2}$,
Richard Kessler${}^{3,4}$,
Rebecca~C.~Chen${}^{5,6,7}$,
Alex~Gagliano${}^{8,9,10}$,
Matthew~Grayling${}^{11}$,
Surhud~More${}^{12}$,
Gautham~Narayan${}^{2}$,
Helen~Qu${}^{13}$,
Srinivasan~Raghunathan${}^{1,14}$,
Alex~I.~Malz${}^{15}$,
Michelle~Lochner${}^{16}$,
\\
The LSST Dark Energy Science Collaboration \\
${}^{1}$ Center for AstroPhysical Surveys, National Center for Supercomputing Applications, Urbana, IL 61801, USA \\
${}^{2}$ Department of Astronomy, University of Illinois at Urbana-Champaign, Urbana, IL 61801, USA \\
${}^{3}$ Kavli Institute for Cosmological Physics, University of Chicago, Chicago, IL 60637, USA \\
${}^{4}$ Department of Astronomy and Astrophysics, University of Chicago, 5640 South Ellis Avenue, Chicago, IL 60637, USA \\
${}^{5}$ Kavli Institute for Particle Astrophysics \& Cosmology, P.O. Box 2450, Stanford University, Stanford, CA 94305, USA \\ 
${}^{6}$ SLAC National Accelerator Laboratory, Menlo Park, CA 94025, USA \\
${}^{7}$ Brinson Prize Fellow \\
${}^{8}$ The NSF AI Institute for Artificial Intelligence and Fundamental Interactions \\
${}^{9}$ Center for Astrophysics \textbar{} Harvard \& Smithsonian, 60 Garden Street, Cambridge, MA 02138, USA \\
${}^{10}$ Department of Physics and Kavli Institute for Astrophysics and Space Research, Massachusetts Institute of Technology, 77 Massachusetts Avenue, Cambridge, MA 02139, USA \\ 
${}^{11}$ Institute of Astronomy and Kavli Institute for Cosmology, Madingley Road, Cambridge CB3 0HA, UK \\
${}^{12}$ The Inter-University Centre for Astronomy and Astrophysics (IUCAA), Post Bag 4, Ganeshkhind, Pune 411007, India \\
${}^{13}$ Center for Computational Astrophysics, Flatiron Institute, 162 5th Ave, New York, NY 10010, USA \\
${}^{14}$ Department of Physics \& Astronomy, University of California, One Shields Avenue, Davis, CA 95616, USA\\
${}^{15}$ Space Telescope Science Institute, Baltimore, MD, USA \\
${}^{16}$ Department of Physics and Astronomy, University of the Western Cape, Bellville, Cape Town 7535, South Africa
}

\date{\today} 

\begin{abstract}
The upcoming Vera C. Rubin Observatory's Legacy Survey of Space and Time (LSST) is expected to discover 
 {nearly a} million  Type Ia supernovae (SNe~Ia), offering an unprecedented opportunity to constrain dark energy. 
The vast majority of these events will lack spectroscopic classification and redshifts, 
necessitating a fully photometric approach  {to maximize cosmology constraining power.}
We present detailed simulations based on the 
Extended LSST Astronomical Time Series Classification Challenge (\acro), 
and a cosmological analysis using photometrically classified SNe~Ia with host galaxy photometric redshifts.
This dataset features realistic multi-band light curves,  {non-SNIa contamination,} 
host mis-associations, and transient-host correlations across the 
high-redshift Deep Drilling Fields (DDF) ($\sim 50$ deg$^2$).
We  {also} include a  {spectroscopically confirmed} 
low-redshift sample based on the Wide Fast Deep (WFD) fields.
We employ a 
joint SN+host photometric redshift fit,  
a neural network based photometric classifier (SCONE),
and BEAMS with Bias Corrections (BBC) methodology to construct a bias-corrected Hubble diagram. 
We produce statistical + systematic covariance matrices, 
and perform cosmology fitting with a prior   {using}    Cosmic Microwave Background    {constraints}.
We fit and present results for the \wCDM\ dark energy model, 
and the more general Chevallier-Polarski-Linder (CPL) \wwCDM\ model.  
With a simulated sample of ${\sim}6000$  {events}, we achieve  {a} 
Figure of Merit (FoM)  {value} of about   {$150$}, 
 {which is significantly larger than the \DESVYR\ FoM of $54$}.
 {Averaging analysis results over 25 independent samples, we find small but significant biases
indicating a need for further analysis testing and development.}
\end{abstract}

\maketitle
\section{Introduction }
\label{sec:intro}

The discovery of dark energy and the accelerating expansion of the Universe, 
first identified through the observation of a few dozen Type Ia supernovae (SNe Ia) \cite{adam,perl}, marked a pivotal moment in modern cosmology. Subsequent SNIa surveys yielded constraints on the dark energy equation of state that  agree well with a cosmological constant (\(w=-1\)) \citep{SNLS:2005qlf,SupernovaSearchTeam:2004lze, 2018ApJ...859..101S, ESSENCE:2007acn,  conley11, kessler3, 2012ApJ...746...85S, Betoule2014}, 
where $w$ is the ratio of pressure to energy density for dark energy 
(for reviews see e.g. \citet{Frieman:2008sn, 2013PhR...530...87W}). 

Few-$\sigma$ hints of evolving dark energy 
have emerged from analyses with higher statistics and better systematics handling from 
Pantheon+ \citep{pantheon_new}
the Dark Energy Survey (DES)  \citep[\DESVYR]{DES:2024jxu},
the Dark Energy Spectroscopic Instrument (DESI) \citep[DESI]{DESI:2024mwx},
and the  UNION {3} sample Bayesian analysis \citep{Rubin:2023ovl}.
While these deviations from a cosmological constant have not reached discovery significance, they have spurred considerable interest in the possibility that dark energy may be dynamic rather than static. 
  {Current sample statistics include $\sim 10^3$ SNe, and }
upcoming projects such as  the 
Rubin Legacy Survey of Space and Time \citep[LSST]{Cahn2009,ivezic} and the  
Roman Space Telescope \citep{roman1,w2} will 
 significantly  improve the statistical   precision on dark energy parameters, and   a corresponding reduction in systematics is needed to make full use of these samples  {\citep{Linder2019, Mitra2021}}.

Historically, SN~Ia cosmology surveys 
relied on intensive spectroscopic follow-up to confirm each transient’s type and to secure a host-galaxy spectroscopic redshift  {(spec-$z$)}. 
These spectroscopically confirmed samples 
 {include  ${\sim}10^3$  SN~Ia, and}  {contribute to}  
composite data sets like the Pantheon compilation \citep{Scolnic2018} and   {the} Pantheon+ extension \citep{pantheon+,pantheon_new}. 

Previous photometric surveys%
\footnote{
CfA: Center for Astrophysics Supernova Program;
CSP: Carnegie Supernova Project;
SNLS: Supernova Legacy Survey;
ESSENCE: Equation of State: SupErNovae trace Cosmic Expansion;
SDSS-II: Sloan Digital Sky Survey-II Supernova Survey;
Pan-STARRS: Panoramic Survey Telescope and Rapid Response System;
DES-SN: Dark Energy Survey Supernova Program.
}
(CfA, CSP, SNLS, ESSENCE, SDSS-II, Pan-STARRS, DES-SN)
acquired spectroscopic information for only a small fraction of their samples, and
this challenge will be even more pronounced in the LSST era, which will have a significantly higher detection rate.  {The 4MOST TiDES \citep{4most_tides} program anticipates $18000$ spectroscopic SNe observations, 
 {which} 
is a very small fraction of the total number of transients.} 
Therefore, for the vast majority of newly discovered SNe  it has become necessary to develop  photometric classification and a formalism to analyse a contaminated Hubble diagram.

 Pan-STARRS1 \citep{Jones2018,FSS:2018cey} used the PSNID classifier \citep{sako}  and the  Bayesian Estimation Applied to Multiple Species (BEAMS) \citep{kunz}  formalism to determine cosmology constraints from a photometrically identified sample. For \DESVYR\ they used more advanced photometric classifiers:  SuperNNova (SNN) \footnote{\url{https://github.com/supernnova/SuperNNova}} \citep{snn} and SCONE \citep{scone} (see Sec.~\ref{subsec:scone}). Both  have exhibited very high levels of purity and accuracy $> 98\%$.
 They also used an extension of the BEAMS formalism, ``BEAMS with Bias Correction {s}'' (BBC), that is described in  \citet{bbc}[hereafter KS17].

While photometric classification has been successfully used to deliver cosmological results, 
photometric redshifts  {(photo-$z$)} have not. 
A photo-$z$ method using the SN+host galaxy has been proposed 
\citep{Kessler2010,Palanque2010,zbeams}, 
and \citet[M23]{Mitra:2022ykq} performed the first rigorous SNIa-cosmology analysis with photo-$z$'s and systematics, but did not include non-SNIa contamination.  {\citet{Chen25} examined the photo-$z$ bias in the \DESVYR\ sample by comparing $w$ results using spec-$z$'s vs photo-$z$. 
\citet{Ruhlmann-Kleider:2022vum} measured $\Omega_m$ using photo-$z$'s in SNLS data, and 
more recently, \citet{2025arXiv250815899K}
analysed simulated data  with photo-$z$ using a simulation based  inference technique.

As part of the DESC\footnote{\href{https://lsstdesc.org/}{Dark Energy Science Collaboration}} effort to prepare  {for} cosmology analyses with the LSST data,
here we simulate LSST data that includes both host galaxy photo-$z$'s and non-SNIa contamination,
and we incorporate SN+host photo-$z$ and BEAMS into the analysis. This analysis supersedes M23, and is applicable to real data. M23 presented a thorough light curves to cosmology analysis with photometric redshift ($\zphot$) using  LSST simulated data. However,  M23 was focused on 
evaluating systematics related to photo-z, and therefore the analysis is not practical on real data because of the simplifying assumption that
every SNIa is perfectly classified (i.e. spectroscopic sample).

A  full-scale analysis involving both photometric classification and 
$\zphot$ with extensive systematics has not yet been done.  In this paper, 
 {we simulate SNIa with host galaxy photometric redshifts and non-SNIa contamination,
and perform an analysis using SN+host photo-$z$s and photometric classification.}
The historical trajectory and evolving methodologies of SNIa cosmology research are summarised in Table~\ref{tab:sn_history}. 
 {While we anticipate spectroscopic resources for a small fraction of the events (eg. 4MOST TiDES \citep{4most_tides}), 
here we exclude all spectroscopic information in the high-$z$ simulation so that the analysis is
more sensitive to potential biases from a photometric analysis.}

\begin{table*}[ht]
\centering
\begin{tabular}{|l|l|l|}
\hline
\textbf{Classification} & \textbf{Redshift} & \textbf{Survey/Project} \\ \hline \hline
Spectroscopic & Spec Only (SN or Host) & \parbox{7cm}{Two decades of results\\
Hi-z \citep{adam}, SCP \citep{perl}\\
SNLS \citep{SNLS:2005qlf}, ESSENCE \citep{ESSENCE:2007acn}, SDSS-II \citep{2009ApJS..185...32K}, PS1 \citep{panstarrs}, RAISIN \citep{Jones:2022mvo},\\
Pantheon \citep{2018ApJ...859..101S, Scolnic2018}, JLA \citep{Betoule2014}, UNION3 \citep{Rubin:2023ovl}, DES-3YR \citep{DES3YR} etc.} \\ \hline
Photometric & Spec Only (SN or Host) & \parbox{7cm}{PS1 \citep{Jones2018}, DES-SN5YR \citep{DES:2024hip}, Amalgame \citep{Popovic24}} \\ \hline
Spectroscopic & Spec or Photo (Host Only) & \parbox{7cm}{DES-5YR + redMaGiC \citep{Chen2022}, LSST forecast [M23] \citep{Mitra:2022ykq}.} \\ \hline
Photometric & Spec or Photo (Host Only) & \parbox{7cm}{Roman forecast \citep{Kessler:2025eib} and \textbf{this work}.} \\ \hline
\end{tabular}
\caption{History of SNIa-based $w$ measurements, including systematics, from previous surveys and from forecasts for future surveys. The history starts with spectroscopic redshift and classifications and evolves to photometric classifications and redshift.}
\label{tab:sn_history}
\end{table*}

This paper is organised as follows, in sec.~\ref{sec:sim} {,} we describe the simulation process
 to generate the data and  bias corrections. 
In sec.~\ref{sec:anaysis} {,} we describe the main analysis of this paper 
including light curve fitting,  selection cuts, systematic uncertainties, photometric classification  using SCONE, 
 {and BBC to produce a bias-corrected Hubble diagram (HD).}
 {In Sec.~\ref{sec:cosmology}, the HD is}  used for  
dark energy  {parameter} estimation. 
Finally in sec.~\ref{sec:conclude} {,} we present the conclusion of this analysis.


\section{Simulated Data}
\label{sec:sim}

\subsection{Overview}
Ideally, we would work with simulated images from the Data Challenge 2 (DC2) \citep{dc2} \citep{2020MNRAS.497..210S} or 
OpenUniverse2024 \citep{LSSTDarkEnergyScience:2025lah},
and obtain light curves from
the LSST difference imaging analysis (DIA)\footnote{\URLDIA} 
based on \citet{DIA1}.
While such an analysis is most realistic, it is very CPU intensive and repeating the analysis is not practical.
Instead,  {we generate catalogue level simulations of lightcurves  corresponding to  }  the output of DIA   {using a two-detection trigger}.

We simulate and analyse \NSAMPLE\ independent samples, where each sample includes two subsets:
\textbf{(1)} a complete low-$z$ $(z\le 0.08)$ sample of {\spec}ally confirmed SNe~Ia with accurate \spec\ redshifts, and \textbf{(2)} a high-$z$  $(z< 1.55)$ SN~Ia sample with non-SNIa contamination from SNII/Ib/Ic and peculiar SNIax and 91bg-like.  The high-$z$ subset includes host-galaxy photo-$z$'s, but does not have \spec\ redshifts nor identification. 

The low-$z$ cadence  is  from the the Wide Fast Deep (WFD) observing strategy from DC2. 
The high-$z$ cadence is from ${\sim5}0$~deg$^2$ Deep Drilling Fields (DDF)\footnote{\URLDDF} 
 {used in the PLaSTiCC data challenge \citep{plasticc_H2020, plasticc_K2019},\footnote{More recent LSST cadences are available \href{https://www.lsst.org/content/charge-survey-cadence-optimization-committee-scoc}{here}}
 {and corresponds to 3 years, between $59570$-$60675$ MJD days.}} 
 {While we anticipate that the number of supernova in WFD will significantly exceed that in DDF, 
here we focus on DDF to better study  {photometric methods}, 
 {and to forecast constraints for a potential early science analysis using a few years of DDF where we
expect templates to be available well before WFD.}
}  


To model host properties and incorrect host associations, we use host galaxy libraries ({\HOSTLIB}s) from the \acro\ (Extended LSST Astronomical Time Series Classification Challenge) \citep{2023AAS...24111701N}. The simulation for SNIa and detection efficiency   is the same in both subsets.

\subsubsection*{ELaSTiCC}
\label{sec:elasticc} 
To help prepare \spec\ follow-up for transients and hosts found by Rubin,
\acro\ was developed to enable early and accurate photometric classification of transients using both light curves and contextual information \citep{Foley:2013jrk,Baldeschi:2020whr, Gagliano:2020ucg}. 
\acro\ includes  $\sim$5 million detected transients and $\sim$50 million alerts across 30+ classes, 
embedding realistic transient–host correlations from the CosmoDC2 catalog \citep{korytov}. 
 {While we don't directly use the final \acro\ simulations here, we leverage the host galaxy library  developed  for \acro\ which includes host-SN correlations, and we include a few 
 modifications
 
to model host mis-associations.}
%


\subsection{Implementation}
\label{sec:snana}
The SNANA  {(Supernova Analysis package)} \footnote{\url{https://github.com/RickKessler/SNANA}} \footnote{We used SNANA version 11-5t} simulation includes  three main steps
as described in \citet{kessler2019}:
\textbf{(a)} estimate the true source magnitude from a rest-frame SED model that incorporates
redshifting and cosmic expansion ($\OM=0.315$, $w=-1$, flatness),
lensing magnification, peculiar velocity, Milky way extinction, and 
host galaxy extinction; 
\textbf{(b)}  model the flux uncertainty and random fluctuations using
sky noise,  zero-point, and PSF 
for each observation;
\textbf{(c)} model DIA detection efficiency vs. SNR,  {and trigger.}

We simulate SNe~Ia  using the SALT3 SED model \citep{salt,salt3}
with a NIR extension from \citet{Pierel:2022pqc} 
 {that covers wavelength range 2,000-20,000\AA\ in the rest-frame. 
To avoid classification artifacts from $u$-band dropouts at high-$z$,
the SED model is linearly extrapolated down to zero flux at 500\AA.}

 {To model instrinsic scatter about the mean SALT3 model, } we use a dust based  model from  \citet{BS20}, along with the stretch and colour populations from  \citet{Popovic:2021yuo}.

The stretch-luminosity parameter is $\alphaTrue=0.15$, and the intrinsic color-luminosity parameter 
$\betaIntTrue=2.12$.
The combination of intrinsic color variation and dust results in an effective fitted $\beta\sim 3$  {in the analysis}  
(sec 3.4  in \citet{DES:2024hip}),
but there is no true effective $\beta$ value to compare with.

Correlations between SNe and host-galaxy mass are modelled using \acro\ {\HOSTLIB}s and probability distributions \citep{lokken}. 
 {For the peculiar velocity, the assumed 
 {distribution is Gaussian with $\sigma=300$~km/s.}
 {In the analysis, we correct redshifts using the true peculiar velocity
with an uncertainty of $200$~km/s.}
}

The average $5\sigma$ limiting magnitudes per co-added night,  and their average time between nights for each of the \bands\ passbands, are listed in Table~\ref{t_lowz} 
for both the low-$z$ (WFD) and the high-$z$ (DDF) subsets. On average the DDF is  $\sim 1.0$ mag deeper  than the WFD, with $g$ and $r$ being the deepest bands. Compared to DDF, the WFD cadence has $30\%$ fewer co-added observations on average.   For  host galaxies  we use a $1$ year co-added depth which corresponds to a depth of $i \sim 25.4$.   

 {
A simulated event is recorded if it passes the 2-detection trigger,
where any two observations, separated by at least 30 minutes, are detected
based on the DIA efficiency vs. detection in Fig. 17 of \citet{Sanchez}.
The magnitudes where the DIA efficiency is 50\% are $23.66, 24.69, 24.06, 23.45, 22.54, 21.62$
for $u,g,r,i,z,Y$ bands, respectively.}

\subsubsection*{Rates}

For each SN type in the high-$z$ sample, we use the redshift range, survey duration, sky area, and
volumetric rate to compute the physical rate defined as the true number of events without instrumental selection.

For SN~Ia, we adopt a redshift-dependent volumetric rate model,
\begin{eqnarray}
     R(z)  & = & 2.5  \times 10^{-5} (1+z)^{1.5}~{\RateUnit} ~~(z<1) \label{eq:rate_lowz} \\
     R(z) & = & 9.7  \times 10^{-5} (1+z)^{-0.5}~{\RateUnit} ~(z>1)~ \label{eq:rate_highz}
\end{eqnarray}
based on \citet{Dilday2008} at $z<1$ and \citet{Hounsell2018} at $z>1$. 
 {The $z>1$ rate is a continuous extension that decreases with increasing redshift and is consistent with the large rate uncertainties in \citet{2014AJ....148...13R}  and  \citet{Strolger:2015kra}}

For the low-$z$ sample, we do not use measured rates; instead, we arbitrarily generate $4200$ SNe~Ia, with a redshift distribution following Eq.~\ref{eq:rate_lowz}, such that the number of selected events is roughly 10\% of the high-$z$ Ia sample.

Peculiar SNe are simulated following \citet{plasticc_K2019} and \citet{Vincenzi2021}. The rate of 91bg-like SNe is set to $\sim$12\% of the Ia rate with the same redshift dependence as Eq.~\ref{eq:rate_lowz}. The Iax rate follows the cosmic star-formation history \citep{Madau:2014bja}, normalized to $6\times 10^{-6}$~yr$^{-1}$~Mpc$^{-3}$ at $z=0$.

Core-collapse SNe rates are based on \citet{Strolger:2015kra} and \citet{Shivvers:2016bnc} (see also \citet{DES:2024hip, Vincenzi2021}). We define $\rcc$ as the green curve in Fig.~6 of \citet{Strolger:2015kra}, with Type II and Type Ibc rates taken as $0.7\times \rcc$ and $0.3\times \rcc$, respectively.

\begin{table}
\begin{center}
\caption{Average depth and time between observations.}
\begin{tabular}{ | c | c  c | c  c |} 
 \hline
          & \multicolumn{2}{c|}{WFD} & \multicolumn{2}{c|}{DDF} \\
 Filter   & depth\tablenote{$5\sigma$ limiting magnitude.}   
          & gap\tablenote{Average time (days) between visits, excluding seasonal gaps.}         & depth   & gap \\   
 \hline
    $u$ & $23.84$ & $10.5$  &  $25.05$ & $5.3$\\
 \hline
   $g$  & $24.80$ & $11.9$&  $25.52$ & $7.3$\\
 \hline
    $r$ &$24.21$ &$8.2$ & $25.60$ &$7.3$ \\
 \hline
   $i$ & $23.57$ & $8.6$&  $25.19$ & $7.3$\\
 \hline
    $z$&$22.65$  &$9.0$      &$24.79$  &$7.3$ \\
 \hline
    $Y$&$21.79$  &$11.2$  &$23.83$  &$7.4$ \\
 \hline
\end{tabular}
\label{t_lowz}
\end{center}
\end{table}

\subsubsection*{Host Galaxies }
\label{sec:wgtmap}
 {We use the host galaxy libraries from \citet{lokken} for our study. Starting from the   {transient-}host  correlations  measured across ~$16,000$ galaxies in Pan-STARRS DR1 \citep{Gagliano:2020ucg, lokken}   {they} employed an approximate nearest neighbour algorithm to construct   {a} synthetic library containing   {a few} million CosmoDC2 galaxies that preserve these observed correlations \citep{korytov}.}
   {This procedure was applied to construct four separate host libraries ({\HOSTLIB}) for: 
SNIa,  II, Ibc and    {peculiar SNIa (Iax and 91bg)}.}
  To select the true host and possible host mismatch, all potential galaxies within a $10^{\prime\prime}$  radius are included for the $d_{DLR}$ analysis described in ~\ref{sec:host_gala_misassociation}. Roughly   half the galaxies have no neighbours within $10^{\prime\prime}$,  $\sim 1/3$  have  a single neighbour and  the remainder have 2 or more neighbours.

  To describe the brightness profile of each galaxy, a single Sersic profile was estimated from \citet{Gagliano:2020ucg}.  The \HOSTLIB\ Sersic  parameters are assumed to represent the true galaxy profile, without instrumental smearing. A true SN location  near a host galaxy is overlaid using this profile. 
To incorporate PSF effects we smear with an average 
 FWHM  $\sim 1.0^{\prime\prime}$.
Thus there are two Sersic profiles: one for selecting SN location and the other for computation of $d_{DLR}$ in the analysis.

 Since each  \HOSTLIB\ is weighted based on the transient type {,} the galaxy density is not accurate and therefore the neighbour fractions are not accurate. To have a host mismatch fraction that is similar to   \desfv\  \citep{DES:2023tfm}, we increase the Sersic size by  a factor of three. To avoid this artifact in future simulations,   an unweighted \HOSTLIB\ should be used eg. the diffsky\footnote{\URLDIFFSKY} catalogue from {\tt OpenUniverse2024}.

  {The photo-$z$ PDFs are true posteriors, derived jointly from the underlying redshift-photometry relation of the transient-specific host catalogs, rather than being the output of any specific photo-z estimators that come bundled with their own assumptions of priors/training sets that may be unrealistically well-matched to the test set relative to what we will have with real, non-simulated data.
This underlying redshift-photometry relation was modeled using a normalizing flow with the \texttt{PZFlow} package \citep{2024AJ....168...80C}.}  
  {For compact storage in the {\tt HOSTLIB},}
the photo-$z$   {PDF}   {is converted to}
$11$ quantiles 
corresponding to an integrated cumulative density function's probabilities of $0,\ 10\% ...100\%$.

  We define three metrics using the quantity $\Dz = |\zphot-\ztrue|/(1+\zphot)$: 
  \begin{enumerate}
      \item $\Dzbias=$ mean bias on $\Dz$,
      \item $\sigIQR=$ RMS  of the inter quartile distribution of $\Dz$ divided by 1.349,
      \item $\fout=$ the outlier fraction    of events satisfying $|\Dz| > 0.1$.
  \end{enumerate}
  {Our metric values are } $\Dzbias=0.001$,  $\sigIQR=0.127$ and $\fout=0.46$   {(see Fig.~\ref{fig:photoz_res}(a))}.

\subsubsection*{Weight Maps}
Unlike the {\tt PLaSTiCC} {\HOSTLIB} used in \citet{Mitra:2022ykq}, the use of \acro\ {\HOSTLIB}s enables modelling of the   host galaxy correlations. Here we weight each transient host as a  function of log of stellar mass ({\tt LOGMASS}) and the star formation rate ({\tt SFR}). The corresponding probability distribution  is called a  {\tt WGTMAP}. 
 {A {\tt WGTMAP}  for each transient class is provided by \citet{Gagliano:2020ucg} who used catalogue data from Pan-STARRS  \citep{ 2016arXiv161205560C}.}

For each transient class, Fig.~\ref{Fig:host_sfr} shows the generated number of events as a function of  
  {observed}\footnote{  {Ideally, the true {\tt SFR} and {\tt LOGMASS} would be used in the simulation,
and observed values used in the analysis (Sec.~\ref{sec:anaysis}). Our simplification of using observed distributions in 
the {\tt WGTMAP} is expected to have a very small effect.} }  
{\tt SFR} and {\tt LOGMASS}  for the host galaxies selected by {\tt SNANA}. The green points  show a representative sample   from the  \acro\ \HOSTLIB, and the yellow  points show the host galaxies  selected by the {\tt WGTMAP}. For all transient classes, the {\tt WGTMAPS} preferentially select higher {\tt SFR}, and for the $91$bg {\tt WGTMAP} preferentially selects higher {\tt LOGMASS}.

\begin{figure*}[htp]
    \centering
    \begin{minipage}[b]{0.29\textwidth}
        \centering
        \includegraphics[width=\linewidth]{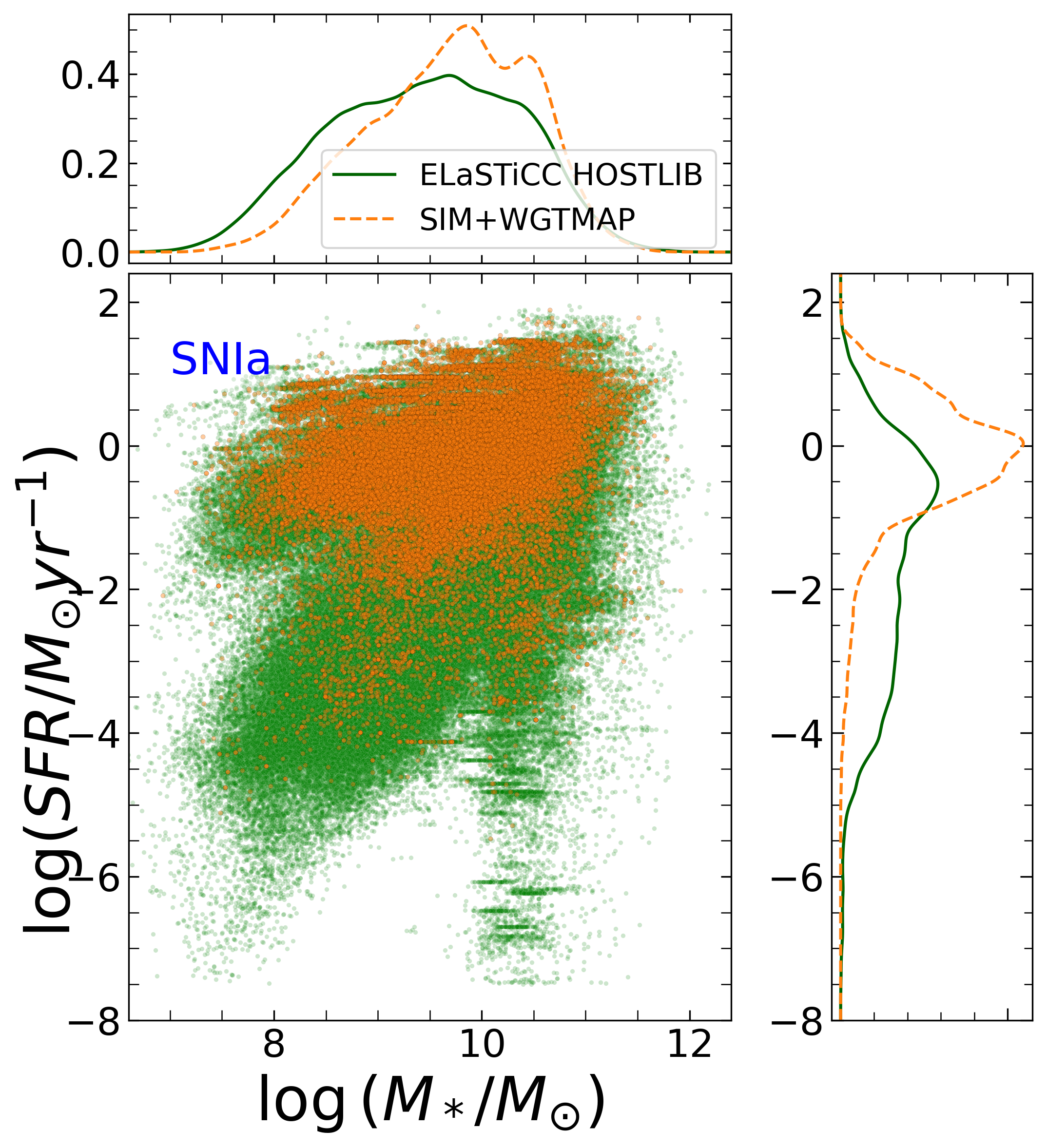}
        \subcaption{SNIa}
    \end{minipage}
    \hspace{0.02\textwidth} 
    \begin{minipage}[b]{0.29\textwidth}
        \centering
        \includegraphics[width=\linewidth]{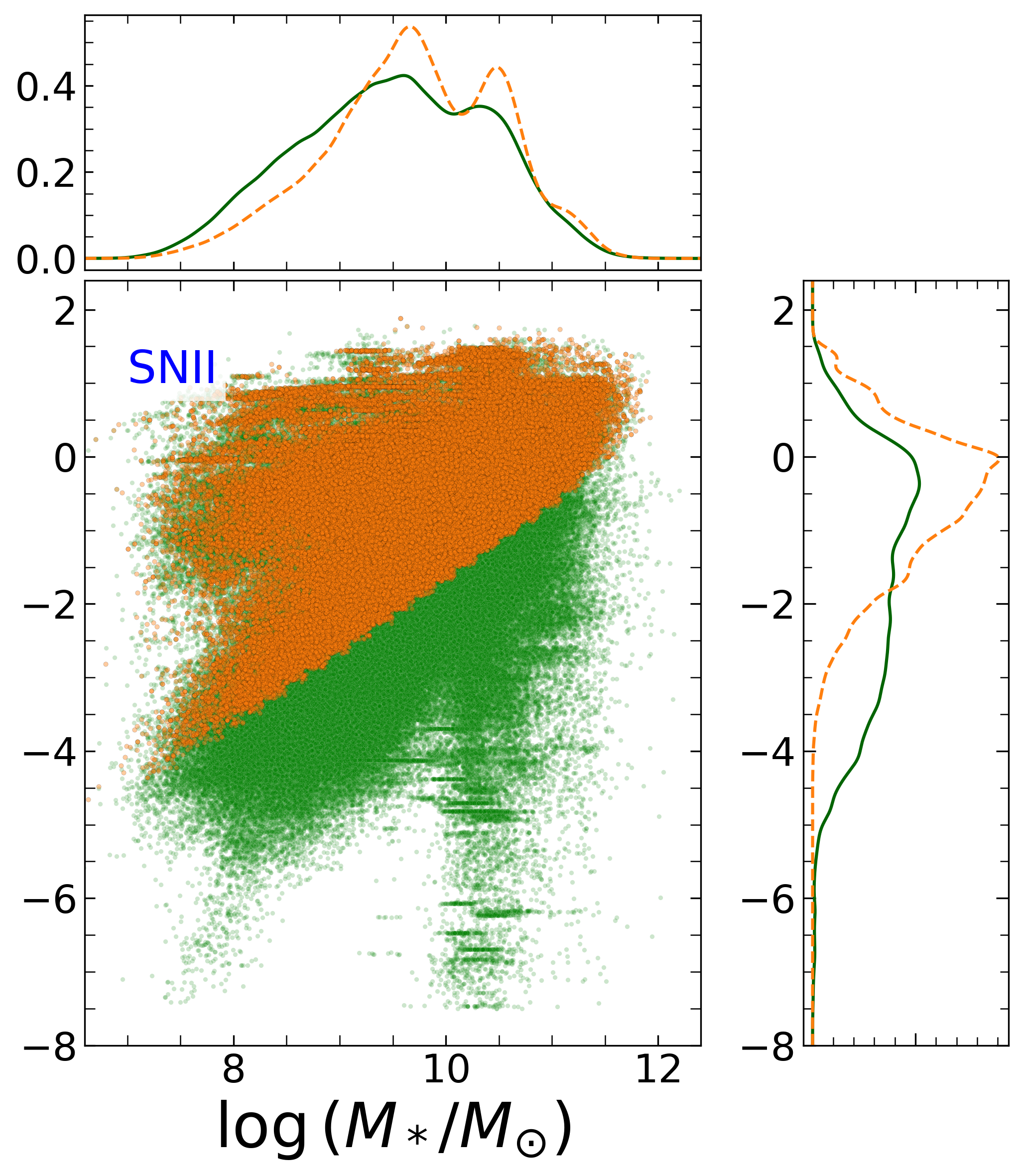}
        \subcaption{SNII}
    \end{minipage}
    \hspace{0.02\textwidth}
    \begin{minipage}[b]{0.335\textwidth}
        \centering
        \includegraphics[width=\linewidth]{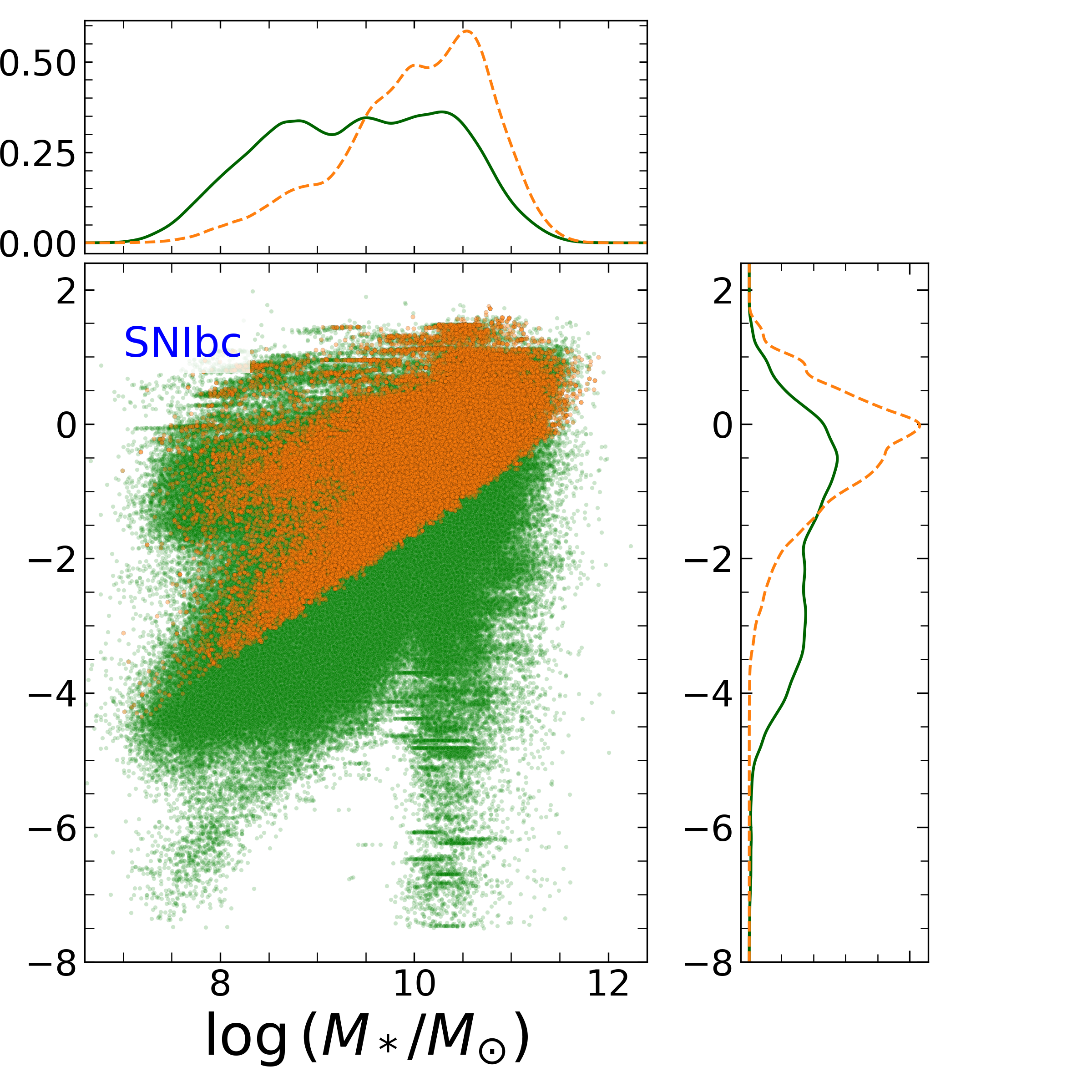}
        \subcaption{SNIbc}
    \end{minipage}

    \vspace{0.02\textwidth} 

    \begin{minipage}[b]{0.31\textwidth}
        \centering
        \includegraphics[width=\linewidth]{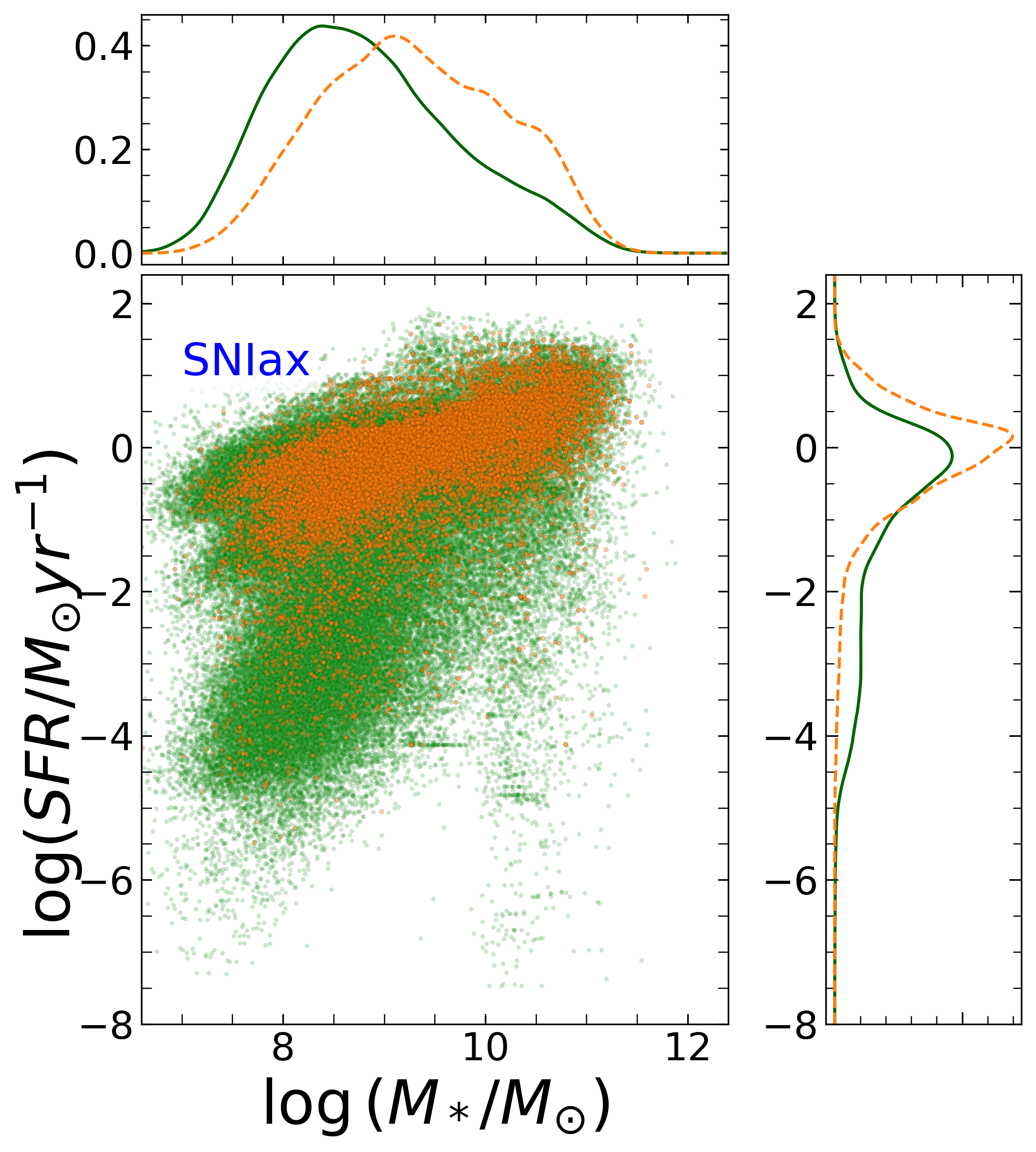}
        \subcaption{SNIax}
    \end{minipage}
    \hspace{0.02\textwidth}
    \begin{minipage}[b]{0.31\textwidth}
        \centering
        \includegraphics[width=\linewidth]{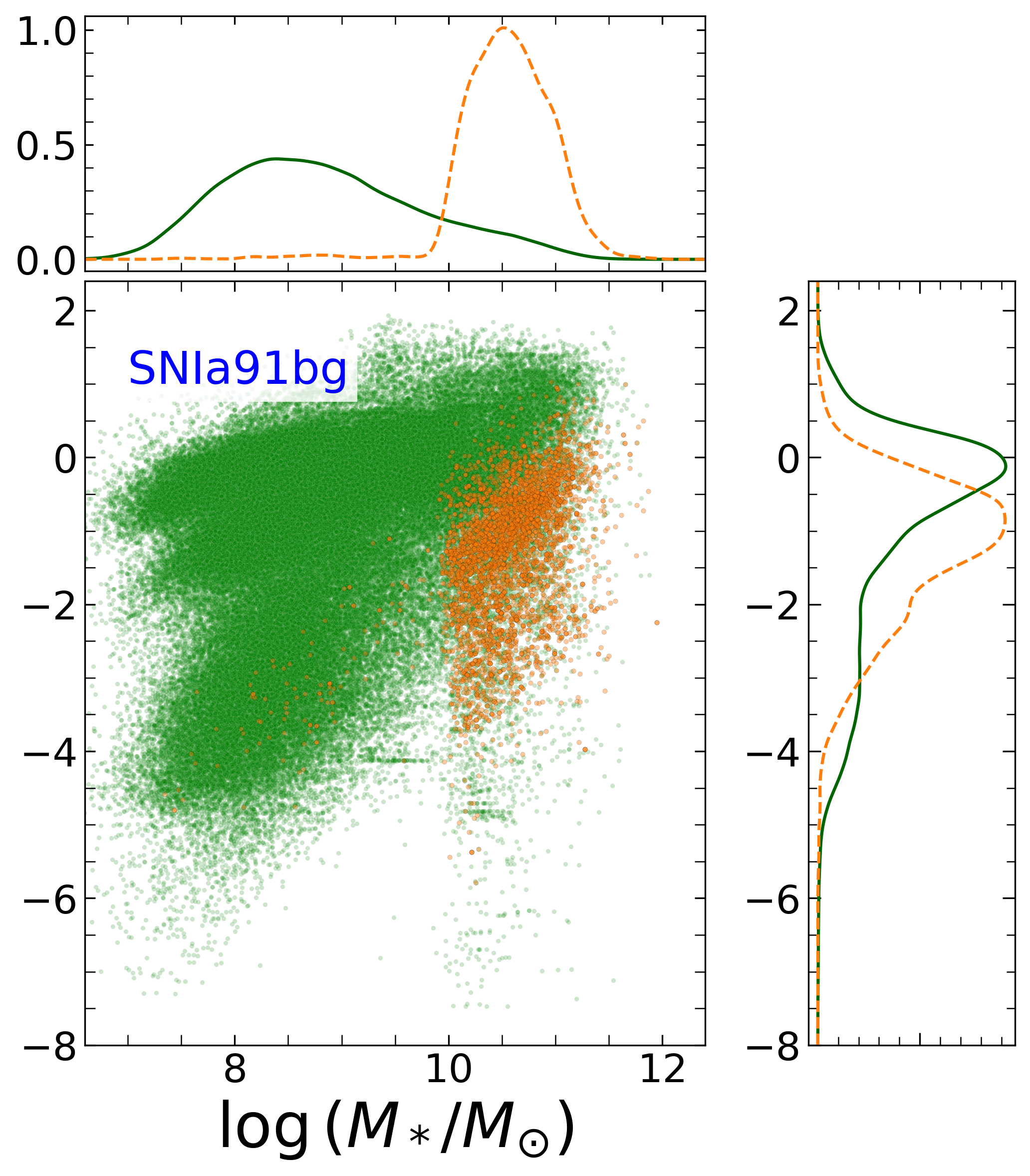}
        \subcaption{SNIa91bg}
    \end{minipage}
\caption{Density of simulated galaxies as a function  {of} {\tt SFR}  {and} {\tt LOGMASS}. 
Each panel shows a different transient class. 
The yellow (green)  points correspond to \acro\ {\tt HOSTLIB} with (without) the {\tt WGTMAP}. 
To avoid saturation we show a random $10\%$ of the \HOSTLIB.  }
\label{Fig:host_sfr}
\end{figure*}

\begin{table}[ht!]
\centering
\caption{Summary of simulation statistics vs. sample and transient type.}
\renewcommand{\arraystretch}{1.2}
\begin{tabular}{|c|c|c|c|c|}
\hline
\multirow{3}{*}{\textbf{Sample}} & \multirow{3}{*}{\textbf{Transient}} & \multicolumn{3}{|c|}{\textbf{Number of events}}\\\cline{3-5}
 &  & \textbf{Total}  & \textbf{Passed} & \textbf{Passed} \\
 & & \textbf{generated}\footnotemark[1] & \textbf{trigger}\footnotemark[2] & \textbf{cuts}\footnotemark[3]\\
\hline\hline
Low-$z$  &  \multirow{2}{*}{Ia}  & \multirow{2}{*}{4200}  & \multirow{2}{*}{1133}   & \multirow{2}{*}{691 (0.165)}\\
 $<0.08$      &   &    &   & \\\hline\hline
\multirow{3}{*}{High-$z$} & All   & 233703 & 23479  & 5034 (0.022)\\
   & Ia & 35719  & 11393 & 4389 (0.13)\\
\multirow{3}{*}{ < 1.55} & Iax & 18909  & 579   & 141 (0.008)\\
   & 91bg & 4798   & 399   & $2\ (\sim 0)$\\
   & II & 121994 & 8643  & 313 (0.003)\\
   & Ibc & 52283  & 2465  & 189 (0.004)\\
\hline\hline
 \textbf{Final} & \multicolumn{4}{c|}{\textbf{5080 (Ia) + 645 (CC) = 5725}} \\
\hline\hline
\end{tabular}
\label{T4}
\footnotetext[1]{ Total number of generated SNe}
\footnotetext[2]{Two or more detections separated by more than $30$ minutes. }
\footnotetext[3]{After selection cuts; numbers in `()' are fractions with respect to  ``Total generated'' column. }
\end{table}

\bigskip

\begin{figure*}[tb]
    \centering
    \makebox[1.\textwidth]{\includegraphics[width=1.0\textwidth]{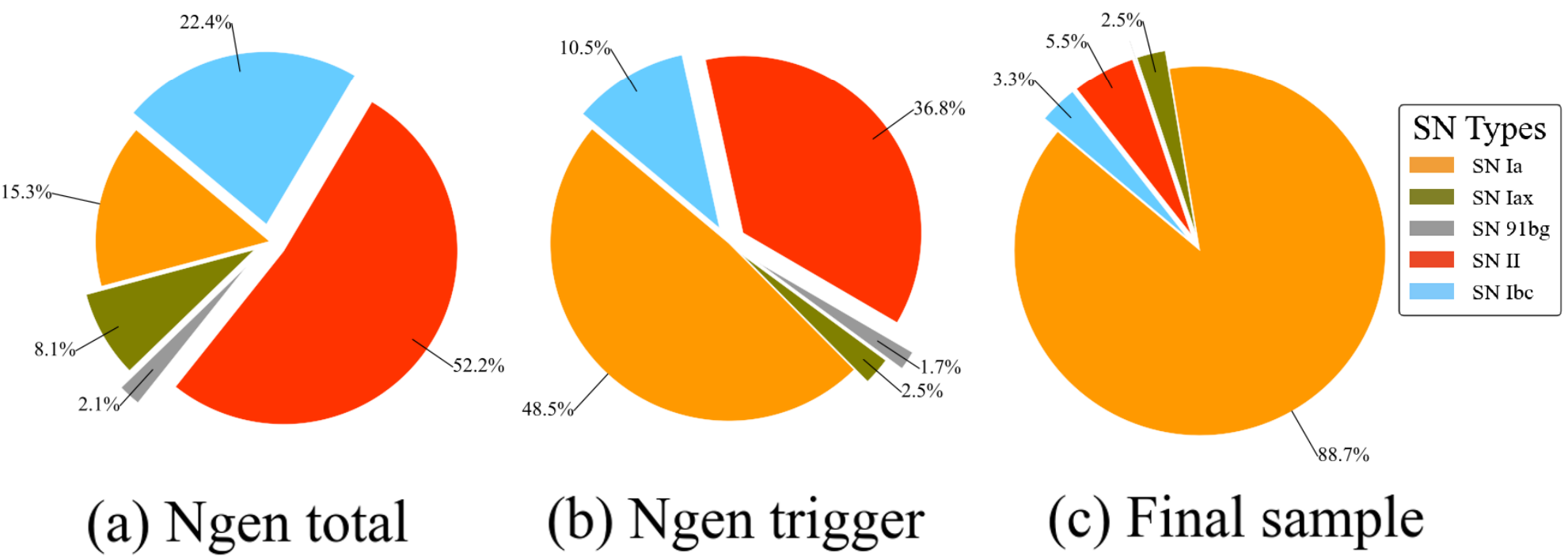}}
    \caption{ Pie chart showing the percentage of true SN  types.  
    \textbf{(a)} Generated  {physical rate}, 
    \textbf{(b)} Passes trigger and 
    \textbf{(c)} After analysis cuts ( last three columns of  Table~\ref{T4}). 
    The evolution of Ia fraction (orange) from left to right illustrates the increasing purity 
    with detection and analysis. }
    \label{fig:sn_pie}
\end{figure*}

\subsubsection*{Sample Sizes }
\label{subsec:ana_simbias}

A summary of average simulation statistics is shown in Table \ref{T4}.  For low-$z$ we generate $4200$ events, and  $\sim 1/4$ satisfy the $2$-detection trigger. For the high-$z$ sample we generate nearly $36,000$ SNIa, and  $\sim 1/3$ satisfy the trigger. The high-$z$ sample also includes  nearly $200,000$ non-SNIa, and  $\sim 6\%$ satisfy the trigger. Compared to the non-SNIa, the SNIa trigger efficiency is much higher because these events are significantly brighter. 

To visualize the relative number of events in the high-$z$ sample Fig.~\ref{fig:sn_pie} shows several pie-charts corresponding to  Table~\ref{T4}. 
Comparing physical rates, SNe~Ia corresponds to $15\%$ of all SNe (Fig.~\ref{fig:sn_pie}a), and after the trigger requirement, nearly $50\%$ of the SNe are type Ia (Fig.~\ref{fig:sn_pie}b). Examples of simulated light curves at different redshifts 
 are shown  in Fig.$1$ of \citet{Mitra:2022ykq}.

To train our photometric classifier (Sec.~\ref{subsec:scone}) and to implement distance bias corrections in BBC   {to account for selection effects} (Sec.~\ref{subsec:ana_bbc}),  {we perform a separate biasCor simulation to generate a large sample of millions of  events.} The simulation procedure is 
 identical to that used for the simulated data. 
When simulated data are replaced with real data, the biasCor simulation is still needed.
 
\section{Analysis}
\label{sec:anaysis}

The SNIa-cosmology analysis steps are shown in Fig.~\ref{fig:analysis},
and described below. This analysis is similar to  M23, with additional complexities, from using  photometric classifier {\tt SCONE}  and the {\tt BEAMS} formalism.

While M23  performed  separate analyses for $\zphot$ and $\zspec$ samples, here we focus our study on  a single $\zphot$ analysis with no spectroscopic redshift or spectroscopic classification
 {in the high-$z$ subset}. 
While we anticipate some spectroscopic resources for LSST-DDF, here we make the most conservative assumption of no resources in order to be more sensitive to potential photo-$z$ related problems in the analysis.

\begin{figure*}[tb]
    \centering
    \makebox[1.\textwidth]
    {\includegraphics[width=1.0\textwidth]{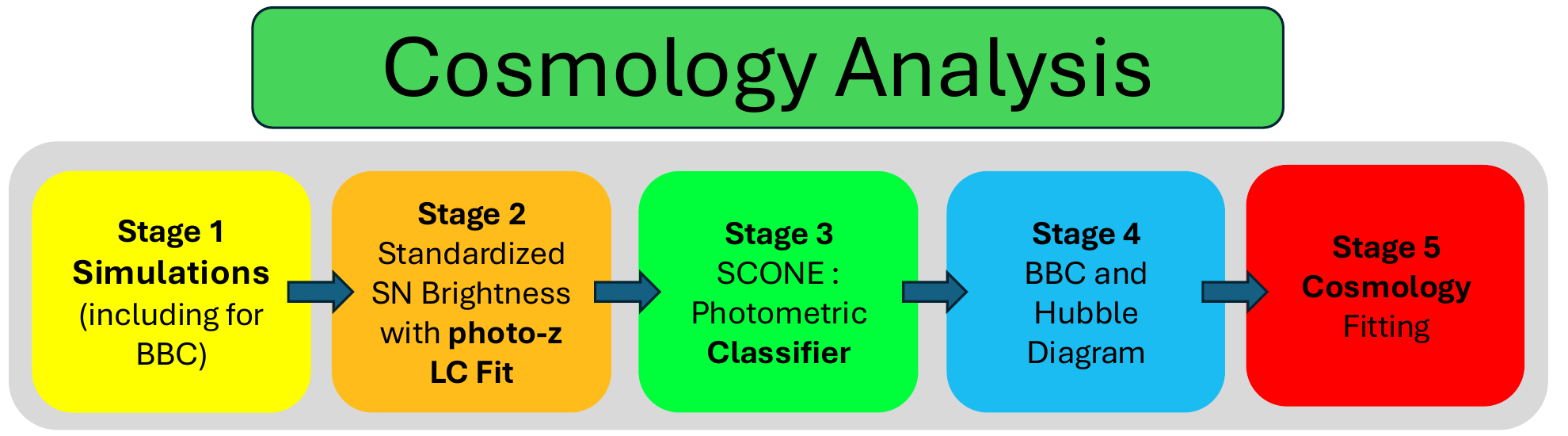}}
    \caption{ Flowchart showing the cosmology analysis steps. }
    \label{fig:analysis}
\end{figure*}

\subsection{Lightcurve Fitting}
\label{subsec:ana_lcfitz} 

To standardize the brightness of SNIa, each light curve is fit
   {with} the same model used in the simulations: 
the SALT3 model in \citet{Pierel:2022pqc} that was trained using the formalism in \citet{salt3}.
This fitting process extracts the following parameters for each event:   time of peak brightness ($t_0$),  amplitude ($x_0$),  stretch ($x_1$),   color ($c$). For the low-$z$ subset we use a fixed redshift $\zspec$. For the high-$z$ subset we float  $\zphot$ as a fifth parameter using the host galaxy $\zphot$ pdf as a prior, and using the fitting formalism in \citet{Kessler2010}    {to measure the SN+Host photo-$z$.} 
SALT3 light curve fits for several events are shown by the smooth curves in Fig. $1$ of M23.

Following Eq.~1 in \citet{Kessler2010}, the 5-parameter SALT3 fit 
uses {\sc minuit} \citep{wfit2} to minimize the following $\chi^2$,

\begin{equation}
    \chi^2 =  \left( \sum_i \chisqflux \right ) + \chisqhost + \chisqsyst ~.
\label{eq:chi2tot}    
\end{equation}

$\chisqflux$  is the flux chi-squared for the $i^{th}$ observation, 
\begin{equation}
\chisqflux =  
       \frac{ \left [\FiData - \FiModel(\xvecSALT) \right ]^2}{\sigFi^2} + \logsigmodel  
    \label{eq:chiflux}
\end{equation}
 where 
 $\FiData$ is the flux for the $i^{th}$ observation,
 $\FiModel$ is the SALT3  model flux computed from the fitted parameters $\xvecSALT=\{t_0,x_0,x_1,c,\zphot\}$,
  {and $\sigFi$ is the quadrature sum of the flux and model uncertainties. 
 Since $\sigFi$ depends on fitted parameters, the 2nd term in eq.~\ref{eq:chiflux} is needed,}
 where $\sigFiTilde = \sigFi$ after  the first fit iteration so that
the $\logsigmodel$ term is close to zero in the second and third fit iteration.

The contribution from the  host galaxy prior is 
\begin{equation}
     \chisqhost = -2\log[\probzhost]
\end{equation}
where $\probzhost$ is the host galaxy probability for the redshift value to be $\zphot$,
  {as determined from the {\tt PZFlow} PDF unpacked from the quantiles.}

The terms $\chisqflux$ and $\chisqhost$ are the conventional terms defined in \citet{Kessler2010} and used in many $\zphot$ studies including M23. Here we add a new term, $\chisqsyst$ to address a subtle artifact 
from fitting with systematics in Sec.~\ref{subsec:ana_cuts_syst}. 
The underlying artifact is that a small systematic perturbation (e.g., changing filter zeropoint) 
can lead to an unphysical large change in SALT3-fitted values that result in either rejecting the
event, or assigning such a large systematic uncertainty that the event is effectively ignored in a cosmology fit.
While this artifact occurs rarely for any one systematic,
the cumulative impact from many systematics can lead to a   significant loss   {(see sec.~\ref{sec:cov})}, 
and this loss is not currently modelled in our BBC method. 
To reduce this loss, we use the reference fitted parameters without a systematic shift as a prior 
in the SALT3 fit for each systematic shift,
\begin{equation}
    \chisqsyst = \left[ (\pk- \pkref ) / \sigkref \right]^2
    \label{eq:chisqsyst}
\end{equation}
where $\pkref$ are the fitted parameters ($k=1,5$) for the reference,
$\pk$ are for fits with a systematic perturbation, and $\sigkref$ is the fitted 
uncertainty from the reference fit.

After each {\sc minuit} fit iteration, the wavelength range for each LSST passband is 
transformed to the rest-frame using fitted $\zphot$.
If the central rest-frame wavelength $\meanlamb$ is outside of the 
SALT3 model range (2800-17000\AA),
the passband is dropped in the next fit iteration;
a previously dropped passband can be added if it is within the model wavelength range.
If any passband is dropped or added, the fit iteration is repeated to ensure that a consistent set of passbands are included in the fit.  

  The fitted photo-$z$ uncertainty  {($\sigma_{\zphot}$)} is not explicitly used in our analysis. Compared to a spec-$z$  fit, the impact of fitting the photo-$z$  is  to increase the uncertainty on the SALT3 fitted parameters.

\subsection{Selection Requirements \& Systematic Uncertainties}
\label{subsec:ana_cuts_syst}

We apply the following selection requirements (cuts) based on previous analyses:
\begin{enumerate}
  \setlength\itemsep{0.1em}
   \item at least three bands with maximum signal to noise ratio SNR$>4$ 
   \item $|x_1| < 3.0$
   \item $|c| < 0.3$
   \item stretch uncertainty $\sigma_{x1} < 1.0$
   \item time of peak brightness uncertainty $\sigma_{t0} < 2.0$~days
   \item $\Pfit > 0.05$\footnote{$\Pfit$ is the SALT3 fit probability 
          computed from $\chi^2$ and number of degrees of freedom.}. No explicit cuts are applied on {\tt SCONE} classification.
   \item  {For low-$z$ sample, $0.01 < \zspec < 0.08$.}     
   \item $0.01 < \zphot < 1.4$ \label{cut_commonlast}
   \item  {SALT3-}fitted bands satisfy $2800 <\meanlamb / (1+\zphot) < 17000 $ the valid range of the rest frame wavelengths ; 
   same cut is applied for  {$\zphot\pm\sigma_{\zphot}$}  \label{cut_fitband}
   \item valid bias correction for all systematic variations (see  Section~\ref{subsec:ana_bbc}). \label{cut_biascor}
\end{enumerate}

 The statistics after cuts are presented in the last column of Table~\ref{T4}, and the transient type fractions  (for high-$z$) are shown in  Fig.~\ref{fig:sn_pie}. The high-$z$ true-Ia fraction is $15\%$ for  physical rates (Fig.~\ref{fig:sn_pie}a),  increases to $\sim 50\%$  after two detection triggers (Fig.~\ref{fig:sn_pie}b), and increases again to $\sim 90\%$ after selection cuts (Fig.~\ref{fig:sn_pie}c).

The  $\zphot$ residual vs. $\ztrue$ is shown in Fig.~\ref{fig:photoz_res}a
for all galaxies in the catalogue, 
and in Fig.~\ref{fig:photoz_res}b for host galaxies
after SN~Ia trigger and selection cuts. 
 {Fig.~\ref{fig:photoz_res}c and Fig.~\ref{fig:photoz_res}d  shows the performance for SN + Host combined fit, broken up into SNIa and non-Ia samples, that are used in the Hubble diagram.}  { 
Both the core resolution and the outlier fraction for Ia  are much  better ($\sim 5$x)  than
for the galaxies in Fig.~\ref{fig:photoz_res}a.}

Figure~\ref{fig:z_dist} shows the redshift distribution in the final cosmology sample for  1 of the \NSAMPLE\ samples, used for the Hubble diagram.  
\begin{figure*}[!ht]
  \begin{minipage}{\textwidth}
    \centering
    \includegraphics[width=.4\textwidth]{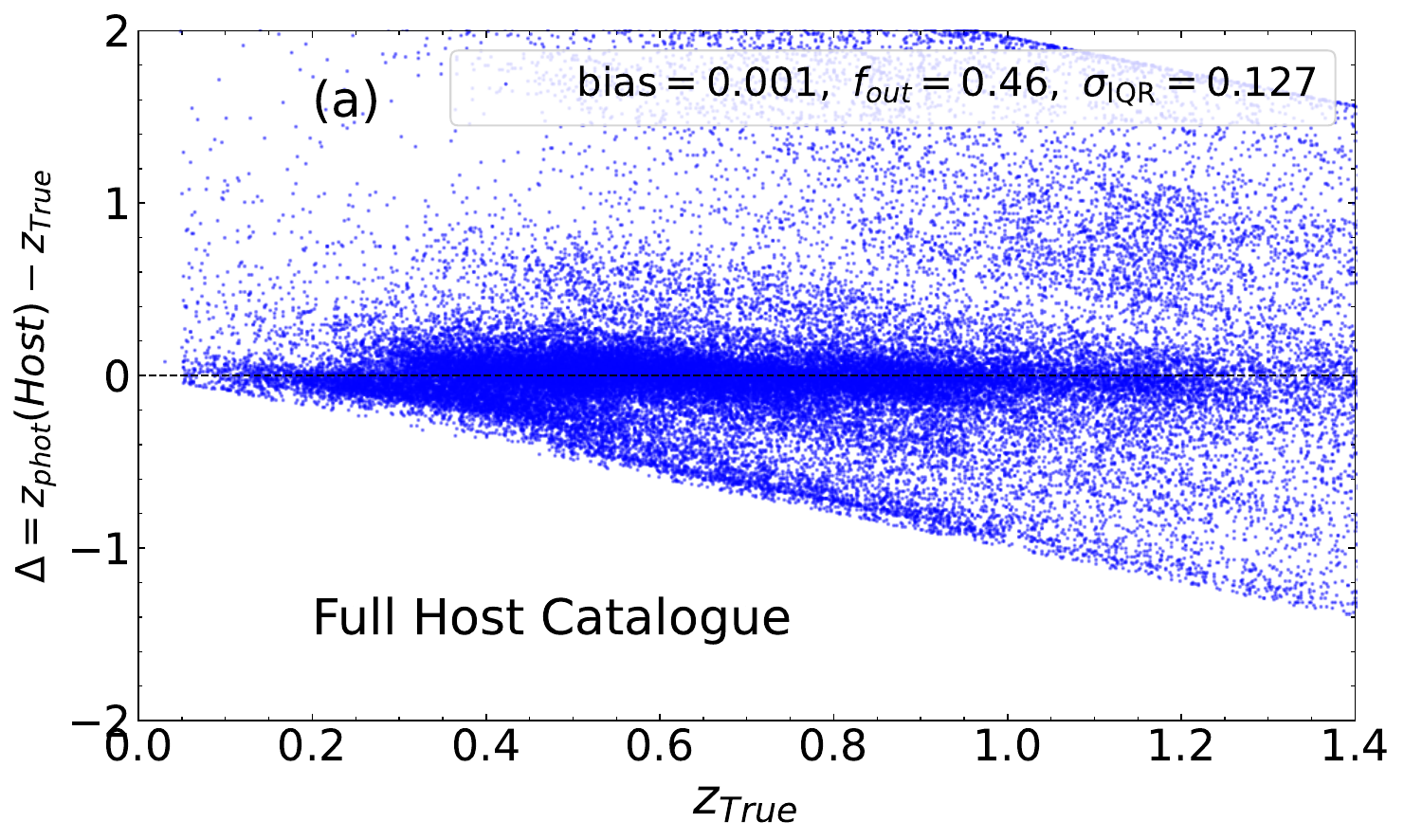}\quad
    \includegraphics[width=.4\textwidth]{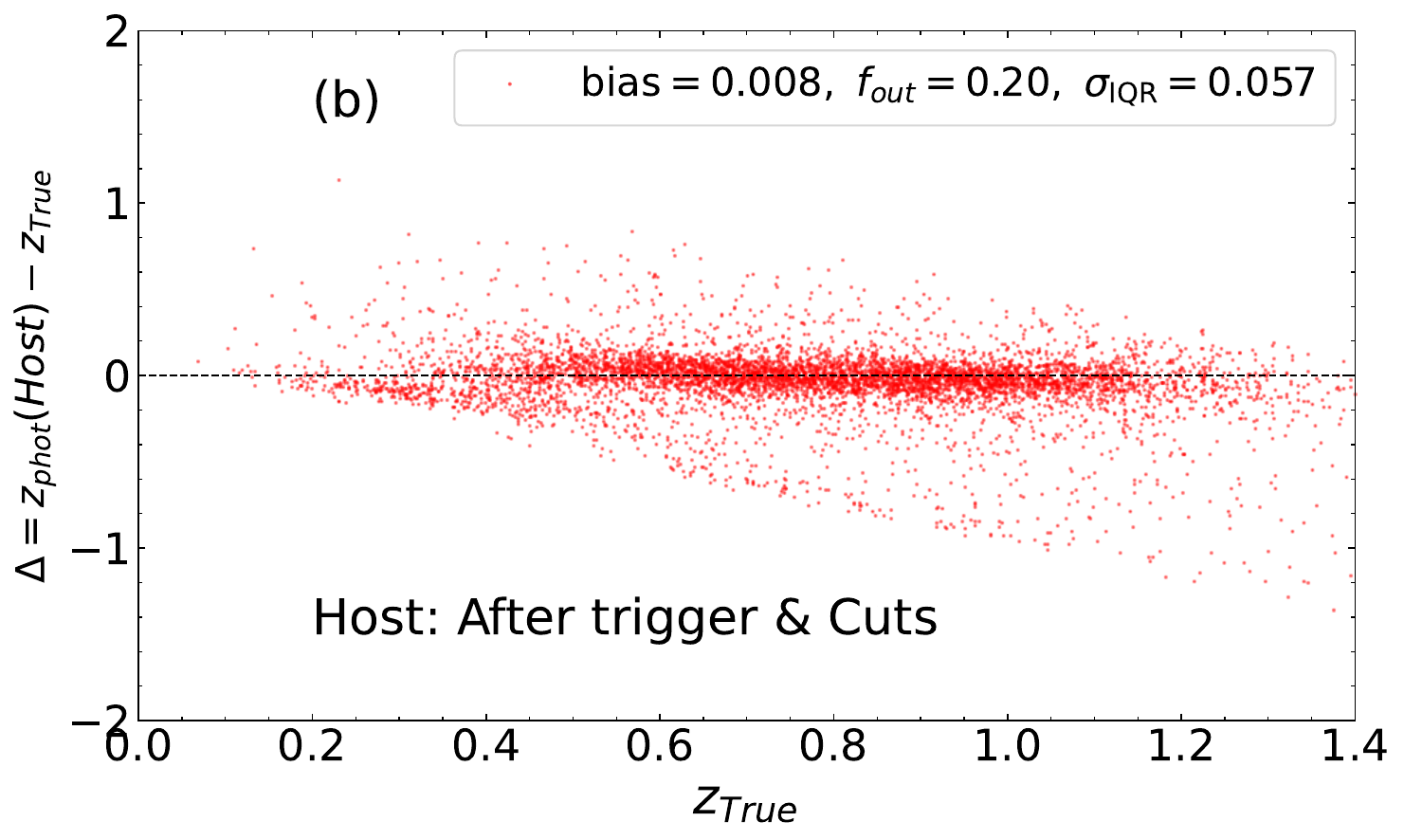}\\
    \includegraphics[width=.4\textwidth]{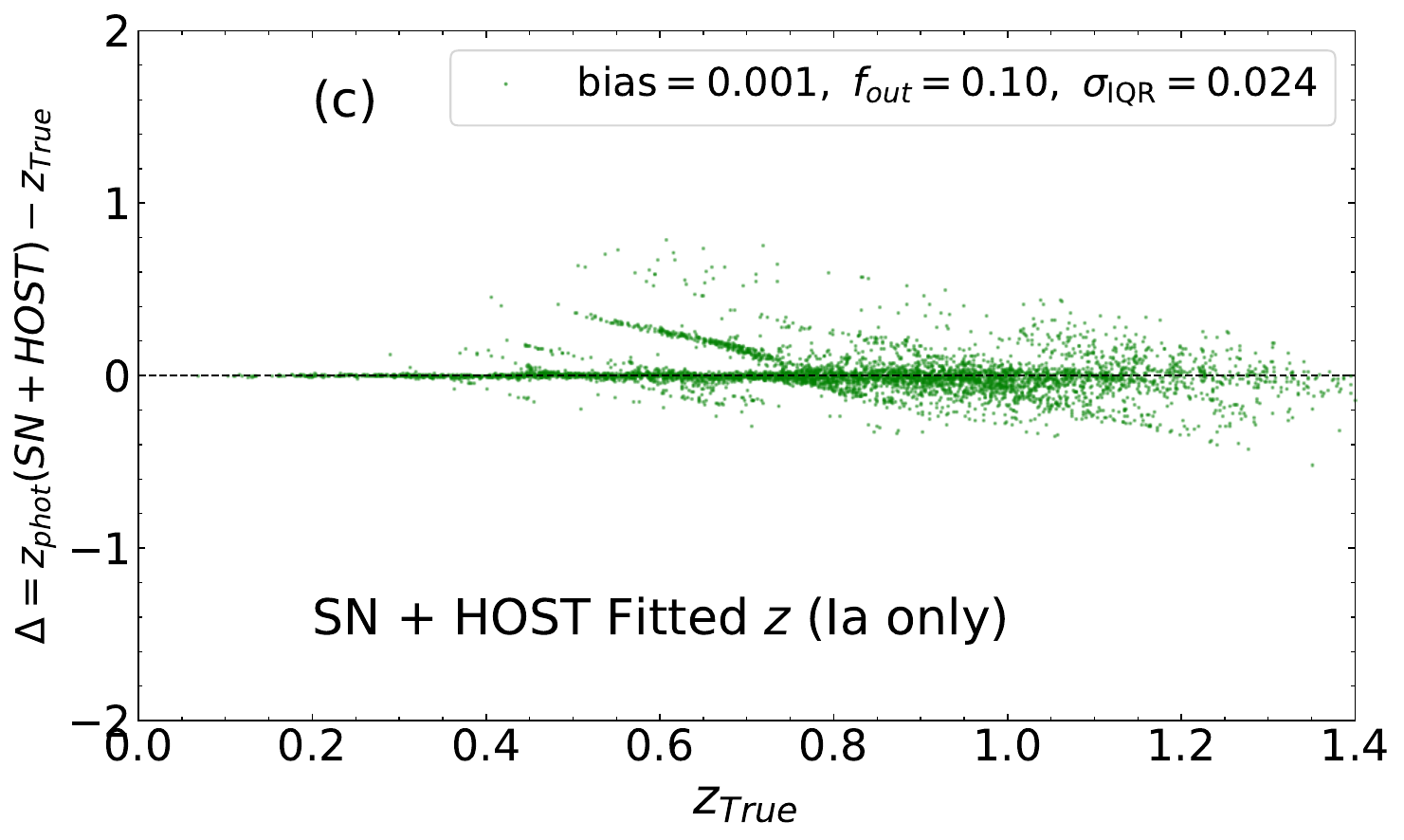}\quad
    \includegraphics[width=.4\textwidth]{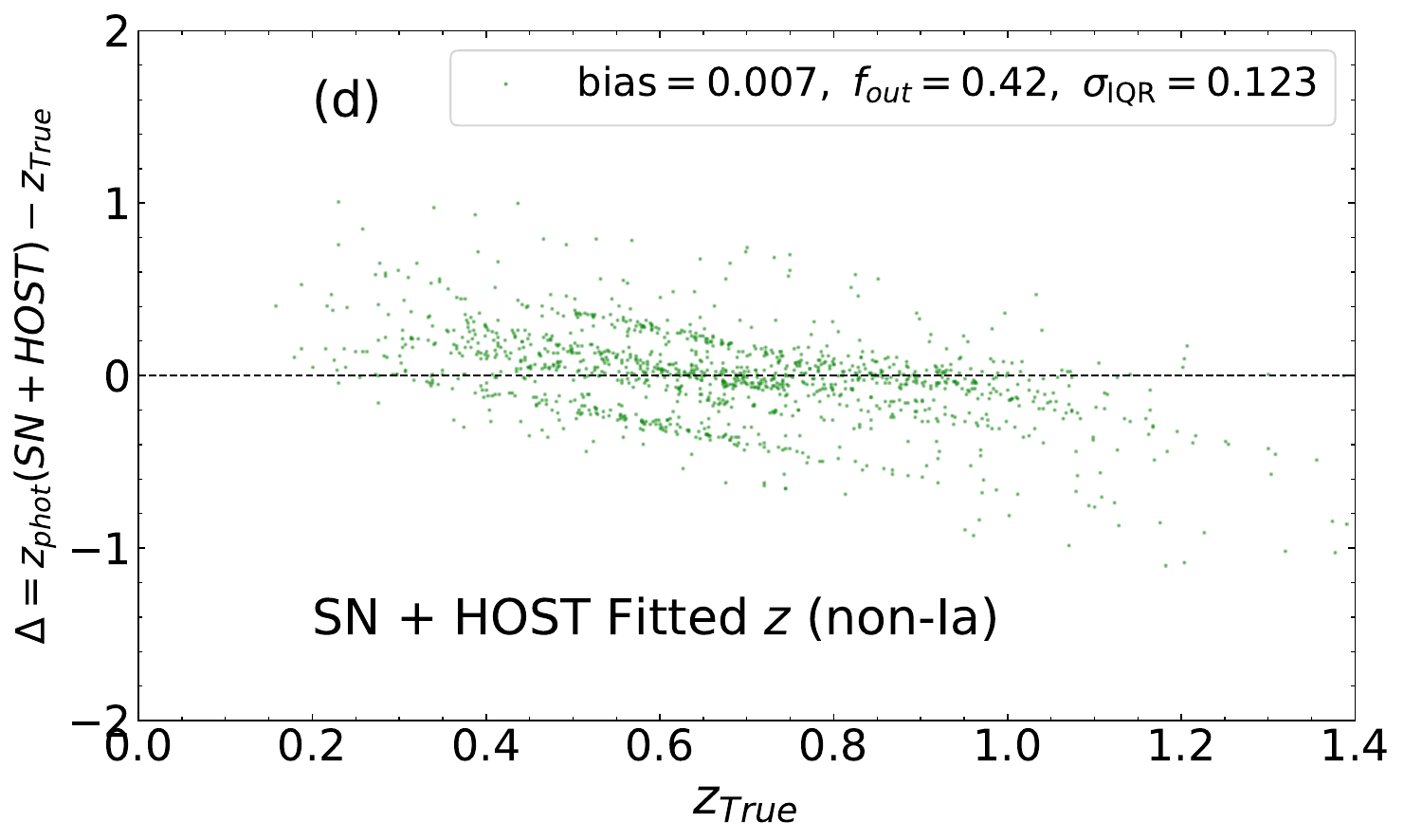}
    \caption{Photo-$z$ residual ($\zphot-\ztrue$) vs. $\ztrue$ for
  (a) full host galaxy catalogue, 
  (b) subset of host galaxy catalogue after trigger and cuts,  
  (c) combined SALT3 fitted SN + HOST photo-$z$ for SNIa only and (d) same but for non-Ia. 
   The $\sigIQR$ and $\fout$ numbers on each panel are computed for $0.4<\ztrue < 1.4$.
  The source of photo-$z$ is indicated on each  {panel.}  } 
    \label{fig:photoz_res}
  \end{minipage}\\[1em]
\end{figure*}

\begin{figure}[h]
    \centering
    \includegraphics[width=\columnwidth]{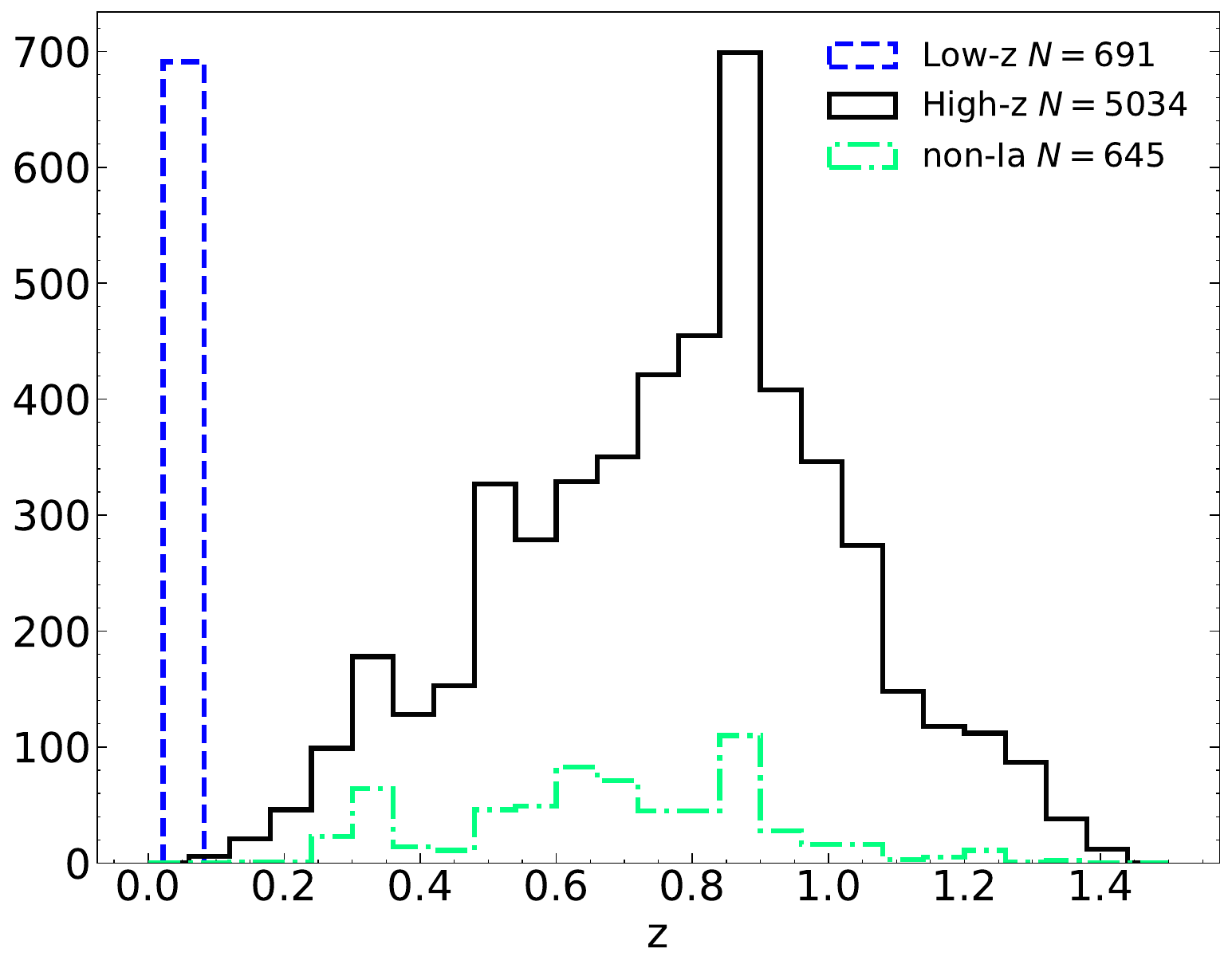}
    \caption{
     {True redshift distributions for 1 of the 25 samples:
    low-$z$ in dashed blue, high-$z$ true SNe~Ia in black, and high-$z$ true non-Ia in green.
    } 
    }
    \label{fig:z_dist}
\end{figure}

 {To evaluate systematic uncertainties, the SALT3 light
curve fits and BBC fit are repeated 7 times, each with
a separate variation shown in Table \ref{tb:syst_sources}. Each variation
results in a distance modulus variation, and we compute
a systematic covariance matrix.}
 {The list of}  systematics considered in this analysis are   the same as those described in Sec.~IV B of \citet{Mitra:2022ykq}.   
The systematics  {applied in this analysis} include variations in galactic extinction, calibration  and host photo-$z$. 
 {The photo-$z$ systematic is a naive coherent shift of $0.01$ based on previous DES weak lensing cosmology  analyses \citep{PZ2021_DES3YR_WL},
but this systematic evaluation needs more effort based on how photo-$z$'s are trained for SN hosts, 
particularly at higher redshifts where the training set is limited. 

Missing systematics include  
\begin{enumerate}
\item   astrophysical model of intrinsic scatter, which dominates the systematic error budget in \desfv ; 
\item   photometric classification (e.g, different classifiers and training sets),
  which contributes $\sim 1/3$  to  the total \DESVYR\ systematic 
  (see Table 7 in \citet{DES:2024hip}); 
\item   {photo-$z$ pdf variations that are more complex than a fixed shift; }
\item   {simulated redshift dependence of volumetric rate;  }
\item   {cosmology parameters used in simulated bias corrections.} 
\end{enumerate}

 {As described in Sec.~\ref{sec:cov}, the events passing cuts from each systematic
must be the same to construct a systematics covariance matrix. To help ensure
common events for each systematic,}
the first \ref{cut_commonlast} cuts  are applied only to the nominal analysis without systematics; for each systematic variation we do not apply these \ref{cut_commonlast} cuts but instead process  the list of events passing cuts from the nominal analysis. For example, if an event has fitted SALT3
color parameter $c=0.299$, and migrates to $c=0.3001$ for a calibration systematic, this event is preserved  without applying cuts that require $|c|<0.3$.

Requiring valid fit bands (cut \ref{cut_fitband}) for each systematic results in 
 {slightly different samples per systematic, and there is}
a $3.3\%$ loss from requiring the same events passing  {this cut.} 
Requiring a valid biasCor (cut \ref{cut_biascor} and Sec.~\ref{sec:bbc}) for each systematic results in a $0.51\%$ loss  {to ensure common events.}

\begin{table}
\caption{Source\rk{s} of Systematic Uncertainty } \vspace{-0.5cm}
\begin{center}
\begin{tabular}{ | l | l | l | c | } 
 \hline
Row  & Label & Decription & Value \footnote{Shift (or scale) applied to simulated data before each re-analysis} \\
 \hline\
 1 & StatOnly & no systematic shifts &  --- \\
 \hline\
 2 &  MWEBV & shift $E(B-V)$ & 5\%  \\ 
 \hline\
 3 &  CAL\_HST  & HST  calibration offset & $0.007\times\lambda$\\ \
  4 & CAL\_ZP & LSST zero point shift & $5$ mmag \\ \
5 & CAL\_WAVE & LSST Filter shift & $5$ \AA\\
  \hline\
6 & zSPEC  & shift $\zspec$ redshifts  &   $5\times 10^{-5}$\\ \
 7 & zPHOT  & shift $\zphot$ redshifts  &   $0.01$ \\ 
 \hline
\end{tabular}
\label{tb:syst_sources}
\end{center}
\end{table}
\subsubsection{Host Galaxy Association}
\label{sec:host_gala_misassociation}
Host galaxy matching is done via the Directional Light Radius (DLR) method 
  { \citep{Sullivan_2010, DES:2016aon}}. 
A normalized, dimensionless parameter \(d_{DLR}\) is defined as the ratio of the angular separation \(\Delta\theta\) between the supernova (SN) and DLR defined as the galaxy's centroid to the galaxy's effective radius in the direction of the SN. DLR is derived from the smeared Sersic profile.
Mathematically, this is expressed as:
\[
d_{DLR} = \frac{\Delta\theta}{\text{DLR}}.
\]
 {The two galaxies with the smallest $d_{DLR}$ are tagged as  $\tt HOST1$ and $\tt HOST2$, where $\tt HOST1$ has the smaller $d_{DLR}$ if $\tt HOST2$ exists.}
A host  match is defined if  {$\tt HOST1$   has $d_{DLR}<4$}.  

 {Defining hostless events for a photo-$z$ analysis is ill-defined because
we cannot perform a photo-$z$ fit on hostless events. Nonetheless, we make a rough
estimate here by applying the SNR and sampling cuts in Sec.~\ref{sec:anaysis}, but no lightcurve fit.
The hostless fraction is   {$2\%$}, which is consistent with the 2\% hostless fraction for \DESVYR\ \citep{DES:2023tfm}.  
If we restrict our
true redshift to the \DESVYR\ range ($z<0.9$), our hostless fraction
stays at 2\% and is still consistent with \DESVYR. For events passing cuts, Table~\ref{tab:host_mis} shows the breakdown of events
with 1 or 2 matches with $\ddlr < 4$ and compares the host galaxy mismatch statistics of 
Host 1 with the numbers reported in \citet{DES:2023tfm}. }
  {Figure~\ref{fig:host_misac} 
illustrates the distributions of DLR and $r$-band magnitude
while showing the Table~\ref{tab:host_mis} nominal (second column) cuts.} 

\begin{table}[htbp]
\centering
\begin{tabular}{|c|c|c|c|}
\hline
 & LSST &LSST&    \\
Condition & SIM &SIM II\footnote{ $z<0.9$,  {to compare with DES} range. 
}
  & DES\footnote{see sec. 5.1 in \citet{Qu:2024fja}.}    \\
\hline
 {$>=1$} Host  {match}\footnote{ {  {atleast one match with DDLR$<4$; for the denominator, } zero matches include events with DDLR$>4$, but does not include
  host-less events}} & 98 &  {98} & 98   \\
$N_{\rm match}{=}2$ all\footnote{ {2 host matches with}  $\ddlr$<4} & 12 & 12  &9  \\ 
$N_{\rm match}{=}2$ wrong\footnote{  Two host matches with $\ddlr <4$, and the galaxy with the smaller $\ddlr$ is the incorrect host. This value is not available for DES. } & 1.0 & 1.1 & ---  \\ 
Host depth  {($i$-band)} & 25.4 && 25.5   \\
\hline
No Host & 2 \% &2 \% & 2 \%   \\ 
\hline
\end{tabular}
\caption{ Statistics of host galaxy (mis)match  (in $\%$) in this analysis (second column).  In the third column we compared it with DES analysis (fourth column) from \citet{DES:2023tfm}.  
}
\label{tab:host_mis}
\end{table}

\begin{figure*}[ht!]
    \centering
    \begin{tikzpicture}
        \node [anchor=south west, inner sep=0] (image) at (0,0) {\includegraphics[width=\textwidth]{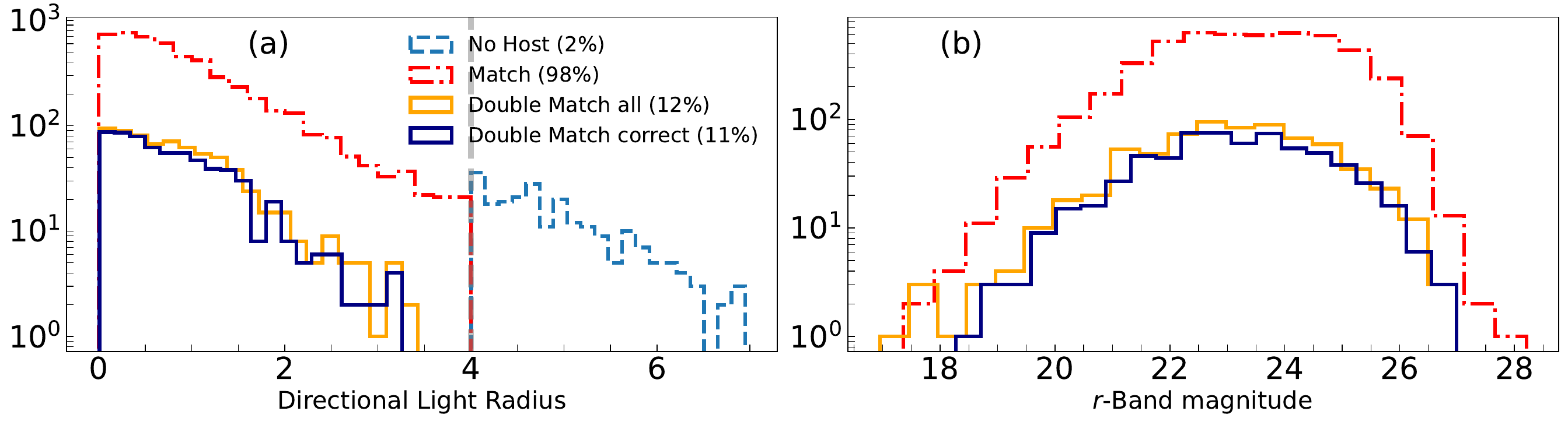}};      
    \end{tikzpicture}
    \caption{  Distributions illustrating
    host galaxy (mis)-association. 
 We find  
 $\sim 98\%$ host galaxy match and $\sim 2\%$ of no host matches.    $12\%$ `double match all',  is the aggregate for all matches for $\rm HOST\_\ddlr < 4$ and $11\%$ for the sum of all correct host matches, satisfying $\rm HOST\_\ddlr < 4$  condition. The grey vertical dashed line in the first figure marks $\ddlr = 4$. 
    }
    \label{fig:host_misac}
\end{figure*}

\subsection{Photometric Classification: SCONE}
\label{subsec:scone}
Deep learning based photometric classifiers have outperformed   {their} predecessors such as  light curve template matching algorithms \citep{p2} or feature extraction with either sequential cuts \citep{campbell} or machine learning methods \citep{Moller2016, Lochner2016, Dai2018}.   {\DESVYR\ used neural network based photometric classifiers, SCONE\footnote{Supernova Classification with a Convolutional Neural Network} \citep{scone} and SuperNNova \cite{Moller2020}}

We use SCONE  as our photometric classifier  {to determine the probability ($\Pia$) that each event is a SN~Ia.} SCONE   requires photometric data only, without the need for  accurate redshift  {estimation},    {and it } has relatively low computational and dataset size requirements for achieving high accuracy.    

For training the network with  $150$ iterations, we generate a simulation consisting of ${\sim}50,000$ Ia and ${\sim}50,000$ non-Ia after selection requirements. Even for real data, the procedure of training on simulated data is the same to ensure that 
the training data has the same redshift range and selection effects as the data.
The training process  took ${\sim}3$ hours. The network attained $98\%$ accuracy for both training and validation  {and a LOSS of 0.07}. Fig.~\ref{fig:scone_pred} shows $\Pia$\   distributions and they match our qualitative expectations; 
for true Ia's the probability is peaked at $p\sim1$ and for true non-Ia the peak is at $p\sim0$.

\begin{figure}[h]
    \centering
    \includegraphics[width=\columnwidth]{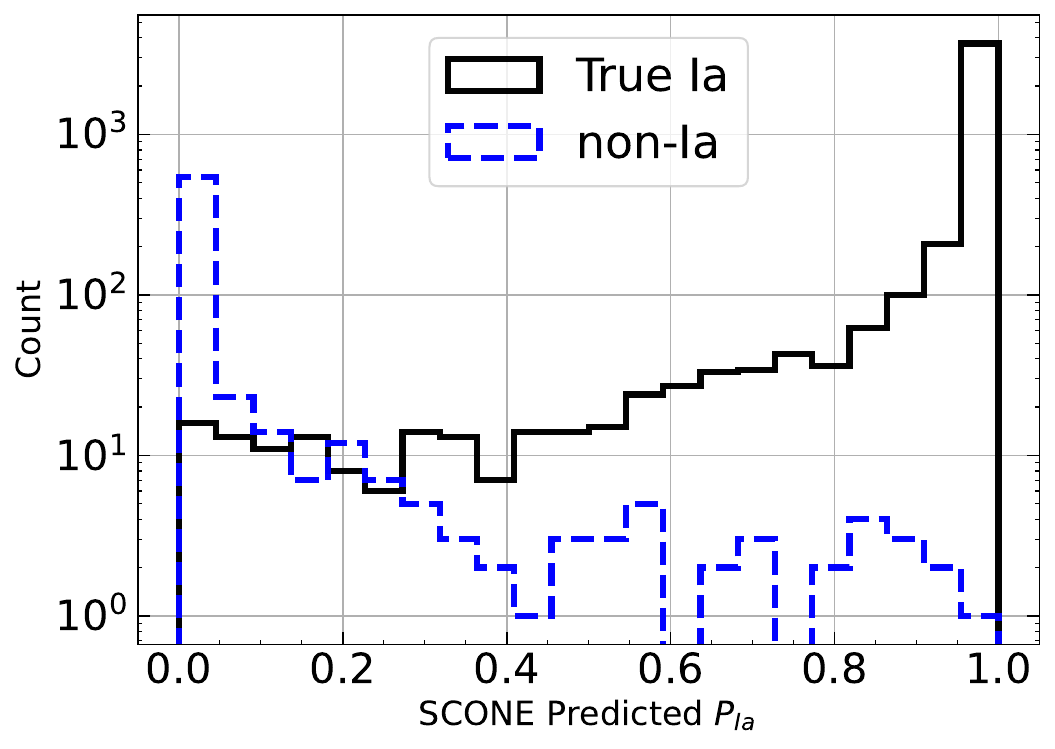}
    \caption{Distribution of SCONE  {$\Pia$}
    for true SNIa (black) and for true non-Ia (dashed blue histogram). 
    }
    \label{fig:scone_pred}
\end{figure}

\subsection{BEAMS with Bias Corrections (BBC)}
\label{subsec:ana_bbc}
 
 BBC  {incorporates} BEAMS  { methodology and} produces a Hubble diagram that is corrected for selection effects and also corrected for non-Ia contamination. 

\subsubsection{ BEAMS}
\label{sec:beams}
The BEAMS framework was introduced by \citet[][]{kunz, 2012MNRAS.421..913N}, \citet{hlozek12, Knights:2012if}. 
 In the BEAMS framework, the total likelihood function includes two terms: one for the SN Ia population, $\mathcal{L}_\mathrm{Ia}$, and another for the contaminant population, $\mathcal{L}_\mathrm{CC}$ \footnote{The CC notation includes all non-Ia, but here we follow the KS17 notation.}, expressed as:
\begin{equation}
    \prod_{i=1}^{N_{\mathrm{SNe}}} \big( \mathcal{L}^i_\text{Ia}+\mathcal{L}^i_\text{CC} \big) .
    \label{eq:SUM}
\end{equation}
The likelihood components for each event $i$  are:
\begin{equation}
  \begin{aligned}
    \mathcal{L}^i_\text{Ia} &=  {\Pia}^i\times D_\text{Ia}(z_i, \mu_{\text{obs},i}, \mu_{\text{ref}}),\\
    \mathcal{L}^i_\text{CC} &= (1-P^i_{\text{Ia}}) \times D_{\text{CC}}(z_i, \mu_{\text{obs},i}, \mu_{\text{ref}}),
  \end{aligned}
  \label{eq:likelihoods}
\end{equation}
where $D_{\text{Ia}}$ and $D_{CC}$  are the SNIa and non-Ia Hubble residual likelihoods.   {A subtle caveat in Eq 8 is that the same light curve data are used for $\Pia$ and $D_{Ia}$, $D_{CC}$ but potential covariances are not accounted for.} The quantities in  Eq.~\ref{eq:likelihoods} can  be   used to determine a  BEAMS probability,
\begin{equation}
   \mathcal{P}_{\tt BEAMS} =  \frac{ {\Pia}^iD^i_{\text{Ia}}}{  P^i_{\text{Ia}}D^i_{\text{Ia}} + (1- P^i_{\text{Ia}})D^i_{\text{CC} }} \label{eq:PBIa},
\end{equation}
which quantifies the probability of a given SN belonging to the Ia population versus the contaminant population, taking into account both the classification probability ($P^i_{\text{Ia}}$) and the Hubble residual information. 

\subsubsection{BBC}
\label{sec:bbc}
  BBC is a fitting framework that  accounts for selection effects using a simulation, incorporates BEAMS to account for non-Ia contamination, and implements the \citet{2011ApJ...740...72M} method of measuring distances independently of cosmological parameters. 
BBC reads the SALT3 fitted parameters (high-$z$ and low-$z$) 
from the data and biasCor simulation (sec.~\ref{subsec:ana_simbias}),
and produces a bias-corrected Hubble diagram, both unbinned and in redshift bins.
For each event, the measured distance modulus is based on the \citet{Trip1998} equation,
\begin{equation}
   \mu = m_B + \alpha x_1 - \beta c + \gamma G_{\rm host} + M_0 + \mubias~,   \label{eq:mu}
\end{equation}
where $m_B \equiv -2.5\log_{10}(x_0)$, 
$\{\alpha,\beta,M_0\}$ are global nuisance parameters, and
$\mubias = \mu - \mutrue$ is determined from the biasCor simulation
in a 4-dimensional space of $\{z,x_1,c,x_M\}$  {\citep{Popovic:2021cwq}},   {where $x_M$ is the host galaxy stellar mass}. A valid bias correction is required for each event, resulting in a  ${<}1 \%$ loss, usually for extreme values of $x_1,c,z$ where the biasCor statistics are very low. The term $\gamma G_{\rm host}$ captures residual dependency between SNIa luminosity and host galaxy magnitude. The distance uncertainty ($\sigmu$) is computed from Eq.~3 of KS17.

There are two subtle issues concerning the use of $\zphot$ and its   {measured} uncertainty $\sigz$.
First, 
the calculated distance error from $\sigma_z$ ($\sigmu^z$ in Eq.~3 of KS17) 
is an overestimate because it does not account for the correlated color error that 
reduces the distance error.
By floating $\zphot$ in the SALT3 fit, redshift correlations propagate to the other SALT3 parameter uncertainties,
and therefore we set $\sigmu^z=0$,   {however the peculiar velocity uncertainty is included}.
The second issue concerns the $\mu_\mathrm{{bias}}$ computation, where
$\mutrue$ is computed at SALT3-fitted $\zphot$ rather than the true redshift. 

To avoid a dependence on cosmological parameters, the BBC fit is performed in \NZBIN\ logarithmically-spaced redshift bins.
The fitted parameters include the global nuisance parameters 
($\alpha,\beta,M_0$) 
and bias-corrected distances
in \NZBIN\ redshift bins. The unbinned Hubble diagram is obtained from Eq.~\ref{eq:mu} using
the fitted parameters, and the distance uncertainty is computed from Eq.~(10) in  \citet{Kessler:2023toi}.

\subsubsection{Covariance Matrix}
\label{sec:cov}
The uncertainties in SN distance measurements are encapsulated in a covariance matrix, which quantifies both statistical and systematic uncertainties. This covariance matrix ($\mathbf{C}$) is 
\begin{equation}
    \mathbf{C} = \mathbf{C}_{\text{stat}}+\mathbf{C}_{\text{syst}}
    \label{eq:cov}
\end{equation}
where $\mathbf{C}_{\text{stat}}$  is the diagonal covariance representing measurement errors.
$\mathbf{C}_{\text{sys}}$  is  the systematic covariance matrix 
\citep{conley11},
\begin{equation}
    C^{ij}_{\text{syst}} = \sum_{S}\left(\Delta \mu_{\text{obs},S}^{i}\right)\left(\Delta \mu_{\text{obs},S}^{j}\right) ,
\end{equation}
where $i$  and $j$ are SN indices,  and  $\Delta\mu_{\text{obs},S}$ is the change in  SN distance moduli resulting from systematic variation $S$. 

Because of cuts~\ref{cut_fitband} and \ref{cut_biascor} in Sec.~\ref{subsec:ana_cuts_syst}, the SN sample for each systematic is slightly different and therefore $\mathbf{C}_{\text{sys}}$ cannot be computed. Following previous analyses we select the common events in all systematics and use this sample for the Hubble diagram and  $\mathbf{C}$.  For redshift systematics that result in migration to another redshift bin, the original (no syst) redshift bin is preserved for the BBC fit.

 {Table~\ref{tb:bbc} summarizes the loss from the common-event cut for two different priors 
($1\sigma$ and $2\sigma$) in Eq.~\ref{eq:chisqsyst}. The  $1\sigma$ prior is used in our analysis, 
while the other prior is a diagnostic to see the impact of a looser prior. 
The final column is a comparison for a flat prior case, which is the standard used for previous analyses 
e.g., \DESVYR. Table~\ref{tb:bbc} breaks down this loss into three sources,}
 {and the dominant source of loss is from analysis cuts.}
This implicit common-event cut is not modelled in the bias correction simulation because we do not perform the CPU-intensive task of evaluating systematics for the large biasCor sample. 
This implicit cut rejects ${\sim}4\%$ of the sample ($2^{nd}$ column of  Table~\ref{tb:bbc}) and  could be a  potential source of unmodelled bias. 
 {In comparison a flat prior results ${\sim}12\%$ loss, illustrating that using a $1\sigma$ prior
significantly reduces this loss.}

\begin{table}[h]
\begin{center}
\caption{
Analysis loss from requiring valid biasCor and requiring that the same events pass cuts for all systematic variations.}
\begin{tabular}{  | c  c | c  c |c |  c| } 
 \hline
    
         Data cuts not applied &       & $\%$ Loss 
          &     & $\%$ Loss& $\%$ Loss\\  
         to biasCor sample &       &  $1\sigma$
          &     &$2\sigma$ &Flat\\   
          &       &   prior\footnote{ {This prior is used in our analysis;  see} $\chisqsyst$ in Eq.~\ref{eq:chisqsyst}.  }
          &     & prior\footnote{ {Additional diagnostic that is not used in the analysis.}}& prior\footnote{All previous analyses using {\tt SNANA}, including M23 and \DESVYR\ , used a flat prior.}\\          
 \hline
      Analysis cuts on systematics &  & $ 3.3\%$ & &$ 3.9\%$& $ 7.5\%$\\
      Valid biasCor in Nominal &  &   $0.44\%$ & &$0.89\%$& $0.86\%$\\
      Valid biasCor in systematics\footnote{for stat-only analysis without systematics, this loss is zero by definition.} &  &   $0.07\%$ & &$0.37\%$& $4.00\%$\\
      \hline
      &  &  &&& \\
     Total &  & $\sim$  $4\%$ & &$\sim 5.2\%$& $\sim 12.4\%$\\
 \hline
\end{tabular}
\label{tb:bbc}
\end{center}
\end{table}

\subsubsection{Cosmology Fitting and Figure of Merit}
\label{subsec:ana_wfit}
In cosmology fitting, 
we minimize the following $\chi^{2}$:
\begin{equation}
    \chi^2 = \Delta\mu^{T}\mathbf{C}^{-1}\Delta\mu,
    \label{eq:chi}
\end{equation}
where $C$ is the covariance matrix (Eq.~\ref{eq:cov}), and $\Delta\mu$ is the vector of differences between observed and theoretical distance moduli.
\newcommand{\urlwfit}{\url{https://github.com/RickKessler/SNANA/blob/master/src/wfit.c}}
We use the same fast minimization code\footnote{\urlwfit} that was used in  \citet{Mitra:2022ykq}. We  approximate a
Cosmic Microwave Background (CMB) prior using the $R$-shift parameter (e.g., see Eq. 69
in \citet{komatsu}) computed from the same cosmological 
parameters that were used to generate the SNe~Ia. The $R$-uncertainty is
$\sigR=0.006$, tuned to approximate the constraining power of \citet{Planck2018}.
We fit with  \wCDM\ and \wwCDM\ (CPL \citep{cpl2}) models, where $w=[w_0+w_a(1-a)]$ and $a = 1/(1+z)$.

  {
Prior to 2021, most cosmology results were based on a redshift binned HD
to reduce computation time compared to an unbinned HD. After \citet{Binning_is_sinning}
showed that fitting an unbinned Hubble diagram gives optimal constraints,
cosmology results have been based on unbinned HDs (e.g.,  {Pantheon+ \citep{pantheon+}} and {\desfv} \citep{DES:2024hip}).
In anticipation of much larger HDs in the future that would require enormous 
memory and CPU resources for the optimal unbinned approach\footnote{For example, a future WFD+DDF HD with $\sim 10^5$ events results in a covariance matrix $\mathrm{C}$ with $\sim 10 $ billion elements, which makes the $\chi^2$ computation (Eq.~\ref{eq:chi}) very CPU intensive.}, 
\citet{Kessler:2023toi} proposed a rebinned approach to optimise constraints with a much smaller Hubble diagram, and hence use much less CPU resources. In our implementation, the rebinned analysis requires approximately four times less wall  time than the corresponding unbinned analysis.
}
We include rebinned results here with \nrebinx\ stretch bins, and \nrebinc\ colour bins. 
The maximum Hubble diagram size is $14 \times \nrebinx 
 \times \nrebinc = \HDSIZEREBIN$; after excluding empty bins the average fitted Hubble diagram size is  $\hdsizerebin$.

For the \wwCDM\ model, the FoM is computed based on the dark energy task force 
(DETF) definition in  {\citet{Albrecht2006_FoM,Wang2008_FoM}},
\begin{equation}
   \text{FoM} \simeq \frac{1}{ \sigma (w_a) \sigma(w_0) \sqrt{1-\rho^2} }~,
    \label{eq:FoM}
\end{equation} 
where $\rho$ is the reduced covariance between $w_0$ and $w_a$. 

We fit the Hubble diagram and evaluate FoM for the three binning methods: binned,  unbinned,  and rebinned. M23 reported that unbinned results were pathological for unknown reasons and therefore not shown. Here we  {no longer } find  {pathological unbinned results, but we cannot trace what caused the problem in M23.}


\section{Cosmology Results  }
\label{sec:cosmology}

For one of the \NSAMPLE\ statistically independent samples, we show the Hubble diagram produced by the BBC fit,
both binned and unbinned, in Fig.~\ref{fig:HD}. The Hubble residuals with respect to the
true cosmology, $\Delta\mu = \mu - \mu_{\rm true}$, are consistent with zero.  
 {There is evidence of a redshift dependence as indicated by the $2.2\sigma$ significance in the 
slope   {and its uncertainty from }  a linear fit of Hubble residual vs redshift. 
  {Fitting a  slope to the combined }  $25$  data sets 
  {results in} a strong $6.5\sigma$  {significance in the slope, which leads to biased} 
cosmological parameters as shown below.
}

\begin{figure*}[!ht]
    \centering
    \includegraphics[width=\textwidth]{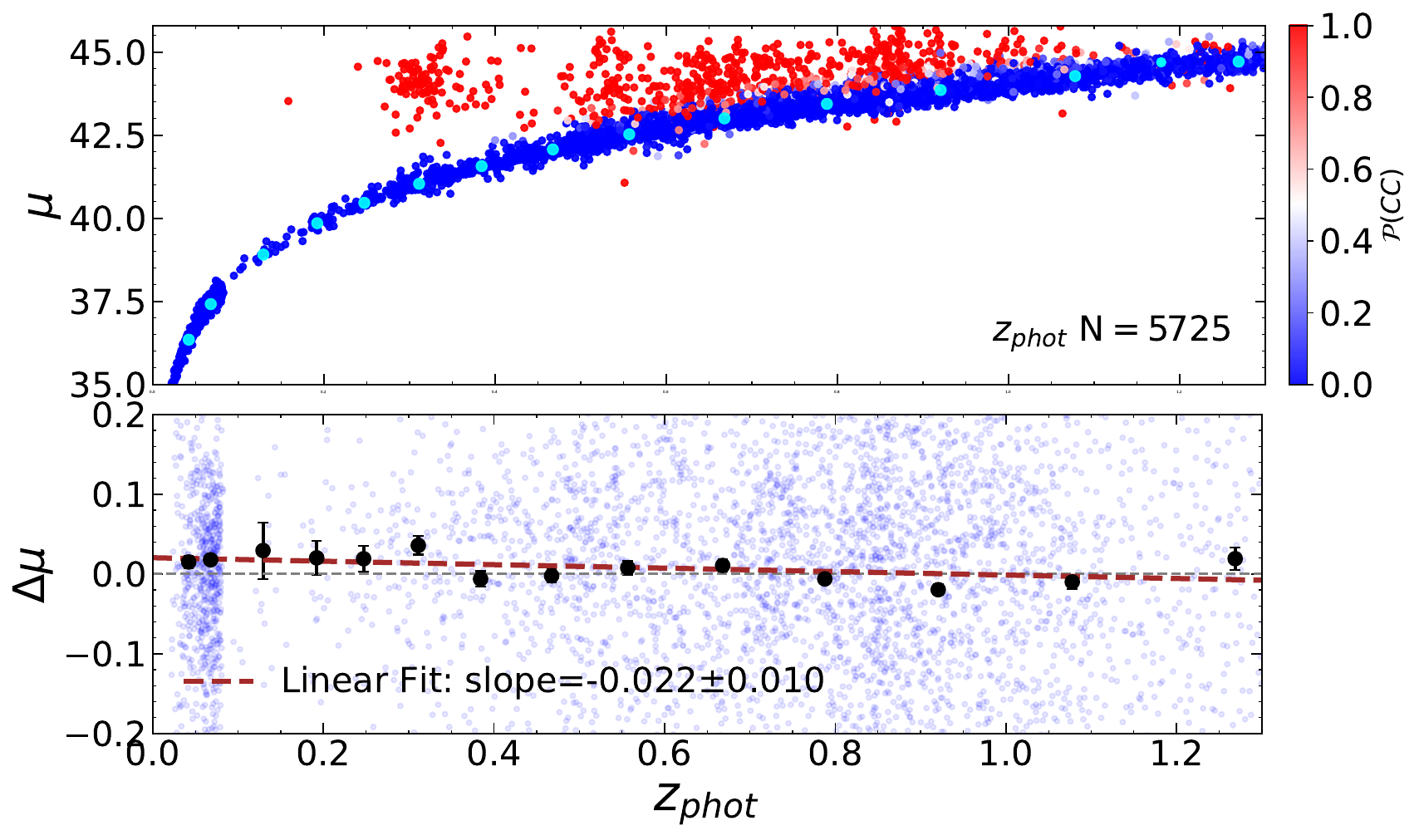}
    \caption{Redshift binned and unbinned Hubble diagram for one of the \NSAMPLE\ statistically independent samples.
   The lower panel shows Hubble residual $\Delta\mu$ with respect  {to} the true cosmology,  {and}
   the error bar shows the  {mean uncertainty}    in each BBC redshift bin (same redshift bins as in Fig.~\ref{fig:musig_compare}). 
   The  red-coloured events show the  {non-Ia} contamination. 
   The corresponding colour bar on the right shows the probability derived from BEAMS. 
   The number of events with $\mathcal{P}$(CC) $>0.75$ is $652$, 
   which is $\sim 11.4\%$ of the  {entire}  $\zphot$ data sample of $5725$ events.  
    }
    \label{fig:HD}
\end{figure*}

\begin{figure}[h]
    \centering
    \includegraphics[width=\columnwidth]{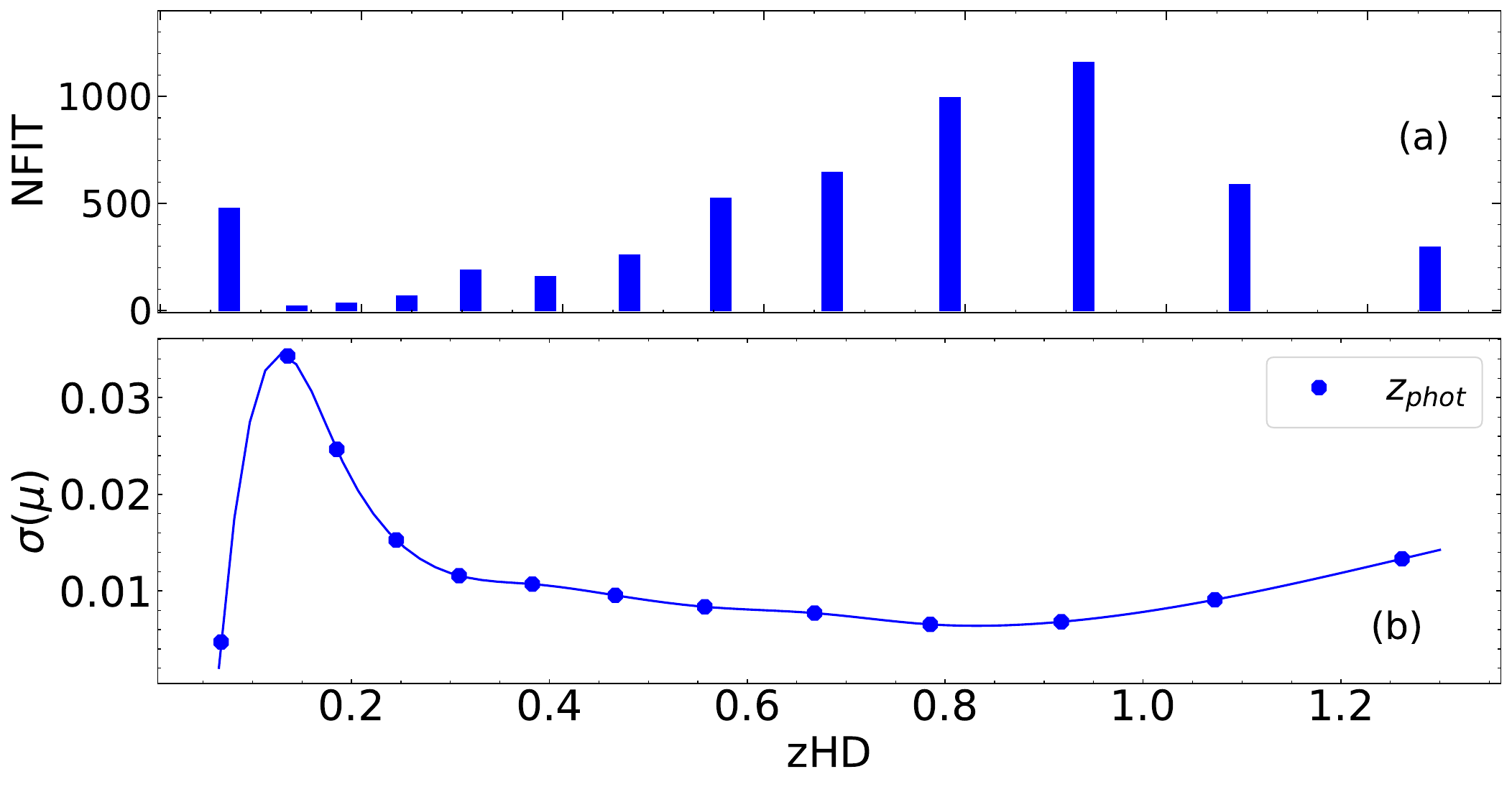}
    \caption{
    Number of events (top) and BBC-fitted distance uncertainty (bottom) per 
     {BBC} redshift bin. 
    }
    \label{fig:musig_compare}
\end{figure}

For BBC-fitted nuisance parameters we find $\alpha-\alphaTrueSym$ = $-0.00095\pm 0.00057$ (averaged over $25$ samples), we cannot however   {evaluate} the bias for $\beta$ since $\beta_{true}$ is 
 {not defined}
as described previously in Sec.~\ref{sec:snana}. We find $\sigint = 
 0.112 $  but there is no true $\sigint$ for comparison. Next, we show the BBC fitted distance uncertainties ($\sigmu$) in redshift bins 
(Fig.~\ref{fig:musig_compare}{b}). 
Owing to the larger statistics in high-$z$, the  {weighted} distance uncertainty 
 {per $0.1$ wide redshift bin} is in  {the} $0.01-0.02$ range at $z>0.6$.

For the subsections below, we define $w$-bias to be   $w - w_{\rm true}$ where $w$ is from the \wCDM\  cosmology fit. 
A similar definition is used for $w_0$ and $w_a$ for the \wwCDM\ model.
{The bias uncertainty is the standard deviation of the  $w$-bias values divided  by $\sqrt{\NSAMPLE}$.}

\subsection{ \wCDM\  Results}
\label{sec:wcdm}
 {For the \wCDM\ cosmology fits, Table~\ref{tb:wcdm_results} summarizes the 
$w$-bias and associated uncertainties for the binned, unbinned, and rebinned analyses, 
each averaged over 25 realizations that include a CMB prior.} 
We find  {for all binning methods that there is a significant stat only bias of $\AVGwbias \sim 0.01$-$0.03$ with a significance of $\sim$ $3$-$8\sigma$ .}  {With systematics the biases are slightly smaller and the significance is in the $\sim$ $2$-$6\sigma$ range.}

 {Overall, the mean $w$-uncertainty across all binning methods is  $\AVGwsig{\sim}0.020$ for the stat only case.
Including systematics the binned method results in $ \AVGwsig{\sim}0.025$, and the unbinned and rebinned methods result in a slightly smaller uncertainty of  $\AVGwsig\sim0.022$.  The larger binned uncertainty is expected.}
 Fig.~\ref{fig:wbias} shows the $w$ bias for all the three binning schemes.  
\begin{figure}[h]
    \centering
    \includegraphics[width=\columnwidth]{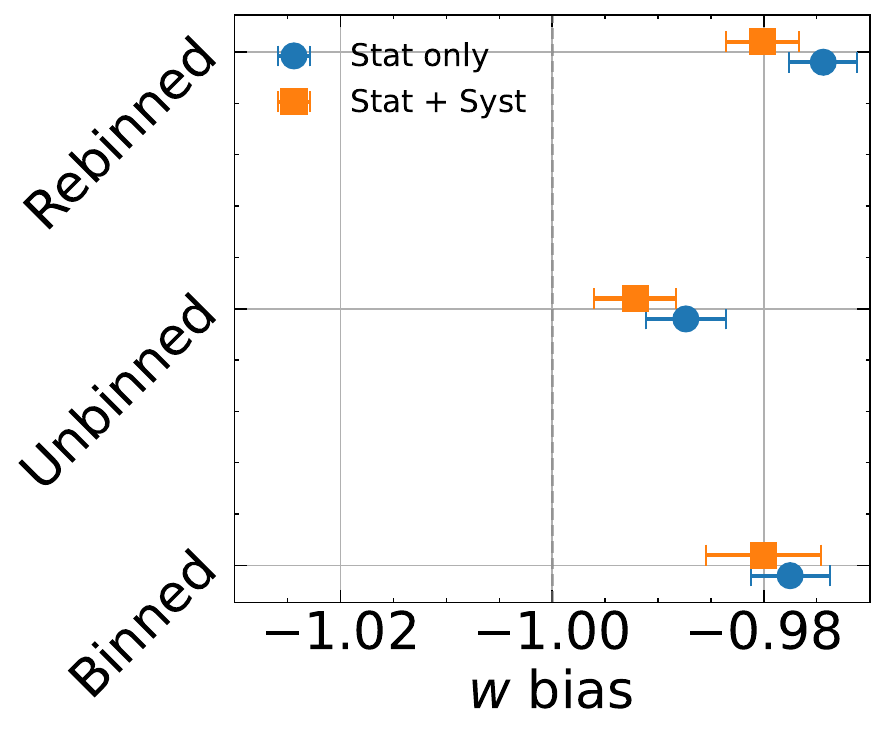}
    \caption{ Bias in $w$CDM cosmology  {for each HD binning method}. 
    The grey dashed vertical line  {shows}  the  {true} reference $w=-1$.   }
    \label{fig:wbias}
\end{figure} 


\begin{table}[ht!]
\begin{center}
\caption{Summary of $w$CDM Cosmology bias for Binned, Unbinned, and Rebinned  {methods.}  }
\begin{tabular}{|c|cc|cc|}
\hline
\multirow{2}{*}{\textbf{Case}} 
 & \multicolumn{2}{c|}{\textbf{Stat only}} 
 & \multicolumn{2}{c|}{\textbf{Stat+Syst}} \\
\cline{2-5}
 & $\AVGwbias$\tablenote{Average bias among $\NSAMPLE$ samples with uncertainty of std/$\sqrt{\NSAMPLE}$} & $\AVGwsig$\tablenote{Average fitted uncertainty among $\NSAMPLE$ samples.}
 & $\AVGwbias$ & $\AVGwsig$ \\
\hline
Binned  & $0.0225 \pm 0.0037$ & $0.0200$  & $0.0200 \pm 0.0054$  & $0.0248$ \\
Unbinned  & $0.0126 \pm 0.0038$ & $0.0205$  & $0.0078 \pm 0.0039$  & $0.0222$ \\
Rebinned  & $0.0256 \pm 0.0032$ & $0.0198$  & $0.0199 \pm 0.0034$  & $0.0216$ \\
\hline
\end{tabular}
\label{tb:wcdm_results}
\end{center}
\end{table}

\subsection{ \wwCDM\ Results}
\label{sec:w0wa}

 {For the \wwCDM\ cosmology fits, Table~\ref{tb:w0wa_results} summarizes the average $w_0$ and $w_a$ biases and associated uncertainties for the binned, unbinned, and rebinned analyses, each averaged over 25 realizations that include a CMB prior. 
The $\sigma_{w_0}$ and $\sigma_{w_a}$  {rows show}
the average fitted uncertainties across the 25 measurements. 
We find for all three binning methods a mild but measurable stat-only bias in $w_0$, ranging from $0.02$ to $0.03$ ($\sim$ $1$-$2$$\sigma$), and in $w_a$, ranging from $0.12$ to $0.24$ ($\sim$2-4$\sigma$). These biases are consistent in sign across the binned, unbinned, and rebinned analyses. 
When including systematics, the $w_0$-bias  slightly increases, varying between $0.02$ and $0.06$ ($\sim$1.5-3.5$\sigma$), while the $w_a$-bias increases modestly to $0.18$--$0.27$ ($\sim$3-4$\sigma$).}

 {For the binned analysis, stat+syst $\AVGFoM=$ \AVGFOMzphotsyst , is a $30\%$ degradation compared  to the stat-only  $\AVGFoM$  \AVGFOMzphotstat. The equivalent rebinned analysis goes  from $196$ to \AVGFOMzphotsystR , making $\sim19\%$ degradation.}  For the unbinned analysis  $\AVGFoM=$ degrades from $196$ to \AVGFOMzphotsystU, giving a degradation of $\sim14\%$. We suspect that the marginal improvement in the FoM for the rebinned analysis over the unbinned analysis arises from Python’s numerical precision handling and the treatment of the covariance matrix during inversion. 
 Therefore on average FoM values are $\sim\!190$ for the stat-only fits and $\sim\!150$ for the stat+syst case, indicating a typical degradation of about 25--30\% 
  {with} systematics. 
 The  {contours and bias}  for a single sample  {are}  shown in Fig.~\ref{fig:de_compare_stat} (stat only) and Fig.~\ref{fig:de_compare} (stat+syst).


\begin{table*}[ht!]
\begin{center}
\caption{Summary of  {bias, uncertainty and FoM for} $w_0w_a$CDM cosmology fits  {using}
binned, unbinned, and rebinned  {methods.}   }
\begin{tabular}{|l|c|c|c|c|c|c|}
\hline
& \multicolumn{3}{c|}{Stat only} & \multicolumn{3}{c|}{Stat+Syst} \\
\cline{2-7}
& Binned & Unbinned & Rebinned & Binned & Unbinned & Rebinned \\
\hline
{$\AVGwwbias$\tablenote{Average bias among $\NSAMPLE$ samples with uncertainty of std/$\sqrt{\NSAMPLE}$}}   & $-0.0221 \pm 0.0152$ & $-0.0175 \pm 0.0153$ & $-0.0314 \pm 0.0152$ & $-0.0242 \pm 0.0154$ & $-0.0573 \pm 0.0161$ & $-0.0307 \pm 0.0150$ \\
$\AVGwabias$   & $0.1874 \pm 0.0623$ & $0.1239 \pm 0.0631$ & $0.2424 \pm 0.0598$ & $0.1787 \pm 0.0646$ & $0.2724 \pm 0.0650$ & $0.2093 \pm 0.0579$ \\
{$\AVGwwsig$\tablenote{Average fitted uncertainty among $\NSAMPLE$ samples.}}   & $0.0632$ & $0.0636$ & $0.0641$ & $0.0767$ & $0.0732$ & $0.0724$ \\
$\AVGwasig$   & $0.2665$ & $0.2739$ & $0.2670$ & $0.3167$ & $0.3008$ & $0.2994$ \\
$\AVGFoM$   & $193$ & $182$ & $196$ & $131$ & $156$ & $159$ \\
\hline
\end{tabular}
\label{tb:w0wa_results}
\end{center}
\end{table*}

\begin{figure}[h]
    \centering
    \includegraphics[width=\columnwidth]{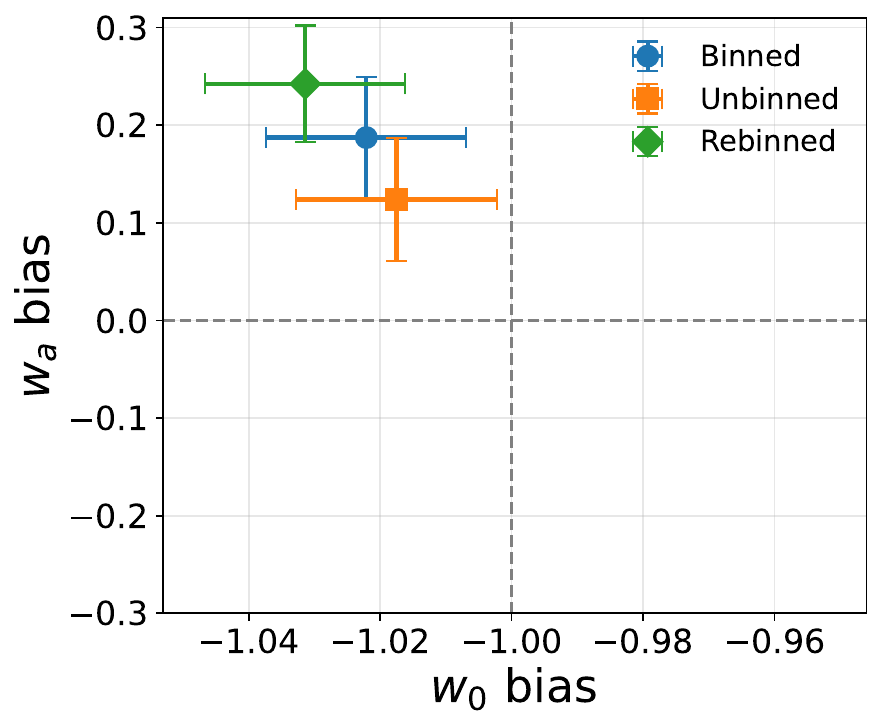}
    \caption{Similar to Fig.~\ref{fig:wbias} but for bias in $w_0 w_a$CDM cosmology. 
    Shown only for stat+systematic case. 
    }
    \label{fig:wabias}
\end{figure}

\begin{figure*}[ht]
    \centering
    \begin{subfigure}[t]{0.48\textwidth}
        \centering
        \includegraphics[width=\columnwidth]{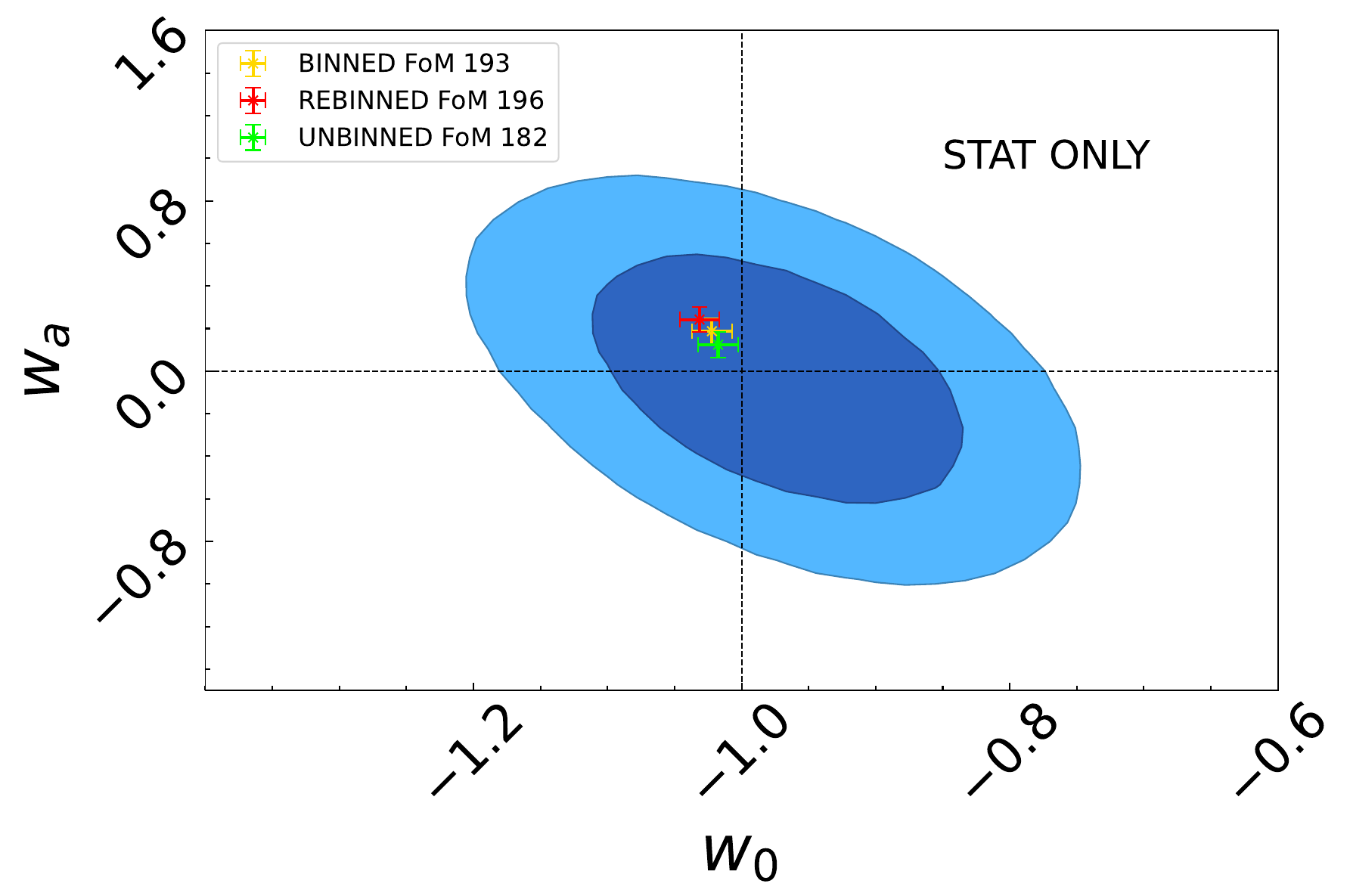}
        \caption{
             {For stat-only analysis,} \ww\  {68\% (dark blue) and 95\% (light blue) confidence }
            contours for a single SNIa data sample combined with CMB prior. 
            The crosses show the  bias  {and $\pm 1\sigma$ uncertainty} 
            from averaging results over the \NSAMPLE\ simulated data samples. 
        }
        \label{fig:de_compare_stat}
    \end{subfigure}
    \hfill
    \begin{subfigure}[t]{0.48\textwidth}
        \centering
        \includegraphics[width=\columnwidth]{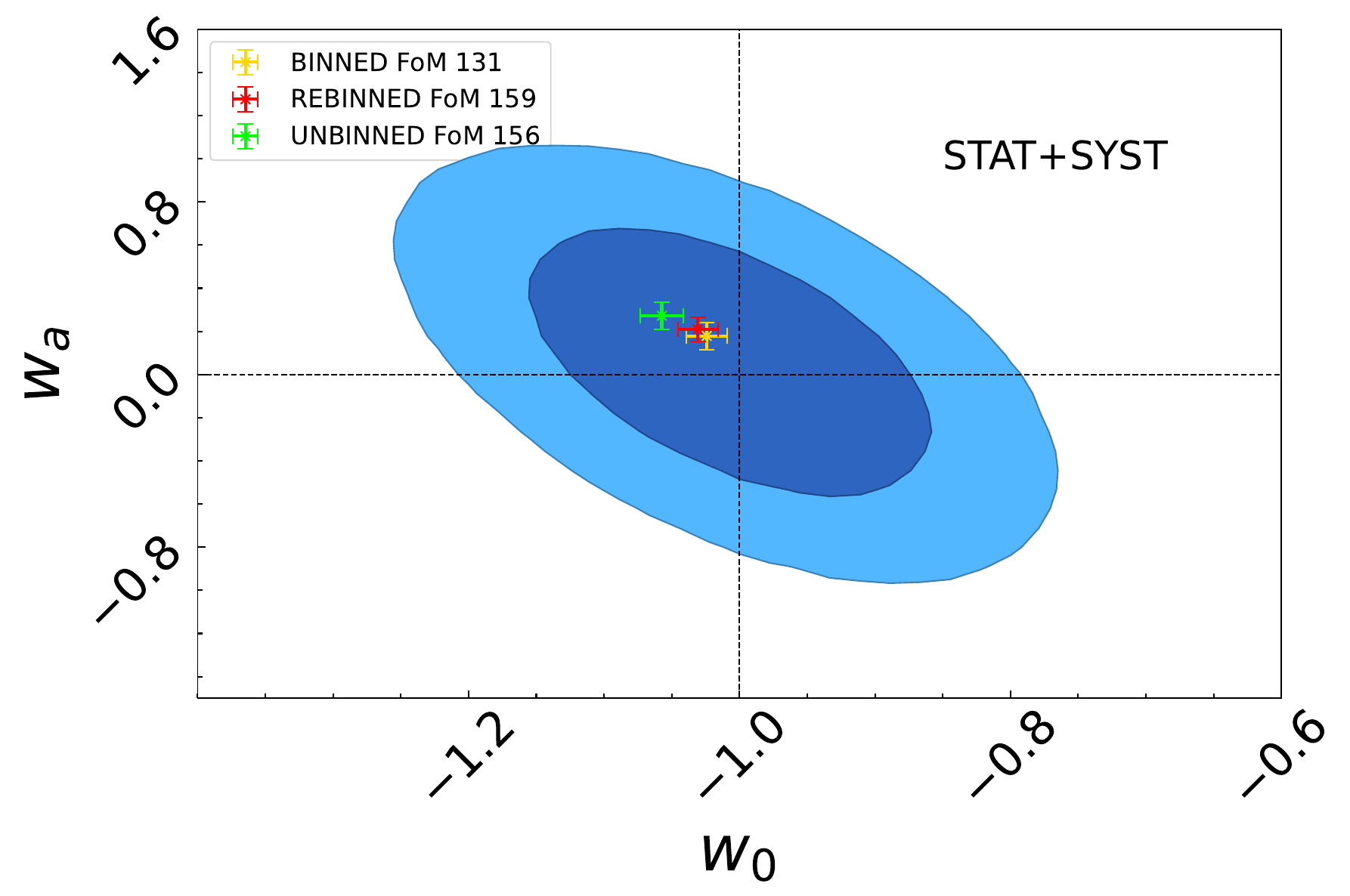}
        \caption{
            Same as in left  {panel},  but for stat+syst case 
             {and unbinned HD.}
        }
        \label{fig:de_compare}
    \end{subfigure}
    \caption{Comparison of $2\sigma$ $(95\%$ confidence$)$ contours for two cases: (a) stat only, and (b) stat+syst.}
    \label{fig:de_compare_sidebyside}
\end{figure*}

Since there are many systematics contributing to the decrease in $\AVGFoM$,
we quantify the impact of each systematic ``$i$'' by 
recomputing the covariance matrix separately for each systematic  ($\mathbf{C}_{\text{syst, i}}$), 
and repeating the cosmology fit for each $\mathbf{C}_{\text{syst, i}}$.
We finally compute the FoM ratios
\begin{equation}
    \RatioFoM = \FoMSysti/\FoMStat~,
\end{equation}
where $\FoMSysti$ is the FoM from including only systematic $i$,
and $\FoMStat$ is the FoM without systematic uncertainties.
Note that $\RatioFoM \leq 1$. 

\begin{table}
\caption{FoM-Ratio $\RatioFoM$ for Each Systematic (\wwCDM\ model) 
 {and for each HD binning method.}
}
\begin{center}
\begin{tabular}{| l | c | c | c |}
 \hline
 Systematic(s) & \multicolumn{3}{c|}{$\RatioFoM$} \\
 \cline{2-4}
               & Binned & Unbinned & Rebinned \\ [0.5ex]
 \hline
None (stat only) & $1.00$ & $1.00$ & $1.00$ \\
zSHIFT & $0.95$ & $0.98$ & $0.98$ \\
Photo\_shift & $0.95$ & $0.99$ & $0.99$ \\
MWEBV & $0.96$ & $0.99$ & $0.99$ \\
CAL\_WAVE & $0.90$ & $0.99$ & $0.95$ \\
Cal\_ZP & $0.80$ & $0.90$ & $0.89$ \\
CAL & $0.78$ & $0.88$ & $0.85$ \\
\hline\hline
Stat + All Syst  & $0.72$ & $0.85$ & $0.84$ \\
 \hline
\end{tabular}
\label{tb:syst_breakdown}
\end{center}
\end{table}

Table~\ref{tb:syst_breakdown}  {shows $\RatioFoM$ for all three binning methods.}
The FoM degradation is  dominated by the calibration systematics. 
Photo-$z$ systematics are not the dominant source of FoM degradation,  {in part because the}
anti-correlation between photo-$z$ and the SALT3 color parameter acts to self-correct distance errors. 
Similar findings were reported by \citet{Chen2022}. 
Thus, the overall impact of photo-$z$ systematics remains modest due to these combined effects.

\subsection{Comparison with \desfv \ and improvements with DESI }
\label{sec:comparison_with_desfv}

In Fig.~\ref{fig_w0wa_desi_des_planck_lsst_data_mock}, we provide constraints on  $w_{0}$ and $w_{a}$ from 
 {\desfv\ (yellow) and 3-year LSST-DDF (blue),}
and also in combination with CMB and baryonic acoustic oscillations (BAO) measurements. 
 {Compared to} the \desfv{} SNe only sample,  the LSST-DDF sample results in
 $\times16$ 
 improvement in the FoM for $w_{0}, w_{a}$. 
We also show the constraints when the SNe are combined with the CMB power spectra measurements from 
{\it Planck} \citep{rosenberg22} and the BAO measurements from the recent DESI-DR2 release \citep{desi_dr2_bao}. 
When combined with CMB and BAO, we find a $\times1.4$ improvement in the FoM 
when 
 {replacing}
DES  {with} 
the LSST sample \citep[Also see][]{raghunathan25}. 
The improvement in this case is less dramatic compared to the SN-only case because the 
 {constraining power is primarily}
from the different directions of degeneracies probed by CMB, BAO, and SN measurements.

To combine the LSST-DDF SN sample with DESI-DR2 measurements, 
we impose a consistent cosmology by replacing the BAO data vectors with those
calculated using the \wwCDM\ cosmology with values set to {\it Planck} 2018 cosmology and $(w_{0}, w_{a}) = (-1, 0)$.
 {The original DESI DR2 covariances are used, 
and we do not  tweak the CMB measurements from {\it Planck}.}
%

\begin{figure*}[h]
    \centering
    {\includegraphics[width=.8\textwidth, keepaspectratio]{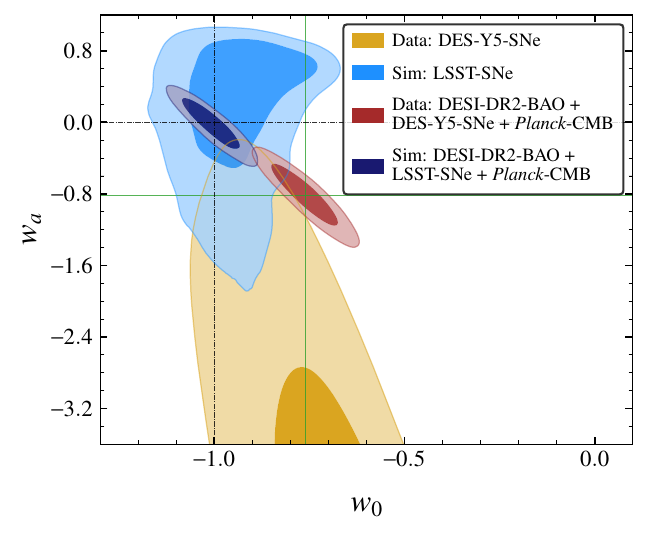}}
    \caption{Comparison of constraints on the dark energy EoS parameters 
$w_{0}$ and $w_{a}$ 
 {using SNe~Ia from a simulated 3-year LSST-DDF sample from this work (light blue)}
and  {from} \DESVYR\ 
 {from \citet{DES:2024hip} (yellow)}.
The joint constraint combining DESI~DR2~BAO and {\it Planck}~CMB with 
\DESVYR\ SNe is shown in maroon,  {and the green lines show the best fit values.
The navy blue contour shows the corresponding joint constraint with LSST-DDF SNe replacing \desfv,
and  BOA  replaced with those calculated using $w_0,w_a = (-1,0).$}
    }
    \label{fig_w0wa_desi_des_planck_lsst_data_mock}
\end{figure*}

\subsection{Bias Tests }
\label{sec:biases}

The cosmology bias led us to implement 3 code improvements and perform 1 test.
Unfortunately this effort did not reduce the bias, but we nonetheless 
describe this effort below to inform future analyses.

 {\begin{itemize}
    \item In previous cosmology analyses using SNANA and pippin, a flat prior on each fitted parameter was used to run SALT3 light curve fitting on both nominal and  systematic variants. As described in sec.~\ref{subsec:ana_lcfitz}, requiring the same events in all systematics results in a $12\%$ loss. 
    Here we update the light curve fitting code to use a prior on systematic variants  
    (see $\chisqsyst$ term in eq.~\ref{eq:chi2tot} and ~\ref{eq:chisqsyst}), 
    which reduces this common-event loss to $4\%$. 
    \item We found and fixed a subtle bug in evaluating the  $D_{CC}$ term in eq. \ref{eq:likelihoods}; 
    the distances were not bias corrected. This fix results in  {few milli-mag}
     changes in the distances. The small impact is not surprising because  the rms of the distance bias correction  is $\stdmubiasccprior$ mag, which is much smaller than the non-Ia Hubble residual rms of $\stddmuccprior$ mag. 
    \item The $D_{CC}$ term in eq.~\ref{eq:likelihoods} has  {previously}     been evaluated in redshift bins. However, we noticed that this term has significant dependence on colour and stretch and therefore we updated the BBC code to evaluate $D_{CC}$ on a $3D$ grid of $z,c,x_1$. 
    \item As part of our bias testing, we found that the bias disappears if we 
      exclude all true non-SNIa events,   {which is equivalent to a perfect classifier}.
    Based on this clue, we ran a test requiring  $\Pia>0.01$ (sec.~\ref{subsec:scone}),
     {which} rejects $75\%$ of the non-SNIa. 
\end{itemize}}


\section{Conclusions}
\label{sec:conclude} 

 {}

This work represents one of the first comprehensive LSST-era cosmology 
 {analyses on a simulated high-$z$ sample}
based solely on photometric supernova data.
 {The analysis} combines photometric Type Ia classification with host-galaxy photometric redshifts. 

However, residual biases in recovered cosmological parameters indicate that further 
 analysis development is needed.  {Potential improvements include better galaxy photo-$z$ algorithm, and using the photometric redshift error (or PDF) in BBC.}
  {In our analysis,} systematics from   {astrophysical and } SN model training, especially limitations of the SALT3 model   {(sec.~\ref{subsec:ana_cuts_syst})},   {are not  included. These effects must be incorporated in future studies. Systematics arising from the photometric classifier (also not studied in this analysis) have been shown to contribute roughly one third of the \DESVYR\ systematic error budget.}
Statistical uncertainties therefore dominate over systematic contributions (sec.~\ref{sec:cosmology})   {in our analysis}.
Finally, the demonstrated efficiency of our rebinned HD approach, with  significant computational savings, highlights a scalable path for analysing the much larger LSST SN sample to come.


\acknowledgments 
Author contributions are listed below. \\

A.~Mitra: co-lead project, SNANA simulations and analysis, writing. 
R.~Kessler: co-lead project, software, analysis, writing. 
R.~Chen: Internal review, and writing assistance.
A.~Gagliano: Analysis of host-galaxy catalogs, manuscript revisions.
M.~Grayling: Internal review, and writing assistance.
S.~More: Conceptualization.
G.~Narayan: Conceptualization and writing assistance.
H.~Qu: SCONE software support. 
S.~Raghunathan: Joint constraints using DESI-DR2. 
A.~I.~Malz: DESC Builder producing photo-$z$ prior information in mock data (data curation, writing).
M.~Lochner: DESC Builder for early contributions to supernova and significant optimization of observing strategy, manuscript revision.
\medskip

RK acknowledges pipeline scientist support from the LSST Dark Energy Science Collaboration. 
ML acknowledges support from the South African Radio Astronomy Observatory and the National Research Foundation (NRF) towards this research. Opinions expressed and conclusions arrived at, are those of the authors and are not necessarily to be attributed to the NRF.
This work was completed in part with resources provided by the University of Chicago’s Research Computing Center. This work is also supported by the National Science Foundation under Cooperative Agreement PHY-2019786 (The NSF AI Institute for Artificial Intelligence and Fundamental Interactions, http://iaifi.org/).

This paper has passed an internal review by the DESC  and we thank the DESC internal reviewers: Matthew Grayling and Rebecca Chen. 

The DESC acknowledges ongoing support from the Institut National de 
Physique Nucl\'eaire et de Physique des Particules in France; the 
Science \& Technology Facilities Council in the United Kingdom; and the
Department of Energy, the National Science Foundation, and the LSST 
Corporation in the United States.  DESC uses resources of the IN2P3 
Computing Center (CC-IN2P3--Lyon/Villeurbanne - France) funded by the 
Centre National de la Recherche Scientifique; the National Energy 
Research Scientific Computing Center, a DOE Office of Science User 
Facility supported by the Office of Science of the U.S.\ Department of
Energy under Contract No.\ DE-AC02-05CH11231; STFC DiRAC HPC Facilities, 
funded by UK BIS National E-infrastructure capital grants; and the UK 
particle physics grid, supported by the GridPP Collaboration.  This 
work was performed in part under DOE Contract DE-AC02-76SF00515.

\appendix

\bibliography{main}

@ARTICLE{raghunathan25,
        author = "Raghunathan, S and Mitra, A a ",
        title = "{----}",
        Year = {2025},
        Journal = {In Preparation},
        Title = {...},
}

@ARTICLE{Chen25,
       author = {{Chen}, R.~C. and {Scolnic}, D. and {Vincenzi}, M. and {Rykoff}, E.~S. and {Myles}, J. and {Kessler}, R. and {Popovic}, B. and {Sako}, M. and {Smith}, M. and {Armstrong}, P. and {Brout}, D. and {Davis}, T.~M. and {Galbany}, L. and {Lee}, J. and {Lidman}, C. and {M{\"o}ller}, A. and {S{\'a}nchez}, B.~O. and {Sullivan}, M. and {Qu}, H. and {Wiseman}, P. and {Abbott}, T.~M.~C. and {Aguena}, M. and {Allam}, S. and {Alves}, O. and {Andrade-Oliveira}, F. and {Annis}, J. and {Bacon}, D. and {Brooks}, D. and {Carnero Rosell}, A. and {Carretero}, J. and {Choi}, A. and {Conselice}, C. and {da Costa}, L.~N. and {Pereira}, M.~E.~S. and {Diehl}, H.~T. and {Doel}, P. and {Everett}, S. and {Ferrero}, I. and {Flaugher}, B. and {Frieman}, J. and {Garc{\'\i}a-Bellido}, J. and {Gatti}, M. and {Gaztanaga}, E. and {Giannini}, G. and {Gruen}, D. and {Gruendl}, R.~A. and {Gutierrez}, G. and {Herner}, K. and {Hinton}, S.~R. and {Hollowood}, D.~L. and {Honscheid}, K. and {Huterer}, D. and {James}, D.~J. and {Kuehn}, K. and {Lewis}, G.~F. and {Lima}, M. and {Marshall}, J.~L. and {Mena-Fern{\'a}ndez}, J. and {Menanteau}, F. and {Miquel}, R. and {Ogando}, R.~L.~C. and {Palmese}, A. and {Pieres}, A. and {Plazas Malag{\'o}n}, A.~A. and {Roodman}, A. and {Samuroff}, S. and {Sanchez}, E. and {Sanchez Cid}, D. and {Sevilla-Noarbe}, I. and {Suchyta}, E. and {Swanson}, M.~E.~C. and {Tarle}, G. and {To}, C. and {Tucker}, D.~L. and {Vikram}, V. and {Weaverdyck}, N. and {Weller}, J. and {DES Collaboration}},
        title = "{Evaluating cosmological biases using photometric redshifts for Type Ia Supernova cosmology with the Dark Energy Survey Supernova Program}",
      journal = {Mon. Not. Roy. Astron. Soc.},
         year = 2025,
        month = jan,
       volume = {536},
       number = {2},
        pages = {1948-1966},
          doi = {10.1093/mnras/stae2703},
archivePrefix = {arXiv},
       eprint = {2407.16744},
 primaryClass = {astro-ph.CO},
       adsurl = {https://ui.adsabs.harvard.edu/abs/2025MNRAS.536.1948C}
}

@ARTICLE{Popovic24,
       author = {{Popovic}, Brodie and {Scolnic}, Daniel and {Vincenzi}, Maria and {Sullivan}, Mark and {Brout}, Dillon and {Chen}, Rebecca and {Patel}, Utsav and {Peterson}, Erik R. and {Kessler}, Richard and {Kelsey}, Lisa and {Sanchez}, Bruno O. and {Bailey}, Ava Claire and {Wiseman}, Phil and {Toy}, Marcus},
        title = "{Amalgame: cosmological constraints from the first combined photometric supernova sample}",
      journal = {"Mon. Not. Roy. Astron. Soc."},
         year = 2024,
        month = apr,
       volume = {529},
       number = {3},
        pages = {2100-2115},
          doi = {10.1093/mnras/stae420},
archivePrefix = {arXiv},
       eprint = {2309.05654},
 primaryClass = {astro-ph.CO},
       adsurl = {https://ui.adsabs.harvard.edu/abs/2024MNRAS.529.2100P}
}

@article{Kessler:2025eib,
    author = "Kessler, Richard and Hounsell, Rebekah and Joshi, Bhavin and Rubin, David and Sako, Masao and Chen, Rebecca and Miranda, Vivian and Rose, Benjamin. M.",
    title = "{Cosmology Constraints from Type Ia Supernova Simulations of the Nancy Grace Roman Space Telescope Strategy Recommended by the High Latitude Time Domain Survey Definition Committee}",
    eprint = "2506.04402",
    journal = "in developement",
    archivePrefix = "arXiv",
    primaryClass = "astro-ph.CO",
    month = "6",
    year = "2025"
}

@article{Ruhlmann-Kleider:2022vum,
    author = {Ruhlmann-Kleider, V. and Lidman, C. and M{\"o}ller, A.},
    title = "{Type Ia supernova Hubble diagrams with host galaxy photometric redshifts}",
    eprint = "2207.03789",
    archivePrefix = "arXiv",
    primaryClass = "astro-ph.CO",
    doi = "10.1088/1475-7516/2022/10/065",
    journal = "JCAP",
    volume = "10",
    pages = "065",
    year = "2022"
}

@ARTICLE{2014AJ....148...13R,
       author = {{Rodney}, Steven A. and {Riess}, Adam G. and {Strolger}, Louis-Gregory and {Dahlen}, Tomas and {Graur}, Or and {Casertano}, Stefano and {Dickinson}, Mark E. and {Ferguson}, Henry C. and {Garnavich}, Peter and {Hayden}, Brian and {Jha}, Saurabh W. and {Jones}, David O. and {Kirshner}, Robert P. and {Koekemoer}, Anton M. and {McCully}, Curtis and {Mobasher}, Bahram and {Patel}, Brandon and {Weiner}, Benjamin J. and {Cenko}, S. Bradley and {Clubb}, Kelsey I. and {Cooper}, Michael and {Filippenko}, Alexei V. and {Frederiksen}, Teddy F. and {Hjorth}, Jens and {Leibundgut}, Bruno and {Matheson}, Thomas and {Nayyeri}, Hooshang and {Penner}, Kyle and {Trump}, Jonathan and {Silverman}, Jeffrey M. and {U}, Vivian and {Azalee Bostroem}, K. and {Challis}, Peter and {Rajan}, Abhijith and {Wolff}, Schuyler and {Faber}, S.~M. and {Grogin}, Norman A. and {Kocevski}, Dale},
        title = "{Type Ia Supernova Rate Measurements to Redshift 2.5 from CANDELS: Searching for Prompt Explosions in the Early Universe}",
      journal = "Astrophys. J.",
         year = 2014,
        month = jul,
       volume = {148},
       number = {1},
          eid = {13},
        pages = {13},
          doi = {10.1088/0004-6256/148/1/13},
archivePrefix = {arXiv},
       eprint = {1401.7978},
 primaryClass = {astro-ph.CO},
       adsurl = {https://ui.adsabs.harvard.edu/abs/2014AJ....148...13R}
}

@article{Strolger:2015kra,
    author = "Strolger, Louis-Gregory and Dahlen, Tomas and Rodney, Steven A. and Graur, Or and Riess, Adam G. and McCully, Curtis and Ravindranath, Swara and Mobasher, Bahram and Shahady, A. Kristin",
    title = "{The Rate of Core Collapse Supernovae to Redshift 2.5 From The CANDELS and CLASH Supernova Surveys}",
    eprint = "1509.06574",
    archivePrefix = "arXiv",
    primaryClass = "astro-ph.GA",
    doi = "10.1088/0004-637X/813/2/93",
    journal = "Astrophys. J.",
    volume = "813",
    number = "2",
    pages = "93",
    year = "2015"
}

@article{Shivvers:2016bnc,
    author = "Shivvers, Isaac and others",
    title = "{Revisiting the Lick Observatory Supernova Search Volume-Limited Sample: Updated Classifications and Revised Stripped-Envelope Supernova Fractions}",
    eprint = "1609.02922",
    archivePrefix = "arXiv",
    primaryClass = "astro-ph.HE",
    doi = "10.1088/1538-3873/aa54a6",
    journal = "Publ. Astron. Soc. Pac.",
    volume = "129",
    number = "975",
    pages = "054201",
    year = "2017"
}

@article{rosenberg22,
    author = "Rosenberg, Erik and Gratton, Steven and Efstathiou, George",
    title = "{CMB power spectra and cosmological parameters from Planck PR4 with CamSpec}",
    eprint = "2205.10869",
    archivePrefix = "arXiv",
    primaryClass = "astro-ph.CO",
    doi = "10.1093/mnras/stac2744",
    journal = "Mon. Not. Roy. Astron. Soc.",
    volume = "517",
    number = "3",
    pages = "4620--4636",
    year = "2022"
}

@ARTICLE{desi_dr2_bao,
       author = {{DESI Collaboration} and {Abdul-Karim}, M. and {Aguilar}, J. and {Ahlen}, S. and {Alam}, S. and {Allen}, L. and {Allende Prieto}, C. and {Alves}, O. and {Anand}, A. and {Andrade}, U. and {Armengaud}, E. and {Aviles}, A. and {Bailey}, S. and {Baltay}, C. and {Bansal}, P. and {Bault}, A. and {Behera}, J. and {BenZvi}, S. and {Bianchi}, D. and {Blake}, C. and {Brieden}, S. and {Brodzeller}, A. and {Brooks}, D. and {Buckley-Geer}, E. and {Burtin}, E. and {Calderon}, R. and {Canning}, R. and {Carnero Rosell}, A. and {Carrilho}, P. and {Casas}, L. and {Castander}, F.~J. and {Cereskaite}, R. and {Charles}, M. and {Chaussidon}, E. and {Chaves-Montero}, J. and {Chebat}, D. and {Chen}, X. and {Claybaugh}, T. and {Cole}, S. and {Cooper}, A.~P. and {Cuceu}, A. and {Dawson}, K.~S. and {de la Macorra}, A. and {de Mattia}, A. and {Deiosso}, N. and {Della Costa}, J. and {Demina}, R. and {Dey}, A. and {Dey}, B. and {Ding}, Z. and {Doel}, P. and {Edelstein}, J. and {Eisenstein}, D.~J. and {Elbers}, W. and {Fagrelius}, P. and {Fanning}, K. and {Fern\textbackslash'andez-Garc\textbackslash'ia}, E. and {Ferraro}, S. and {Font-Ribera}, A. and {Forero-Romero}, J.~E. and {Frenk}, C.~S. and {Garcia-Quintero}, C. and {Garrison}, L.~H. and {Gazta\textbackslash\raisebox{-0.5ex}\textasciitildenaga}, E. and {Gil-Mar\textbackslash'in}, H. and {Gontcho}, S. Gontcho A and {Gonzalez}, D. and {Gonzalez-Morales}, A.~X. and {Gordon}, C. and {Green}, D. and {Gutierrez}, G. and {Guy}, J. and {Hadzhiyska}, B. and {Hahn}, C. and {He}, S. and {Herbold}, M. and {Herrera-Alcantar}, H.~K. and {Ho}, M. and {Honscheid}, K. and {Howlett}, C. and {Huterer}, D. and {Ishak}, M. and {Juneau}, S. and {Kamble}, N.~V. and {Kara\textbackslashc\{c\}ayl\{\textbackslashi\}}, N.~G. and {Kehoe}, R. and {Kent}, S. and {Kim}, A.~G. and {Kirkby}, D. and {Kisner}, T. and {Koposov}, S.~E. and {Kremin}, A. and {Krolewski}, A. and {Lahav}, O. and {Lamman}, C. and {Landriau}, M. and {Lang}, D. and {Lasker}, J. and {Le Goff}, J.~M. and {Le Guillou}, L. and {Leauthaud}, A. and {Levi}, M.~E. and {Li}, Q. and {Li}, T.~S. and {Lodha}, K. and {Lokken}, M. and {Lozano-Rodr\textbackslash'iguez}, F. and {Magneville}, C. and {Manera}, M. and {Martini}, P. and {Matthewson}, W.~L. and {Meisner}, A. and {Mena-Fern\textbackslash'andez}, J. and {Menegas}, A. and {Mergulh\textbackslash\raisebox{-0.5ex}\textasciitildeao}, T. and {Miquel}, R. and {Moustakas}, J. and {Mu\textbackslash\raisebox{-0.5ex}\textasciitildenoz-Guti\textbackslash'errez}, A. and {Mu\textbackslash\raisebox{-0.5ex}\textasciitildenoz-Santos}, D. and {Myers}, A.~D. and {Nadathur}, S. and {Naidoo}, K. and {Napolitano}, L. and {Newman}, J.~A. and {Niz}, G. and {Noriega}, H.~E. and {Paillas}, E. and {Palanque-Delabrouille}, N. and {Pan}, J. and {Peacock}, J. and {Pellejero Ibanez}, Marcos and {Percival}, W.~J. and {P\textbackslash'erez-Fern\textbackslash'andez}, A. and {P\textbackslash'erez-R\textbackslash`afols}, I. and {Pieri}, M.~M. and {Poppett}, C. and {Prada}, F. and {Rabinowitz}, D. and {Raichoor}, A. and {Ram\textbackslash'irez-P\textbackslash'erez}, C. and {Rashkovetskyi}, M. and {Ravoux}, C. and {Rich}, J. and {Rocher}, A. and {Rockosi}, C. and {Rohlf}, J. and {Rom\textbackslash'an-Herrera}, J.~O. and {Ross}, A.~J. and {Rossi}, G. and {Ruggeri}, R. and {Ruhlmann-Kleider}, V. and {Samushia}, L. and {Sanchez}, E. and {Sanders}, N. and {Schlegel}, D. and {Schubnell}, M. and {Seo}, H. and {Shafieloo}, A. and {Sharples}, R. and {Silber}, J. and {Sinigaglia}, F. and {Sprayberry}, D. and {Tan}, T. and {Tarl\textbackslash'e}, G. and {Taylor}, P. and {Turner}, W. and {Ure\textbackslash\raisebox{-0.5ex}\textasciitildena-L\textbackslash'opez}, L.~A. and {Vaisakh}, R. and {Valdes}, F. and {Valogiannis}, G. and {Vargas-Maga\textbackslash\raisebox{-0.5ex}\textasciitildena}, M. and {Verde}, L. and {Walther}, M. and {Weaver}, B.~A. and {Weinberg}, D.~H. and {White}, M. and {Wolfson}, M. and {Y\textbackslash`eche}, C. and {Yu}, J. and {Zaborowski}, E.~A. and {Zarrouk}, P. and {Zhai}, Z. and {Zhang}, H. and {Zhao}, C. and {Zhao}, G.~B. and {Zhou}, R. and {Zou}, H.},
        title = "{DESI DR2 Results II: Measurements of Baryon Acoustic Oscillations and Cosmological Constraints}",
      journal = {arXiv e-prints},
         year = 2025,
        month = mar,
          eid = {arXiv:2503.14738},
        pages = {arXiv:2503.14738},
          doi = {10.48550/arXiv.2503.14738},
archivePrefix = {arXiv},
       eprint = {2503.14738},
 primaryClass = {astro-ph.CO},
       adsurl = {https://ui.adsabs.harvard.edu/abs/2025arXiv250314738D}
}

@ARTICLE{Binning_is_sinning,
       author = {{Brout}, Dillon and {Hinton}, Samuel R. and {Scolnic}, Dan},
        title = "{Binning is Sinning (Supernova Version): The Impact of Self-calibration in Cosmological Analyses with Type Ia Supernovae}",
      journal = {Astrophysical Journal, Letters},
         year = 2021,
        month = may,
       volume = {912},
       number = {2},
          eid = {L26},
        pages = {L26},
          doi = {10.3847/2041-8213/abf4db},
archivePrefix = {arXiv},
       eprint = {2012.05900},
 primaryClass = {astro-ph.CO},
       adsurl = {https://ui.adsabs.harvard.edu/abs/2021ApJ...912L..26B}
}

@article{Jones:2022mvo,
    author = "Jones, D. O. and others",
    title = "{Cosmological Results from the RAISIN Survey: Using Type Ia Supernovae in the Near Infrared as a Novel Path to Measure the Dark Energy Equation of State}",
    eprint = "2201.07801",
    archivePrefix = "arXiv",
    primaryClass = "astro-ph.CO",
    doi = "10.3847/1538-4357/ac755b",
    journal = "Astrophys. J.",
    volume = "933",
    number = "2",
    pages = "172",
    year = "2022"
}

@ARTICLE{PZ2021_DES3YR_WL,
       author = {{Myles}, J. and {Alarcon}, A. and {Amon}, A. and {S{\'a}nchez}, C. and {Everett}, S. and {DeRose}, J. and {McCullough}, J. and {Gruen}, D. and {Bernstein}, G.~M. and {Troxel}, M.~A. and {Dodelson}, S. and {Campos}, A. and {MacCrann}, N. and {Yin}, B. and {Raveri}, M. and {Amara}, A. and {Becker}, M.~R. and {Choi}, A. and {Cordero}, J. and {Eckert}, K. and {Gatti}, M. and {Giannini}, G. and {Gschwend}, J. and {Gruendl}, R.~A. and {Harrison}, I. and {Hartley}, W.~G. and {Huff}, E.~M. and {Kuropatkin}, N. and {Lin}, H. and {Masters}, D. and {Miquel}, R. and {Prat}, J. and {Roodman}, A. and {Rykoff}, E.~S. and {Sevilla-Noarbe}, I. and {Sheldon}, E. and {Wechsler}, R.~H. and {Yanny}, B. and {Abbott}, T.~M.~C. and {Aguena}, M. and {Allam}, S. and {Annis}, J. and {Bacon}, D. and {Bertin}, E. and {Bhargava}, S. and {Bridle}, S.~L. and {Brooks}, D. and {Burke}, D.~L. and {Carnero Rosell}, A. and {Carrasco Kind}, M. and {Carretero}, J. and {Castander}, F.~J. and {Conselice}, C. and {Costanzi}, M. and {Crocce}, M. and {da Costa}, L.~N. and {Pereira}, M.~E.~S. and {Desai}, S. and {Diehl}, H.~T. and {Eifler}, T.~F. and {Elvin-Poole}, J. and {Evrard}, A.~E. and {Ferrero}, I. and {Fert{\'e}}, A. and {Flaugher}, B. and {Fosalba}, P. and {Frieman}, J. and {Garc{\'\i}a-Bellido}, J. and {Gaztanaga}, E. and {Giannantonio}, T. and {Hinton}, S.~R. and {Hollowood}, D.~L. and {Honscheid}, K. and {Hoyle}, B. and {Huterer}, D. and {James}, D.~J. and {Krause}, E. and {Kuehn}, K. and {Lahav}, O. and {Lima}, M. and {Maia}, M.~A.~G. and {Marshall}, J.~L. and {Martini}, P. and {Melchior}, P. and {Menanteau}, F. and {Mohr}, J.~J. and {Morgan}, R. and {Muir}, J. and {Ogando}, R.~L.~C. and {Palmese}, A. and {Paz-Chinch{\'o}n}, F. and {Plazas}, A.~A. and {Rodriguez-Monroy}, M. and {Samuroff}, S. and {Sanchez}, E. and {Scarpine}, V. and {Secco}, L.~F. and {Serrano}, S. and {Smith}, M. and {Soares-Santos}, M. and {Suchyta}, E. and {Swanson}, M.~E.~C. and {Tarle}, G. and {Thomas}, D. and {To}, C. and {Varga}, T.~N. and {Weller}, J. and {Wester}, W.},
        title = "{Dark Energy Survey Year 3 results: redshift calibration of the weak lensing source galaxies}",
      journal = {Mon. Not. Roy. Astron. Soc.},
         year = 2021,
        month = aug,
       volume = {505},
       number = {3},
        pages = {4249-4277},
          doi = {10.1093/mnras/stab1515},
archivePrefix = {arXiv},
       eprint = {2012.08566},
 primaryClass = {astro-ph.CO},
       adsurl = {https://ui.adsabs.harvard.edu/abs/2021MNRAS.505.4249M}
}

@article{Madau:2014bja,
    author = "Madau, Piero and Dickinson, Mark",
    title = "{Cosmic Star Formation History}",
    eprint = "1403.0007",
    archivePrefix = "arXiv",
    primaryClass = "astro-ph.CO",
    doi = "10.1146/annurev-astro-081811-125615",
    journal = "Ann. Rev. Astron. Astrophys.",
    volume = "52",
    pages = "415--486",
    year = "2014"
}

@ARTICLE{2016arXiv161205560C,
       author = {{Chambers}, K.~C. and {Magnier}, E.~A. and {Metcalfe}, N. and {Flewelling}, H.~A. and {Huber}, M.~E. and {Waters}, C.~Z. and {Denneau}, L. and {Draper}, P.~W. and {Farrow}, D. and {Finkbeiner}, D.~P. and {Holmberg}, C. and {Koppenhoefer}, J. and {Price}, P.~A. and {Rest}, A. and {Saglia}, R.~P. and {Schlafly}, E.~F. and {Smartt}, S.~J. and {Sweeney}, W. and {Wainscoat}, R.~J. and {Burgett}, W.~S. and {Chastel}, S. and {Grav}, T. and {Heasley}, J.~N. and {Hodapp}, K.~W. and {Jedicke}, R. and {Kaiser}, N. and {Kudritzki}, R. -P. and {Luppino}, G.~A. and {Lupton}, R.~H. and {Monet}, D.~G. and {Morgan}, J.~S. and {Onaka}, P.~M. and {Shiao}, B. and {Stubbs}, C.~W. and {Tonry}, J.~L. and {White}, R. and {Ba{\~n}ados}, E. and {Bell}, E.~F. and {Bender}, R. and {Bernard}, E.~J. and {Boegner}, M. and {Boffi}, F. and {Botticella}, M.~T. and {Calamida}, A. and {Casertano}, S. and {Chen}, W. -P. and {Chen}, X. and {Cole}, S. and {Deacon}, N. and {Frenk}, C. and {Fitzsimmons}, A. and {Gezari}, S. and {Gibbs}, V. and {Goessl}, C. and {Goggia}, T. and {Gourgue}, R. and {Goldman}, B. and {Grant}, P. and {Grebel}, E.~K. and {Hambly}, N.~C. and {Hasinger}, G. and {Heavens}, A.~F. and {Heckman}, T.~M. and {Henderson}, R. and {Henning}, T. and {Holman}, M. and {Hopp}, U. and {Ip}, W. -H. and {Isani}, S. and {Jackson}, M. and {Keyes}, C.~D. and {Koekemoer}, A.~M. and {Kotak}, R. and {Le}, D. and {Liska}, D. and {Long}, K.~S. and {Lucey}, J.~R. and {Liu}, M. and {Martin}, N.~F. and {Masci}, G. and {McLean}, B. and {Mindel}, E. and {Misra}, P. and {Morganson}, E. and {Murphy}, D.~N.~A. and {Obaika}, A. and {Narayan}, G. and {Nieto-Santisteban}, M.~A. and {Norberg}, P. and {Peacock}, J.~A. and {Pier}, E.~A. and {Postman}, M. and {Primak}, N. and {Rae}, C. and {Rai}, A. and {Riess}, A. and {Riffeser}, A. and {Rix}, H.~W. and {R{\"o}ser}, S. and {Russel}, R. and {Rutz}, L. and {Schilbach}, E. and {Schultz}, A.~S.~B. and {Scolnic}, D. and {Strolger}, L. and {Szalay}, A. and {Seitz}, S. and {Small}, E. and {Smith}, K.~W. and {Soderblom}, D.~R. and {Taylor}, P. and {Thomson}, R. and {Taylor}, A.~N. and {Thakar}, A.~R. and {Thiel}, J. and {Thilker}, D. and {Unger}, D. and {Urata}, Y. and {Valenti}, J. and {Wagner}, J. and {Walder}, T. and {Walter}, F. and {Watters}, S.~P. and {Werner}, S. and {Wood-Vasey}, W.~M. and {Wyse}, R.},
        title = "{The Pan-STARRS1 Surveys}",
      journal = {arXiv e-prints},
         year = 2016,
        month = dec,
          eid = {arXiv:1612.05560},
        pages = {arXiv:1612.05560},
          doi = {10.48550/arXiv.1612.05560},
archivePrefix = {arXiv},
       eprint = {1612.05560},
 primaryClass = {astro-ph.IM},
       adsurl = {https://ui.adsabs.harvard.edu/abs/2016arXiv161205560C}
}

@ARTICLE{plasticc_K2019,
       author = {{Kessler}, R. and {Narayan}, G. and {Avelino}, A. and {Bachelet}, E. and {Biswas}, R. and {Brown}, P.~J. and {Chernoff}, D.~F. and {Connolly}, A.~J. and {Dai}, M. and {Daniel}, S. and {Di Stefano}, R. and {Drout}, M.~R. and {Galbany}, L. and {Gonz{\'a}lez-Gait{\'a}n}, S. and {Graham}, M.~L. and {Hlo{\v{z}}ek}, R. and {Ishida}, E.~E.~O. and {Guillochon}, J. and {Jha}, S.~W. and {Jones}, D.~O. and {Mandel}, K.~S. and {Muthukrishna}, D. and {O'Grady}, A. and {Peters}, C.~M. and {Pierel}, J.~R. and {Ponder}, K.~A. and {Pr{\v{s}}a}, A. and {Rodney}, S. and {Villar}, V.~A. and {LSST Dark Energy Science Collaboration} and {Transient and Variable Stars Science Collaboration}},
        title = "{Models and Simulations for the Photometric LSST Astronomical Time Series Classification Challenge (PLAsTiCC)}",
      journal = {Publ. Astron. Soc. Pac.},
         year = 2019,
        month = sep,
       volume = {131},
       number = {1003},
        pages = {094501},
          doi = {10.1088/1538-3873/ab26f1},
archivePrefix = {arXiv},
       eprint = {1903.11756},
 primaryClass = {astro-ph.HE},
       adsurl = {https://ui.adsabs.harvard.edu/abs/2019PASP..131i4501K}
}

@ARTICLE{BS20,
       author = {{Brout}, Dillon and {Scolnic}, Daniel},
        title = "{It's Dust: Solving the Mysteries of the Intrinsic Scatter and Host-galaxy Dependence of Standardized Type Ia Supernova Brightnesses}",
      journal = {Astrophys. J.s},
         year = 2021,
        month = mar,
       volume = {909},
       number = {1},
          eid = {26},
        pages = {26},
          doi = {10.3847/1538-4357/abd69b},
archivePrefix = {arXiv},
       eprint = {2004.10206},
 primaryClass = {astro-ph.CO},
       adsurl = {https://ui.adsabs.harvard.edu/abs/2021ApJ...909...26B}
}

@INBOOK{Cahn2009,
       author = {{Cahn}, Robert N.},
        title = "{Dark Energy Task Force}",
    booktitle = {From Quantum to Cosmos: Fundamental Physics Research in Space. Edited by TURYSHEV SLAVA G. Published by World Scientific Publishing Co. Pte. Ltd},
         year = 2009,
        pages = {685-695},
          doi = {10.1142/9789814261210\_0058},
          publisher={World Scientific},
       adsurl = {https://ui.adsabs.harvard.edu/abs/2009fqcf.book..685C}
}

@ARTICLE{Dilday2008,
       author = {{Dilday}, Benjamin and {Kessler}, Richard and {Frieman}, Joshua A. and {Holtzman}, Jon and {Marriner}, John and {Miknaitis}, Gajus and {Nichol}, Robert C. and {Romani}, Roger and {Sako}, Masao and {Bassett}, Bruce and {Becker}, Andrew and {Cinabro}, David and {DeJongh}, Fritz and {Depoy}, Darren L. and {Doi}, Mamoru and {Garnavich}, Peter M. and {Hogan}, Craig J. and {Jha}, Saurabh and {Konishi}, Kohki and {Lampeitl}, Hubert and {Marshall}, Jennifer L. and {McGinnis}, David and {Prieto}, Jose Luis and {Riess}, Adam G. and {Richmond}, Michael W. and {Schneider}, Donald P. and {Smith}, Mathew and {Takanashi}, Naohiro and {Tokita}, Kouichi and {van der Heyden}, Kurt and {Yasuda}, Naoki and {Zheng}, Chen and {Barentine}, John and {Brewington}, Howard and {Choi}, Changsu and {Crotts}, Arlin and {Dembicky}, Jack and {Harvanek}, Michael and {Im}, Myunshin and {Ketzeback}, William and {Kleinman}, Scott J. and {Krzesi{\'n}ski}, Jurek and {Long}, Daniel C. and {Malanushenko}, Elena and {Malanushenko}, Viktor and {McMillan}, Russet J. and {Nitta}, Atsuko and {Pan}, Kaike and {Saurage}, Gabrelle and {Snedden}, Stephanie A. and {Watters}, Shannon and {Wheeler}, J. Craig and {York}, Donald},
        title = "{A Measurement of the Rate of Type Ia Supernovae at Redshift z {\ensuremath{\approx}} 0.1 from the First Season of the SDSS-II Supernova Survey}",
      journal = {Astrophys. J.},
         year = 2008,
        month = jul,
       volume = {682},
       number = {1},
        pages = {262-282},
          doi = {10.1086/587733},
archivePrefix = {arXiv},
       eprint = {0801.3297},
 primaryClass = {astro-ph},
       adsurl = {https://ui.adsabs.harvard.edu/abs/2008ApJ...682..262D}
}

@article{SNLS:2005qlf,
    author = "Astier, P. and others",
    collaboration = "SNLS",
    title = "{The Supernova Legacy Survey: Measurement of $\Omega_M$, $\Omega_\Lambda$ and ${\cal w}$ from the first year data set}",
    eprint = "astro-ph/0510447",
    archivePrefix = "arXiv",
    doi = "10.1051/0004-6361:20054185",
    journal = "Astron. Astrophys.",
    volume = "447",
    pages = "31--48",
    year = "2006"
}

@ARTICLE{Hounsell2018,
       author = {{Hounsell}, R. and {Scolnic}, D. and {Foley}, R.~J. and {Kessler}, R. and {Miranda}, V. and {Avelino}, A. and {Bohlin}, R.~C. and {Filippenko}, A.~V. and {Frieman}, J. and {Jha}, S.~W. and {Kelly}, P.~L. and {Kirshner}, R.~P. and {Mandel}, K. and {Rest}, A. and {Riess}, A.~G. and {Rodney}, S.~A. and {Strolger}, L.},
        title = "{Simulations of the WFIRST Supernova Survey and Forecasts of Cosmological Constraints}",
      journal = {Astrophys. J.},
         year = 2018,
        month = nov,
       volume = {867},
       number = {1},
          eid = {23},
        pages = {23},
          doi = {10.3847/1538-4357/aac08b},
archivePrefix = {arXiv},
       eprint = {1702.01747},
 primaryClass = {astro-ph.IM},
       adsurl = {https://ui.adsabs.harvard.edu/abs/2018ApJ...867...23H}
}

@article{Planck2018,
    author = "{Planck Collaboration} and {Aghanim}, N. and {Akrami}, Y. and {Ashdown}, M. and {Aumont}, J. and {Baccigalupi}, C. and others",
    collaboration = "Planck",
    title = "{Planck 2018 results. VI. Cosmological parameters}",
    eprint = "1807.06209",
    archivePrefix = "arXiv",
    primaryClass = "astro-ph.CO",
    doi = "10.1051/0004-6361/201833910",
    journal = "Astron. Astrophys.",
    volume = "641",
    pages = "A6",
    year = "2020",
    note = "[Erratum: Astron.Astrophys. 652, C4 (2021)]"
}

@ARTICLE{bbc,
       author = {{Kessler}, R. and {Scolnic}, D.},
        title = "{Correcting Type Ia Supernova Distances for Selection Biases and Contamination in Photometrically Identified Samples}",
      journal = {Astrophys. J.},
         year = 2017,
        month = feb,
       volume = {836},
       number = {1},
          eid = {56},
        pages = {56},
          doi = {10.3847/1538-4357/836/1/56},
archivePrefix = {arXiv},
       eprint = {1610.04677},
 primaryClass = {astro-ph.CO},
       adsurl = {https://ui.adsabs.harvard.edu/abs/2017ApJ...836...56K}
}

@ARTICLE{korytov,
       author = {{Korytov}, Danila and {Hearin}, Andrew and {Kovacs}, Eve and {Larsen}, Patricia and {Rangel}, Esteban and {Hollowed}, Joseph and {Benson}, Andrew J. and {Heitmann}, Katrin and {Mao}, Yao-Yuan and {Bahmanyar}, Anita and {Chang}, Chihway and {Campbell}, Duncan and {DeRose}, Joseph and {Finkel}, Hal and {Frontiere}, Nicholas and {Gawiser}, Eric and {Habib}, Salman and {Joachimi}, Benjamin and {Lanusse}, Fran{\c{c}}ois and {Li}, Nan and {Mandelbaum}, Rachel and {Morrison}, Christopher and {Newman}, Jeffrey A. and {Pope}, Adrian and {Rykoff}, Eli and {Simet}, Melanie and {To}, Chun-Hao and {Vikraman}, Vinu and {Wechsler}, Risa H. and {White}, Martin and {(The LSST Dark Energy Science Collaboration}},
        title = "{CosmoDC2: A Synthetic Sky Catalog for Dark Energy Science with LSST}",
      journal = {Astrophys. J., Suppl. Ser.},
         year = 2019,
        month = dec,
       volume = {245},
       number = {2},
          eid = {26},
        pages = {26},
          doi = {10.3847/1538-4365/ab510c},
archivePrefix = {arXiv},
       eprint = {1907.06530},
 primaryClass = {astro-ph.CO},
       adsurl = {https://ui.adsabs.harvard.edu/abs/2019ApJS..245...26K}
}

@ARTICLE{dc2,
       author = {{LSST DESC}},
        title = "{The LSST DESC DC2 Simulated Sky Survey}",
      journal = {Astrophys. J., Suppl. Ser.},
         year = 2021,
        month = mar,
       volume = {253},
       number = {1},
          eid = {31},
        pages = {31},
          doi = {10.3847/1538-4365/abd62c},
archivePrefix = {arXiv},
       eprint = {2010.05926},
 primaryClass = {astro-ph.IM},
       adsurl = {https://ui.adsabs.harvard.edu/abs/2021ApJS..253...31L}
}

@article{ESSENCE:2007acn,
    author = "Wood-Vasey, W. M. and others",
    collaboration = "ESSENCE",
    title = "{Observational Constraints on the Nature of the Dark Energy: First Cosmological Results from the ESSENCE Supernova Survey}",
    eprint = "astro-ph/0701041",
    archivePrefix = "arXiv",
    reportNumber = "SLAC-PUB-12281",
    doi = "10.1086/518642",
    journal = "Astrophys. J.",
    volume = "666",
    pages = "694--715",
    year = "2007"
}

@ARTICLE{plasticc_H2020,
       author = {{Hlo{\v{z}}ek}, R. and {Ponder}, K.~A. and {Malz}, A.~I. and {Dai}, M. and {Narayan}, G. and {Ishida}, E.~E.~O. and {Allam}, T., Jr and {Bahmanyar}, A. and {Biswas}, R. and {Galbany}, L. and {Jha}, S.~W. and {Jones}, D.~O. and {Kessler}, R. and {Lochner}, M. and {Mahabal}, A.~A. and {Mandel}, K.~S. and {Mart{\'\i}nez-Galarza}, J.~R. and {McEwen}, J.~D. and {Muthukrishna}, D. and {Peiris}, H.~V. and {Peters}, C.~M. and {Setzer}, C.~N.},
        title = "{Results of the Photometric LSST Astronomical Time-series Classification Challenge (PLAsTiCC)}",
      journal = {arXiv e-prints},
         year = 2020,
        month = dec,
          eid = {arXiv:2012.12392},
        pages = {arXiv:2012.12392},
archivePrefix = {arXiv},
       eprint = {2012.12392},
 primaryClass = {astro-ph.IM},
       adsurl = {https://ui.adsabs.harvard.edu/abs/2020arXiv201212392H}
}

@ARTICLE{hlozek12,
       author = {{Hlozek}, Ren{\'e}e and {Kunz}, Martin and {Bassett}, Bruce and {Smith}, Mat and {Newling}, James and others},
        title = "{Photometric Supernova Cosmology with BEAMS and SDSS-II}",
      journal = {Astrophys. J.},
         year = 2012,
        month = jun,
       volume = {752},
       number = {2},
          eid = {79},
        pages = {79},
          doi = {10.1088/0004-637X/752/2/79},
archivePrefix = {arXiv},
       eprint = {1111.5328},
 primaryClass = {astro-ph.CO},
       adsurl = {https://ui.adsabs.harvard.edu/abs/2012ApJ...752...79H}
}

@ARTICLE{2011ApJ...740...72M,
       author = {{Marriner}, John and {Bernstein}, J.~P. and {Kessler}, Richard and {Lampeitl}, Hubert and {Miquel}, Ramon and {Mosher}, Jennifer and {Nichol}, Robert C. and {Sako}, Masao and {Schneider}, Donald P. and {Smith}, Mathew},
        title = "{A More General Model for the Intrinsic Scatter in Type Ia Supernova Distance Moduli}",
      journal = {Astrophys. J.s},
         year = 2011,
        month = oct,
       volume = {740},
       number = {2},
          eid = {72},
        pages = {72},
          doi = {10.1088/0004-637X/740/2/72},
archivePrefix = {arXiv},
       eprint = {1107.4631},
 primaryClass = {astro-ph.CO},
       adsurl = {https://ui.adsabs.harvard.edu/abs/2011ApJ...740...72M}
}

@article{p2,
    author = "Sako, Masao and others",
    collaboration = "SDSS",
    title = "{The Sloan Digital Sky Survey-II Supernova Survey: Search Algorithm and Follow-up Observations}",
    eprint = "0708.2750",
    archivePrefix = "arXiv",
    primaryClass = "astro-ph",
    reportNumber = "FERMILAB-PUB-07-791-A, SLAC-PUB-12781",
    doi = "10.1088/0004-6256/135/1/348",
    journal = "Astron. J.",
    volume = "135",
    pages = "348--373",
    year = "2008"
}

@article{Lochner2016,
    author = "Lochner, Michelle and others",
    title = "{Photometric Supernova Classification With Machine Learning}",
    eprint = "1603.00882",
    archivePrefix = "arXiv",
    primaryClass = "astro-ph.IM",
    doi = "10.3847/0067-0049/225/2/31",
    journal = "Astrophys. J. Suppl.",
    volume = "225",
    number = "2",
    pages = "31",
    year = "2016"
}

@ARTICLE{Moller2020,
       author = {{M{\"o}ller}, A. and {de Boissi{\`e}re}, T.},
        title = "{SuperNNova: an open-source framework for Bayesian, neural network-based supernova classification}",
      journal = {Mon. Not. Roy. Astron. Soc.},
         year = 2020,
        month = jan,
       volume = {491},
       number = {3},
        pages = {4277-4293},
          doi = {10.1093/mnras/stz3312},
archivePrefix = {arXiv},
       eprint = {1901.06384},
 primaryClass = {astro-ph.IM},
       adsurl = {https://ui.adsabs.harvard.edu/abs/2020MNRAS.491.4277M}
}

@article{Moller2016,
    author = {M\"oller, A. and others},
    title = "{Photometric classification of type Ia supernovae in the SuperNova Legacy Survey with supervised learning}",
    eprint = "1608.05423",
    archivePrefix = "arXiv",
    primaryClass = "astro-ph.IM",
    doi = "10.1088/1475-7516/2016/12/008",
    journal = "J. Cosmol. Astropart. Phys.",
    volume = "12",
    pages = "008",
    year = "2016"
}

@article{Pierel:2022pqc,
    author = "Pierel, J. D. R. and others",
    title = "{SALT3-NIR: Taking the Open-source Type Ia Supernova Model to Longer Wavelengths for Next-generation Cosmological Measurements}",
    eprint = "2209.05594",
    archivePrefix = "arXiv",
    primaryClass = "astro-ph.CO",
    doi = "10.3847/1538-4357/ac93f9",
    journal = "Astrophys. J.",
    volume = "939",
    number = "1",
    pages = "11",
    year = "2022"
}

@article{Kessler:2023toi,
    author = "Kessler, Richard and Vincenzi, Maria and Armstrong, Patrick",
    title = "{Binning is Sinning: Redemption for Hubble Diagram Using Photometrically Classified Type Ia Supernovae}",
    eprint = "2306.05819",
    archivePrefix = "arXiv",
    primaryClass = "astro-ph.CO",
    doi = "10.3847/2041-8213/ace34d",
    journal = "Astrophys. J. Lett.",
    volume = "952",
    number = "1",
    pages = "L8",
    year = "2023"
}

@ARTICLE{DES:2024hip,
       author = {{Vincenzi}, M. and {Brout}, D. and {Armstrong}, P. and {Popovic}, B. and {Taylor}, G. and {Acevedo}, M. and {Camilleri}, R. and {Chen}, R. and {Davis}, T.~M. and {Lee}, J. and {Lidman}, C. and {Hinton}, S.~R. and {Kelsey}, L. and {Kessler}, R. and {M{\"o}ller}, A. and {Qu}, H. and {Sako}, M. and {Sanchez}, B. and {Scolnic}, D. and {Smith}, M. and {Sullivan}, M. and {Wiseman}, P. and {Asorey}, J. and {Bassett}, B.~A. and {Carollo}, D. and {Carr}, A. and {Foley}, R.~J. and {Frohmaier}, C. and {Galbany}, L. and {Glazebrook}, K. and {Graur}, O. and {Kovacs}, E. and {Kuehn}, K. and {Malik}, U. and {Nichol}, R.~C. and {Rose}, B. and {Tucker}, B.~E. and {Toy}, M. and {Tucker}, D.~L. and {Yuan}, F. and {Abbott}, T.~M.~C. and {Aguena}, M. and {Alves}, O. and {Allam}, S.~S. and {Andrade-Oliveira}, F. and {Annis}, J. and {Bacon}, D. and {Bechtol}, K. and {Bernstein}, G.~M. and {Brooks}, D. and {Burke}, D.~L. and {Carnero Rosell}, A. and {Carretero}, J. and {Castander}, F.~J. and {Conselice}, C. and {da Costa}, L.~N. and {Pereira}, M.~E.~S. and {Desai}, S. and {Diehl}, H.~T. and {Doel}, P. and {Ferrero}, I. and {Flaugher}, B. and {Friedel}, D. and {Frieman}, J. and {Garc{\'\i}a-Bellido}, J. and {Gatti}, M. and {Giannini}, G. and {Gruen}, D. and {Gruendl}, R.~A. and {Hollowood}, D.~L. and {Honscheid}, K. and {Huterer}, D. and {James}, D.~J. and {Kuropatkin}, N. and {Lahav}, O. and {Lee}, S. and {Lin}, H. and {Marshall}, J.~L. and {Mena-Fern{\'a}ndez}, J. and {Menanteau}, F. and {Miquel}, R. and {Palmese}, A. and {Pieres}, A. and {Plazas Malag{\'o}n}, A.~A. and {Porredon}, A. and {Romer}, A.~K. and {Roodman}, A. and {Sanchez}, E. and {Sanchez Cid}, D. and {Schubnell}, M. and {Sevilla-Noarbe}, I. and {Suchyta}, E. and {Swanson}, M.~E.~C. and {Tarle}, G. and {To}, C. and {Walker}, A.~R. and {Weaverdyck}, N. and {Yamamoto}, M.},
        title = "{The Dark Energy Survey Supernova Program: Cosmological Analysis and Systematic Uncertainties}",
      journal = {\apj},
         year = 2024,
        month = nov,
       volume = {975},
       number = {1},
          eid = {86},
        pages = {86},
          doi = {10.3847/1538-4357/ad5e6c},
archivePrefix = {arXiv},
       eprint = {2401.02945},
 primaryClass = {astro-ph.CO},
       adsurl = {https://ui.adsabs.harvard.edu/abs/2024ApJ...975...86V}
}

@article{scone,
    author = {Qu, Helen and Sako, Masao and M\"oller, Anais and Doux, Cyrille},
    title = "{SCONE: Supernova Classification with a Convolutional Neural Network}",
    eprint = "2106.04370",
    archivePrefix = "arXiv",
    primaryClass = "astro-ph.IM",
    doi = "10.3847/1538-3881/ac0824",
    journal = "Astron. J.",
    volume = "162",
    number = "2",
    pages = "67",
    year = "2021"
}

@article{zbeams,
    author = "Roberts, Ethan and Lochner, Michelle and Fonseca, Jos\'e and Bassett, Bruce A. and Lablanche, Pierre-Yves and Agarwal, Shankar",
    title = "{zBEAMS: A unified solution for supernova cosmology with redshift uncertainties}",
    eprint = "1704.07830",
    archivePrefix = "arXiv",
    primaryClass = "astro-ph.CO",
    doi = "10.1088/1475-7516/2017/10/036",
    journal = "J. Cosmol. Astropart. Phys.",
    volume = "10",
    pages = "036",
    year = "2017"
}

@article{Dai2018,
    author = "Dai, Mi and others",
    title = "{Photometric classification and redshift estimation of LSST Supernovae}",
    eprint = "1701.05689",
    archivePrefix = "arXiv",
    primaryClass = "astro-ph.CO",
    doi = "10.1093/mnras/sty965",
    journal = "Mon. Not. Roy. Astron. Soc.",
    volume = "477",
    number = "3",
    pages = "4142--4151",
    year = "2018"
}

@article{salt3,
    author = "Kenworthy, W. D. and others",
    title = "{SALT3: An Improved Type Ia Supernova Model for Measuring Cosmic Distances}",
    eprint = "2104.07795",
    archivePrefix = "arXiv",
    primaryClass = "astro-ph.CO",
    doi = "10.3847/1538-4357/ac30d8",
    journal = "Astrophys. J.",
    volume = "923",
    number = "2",
    pages = "265",
    year = "2021"
}

@article{salt,
    author = "Guy, Julien and others",
    collaboration = "SNLS",
    title = "{SALT: A Spectral adaptive Light curve Template for Type Ia supernovae}",
    eprint = "astro-ph/0506583",
    archivePrefix = "arXiv",
    doi = "10.1051/0004-6361:20053025",
    journal = "Astron. Astrophys.",
    volume = "443",
    pages = "781--791",
    year = "2005"
}

@ARTICLE{Jones2018,
       author = {{Jones}, D.~O. and {Scolnic}, D.~M. and {Riess}, A.~G. and {Rest}, A. and {Kirshner}, R.  and others},
        title = "{Measuring Dark Energy Properties with Photometrically Classified Pan-STARRS Supernovae. II. Cosmological Parameters}",
      journal = {Astrophys. J.},
         year = 2018,
        month = apr,
       volume = {857},
       number = {1},
          eid = {51},
        pages = {51},
          doi = {10.3847/1538-4357/aab6b1},
archivePrefix = {arXiv},
       eprint = {1710.00846},
 primaryClass = {astro-ph.CO},
       adsurl = {https://ui.adsabs.harvard.edu/abs/2018ApJ...857...51J}
}

@article{FSS:2018cey,
    author = "Jones, D. O. and others",
    collaboration = "FSS",
    title = "{The Foundation Supernova Survey: Measuring Cosmological Parameters with Supernovae from a Single Telescope}",
    eprint = "1811.09286",
    archivePrefix = "arXiv",
    primaryClass = "astro-ph.CO",
    doi = "10.3847/1538-4357/ab2bec",
    journal = "Astrophys. J.",
    volume = "881",
    pages = "19",
    year = "2019"
}

@article{lokken,
    author = "Lokken, Martine and others",
    collaboration = "LSST Dark Energy Science",
    title = "{The simulated catalogue of optical transients and correlated hosts (SCOTCH)}",
    eprint = "2206.02815",
    archivePrefix = "arXiv",
    primaryClass = "astro-ph.IM",
    doi = "10.1093/mnras/stad302",
    journal = "Mon. Not. Roy. Astron. Soc.",
    volume = "520",
    number = "2",
    pages = "2887--2912",
    year = "2023"
}

@article{DES:2016aon,
    author = "Gupta, Ravi R. and others",
    collaboration = "DES",
    title = "{Host Galaxy Identification for Supernova Surveys}",
    eprint = "1604.06138",
    archivePrefix = "arXiv",
    primaryClass = "astro-ph.CO",
    reportNumber = "FERMILAB-PUB-16-124-AE-PPD",
    doi = "10.3847/0004-6256/152/6/154",
    journal = "Astron. J.",
    volume = "152",
    pages = "154",
    year = "2016"
}

@article{Vincenzi2021,
    author = "{Vincenzi}, M. and {Sullivan}, M. and {Graur}, O. and {Brout}, D. and {Davis}, T. and others",
    collaboration = "DES",
    title = "{The Dark Energy Survey supernova programme: modelling selection efficiency and observed core-collapse supernova contamination}",
    eprint = "2012.07180",
    archivePrefix = "arXiv",
    primaryClass = "astro-ph.CO",
    reportNumber = "FERMILAB-PUB-20-629-AE",
    doi = "10.1093/mnras/stab1353",
    journal = "Mon. Not. Roy. Astron. Soc.",
    volume = "505",
    number = "2",
    pages = "2819--2839",
    year = "2021"
}

@ARTICLE{Chen2022,
       author = {{Chen}, R. and {Scolnic}, D. and {Rozo}, E. and {Rykoff}, E.~S. and {Popovic}, B. and {Kessler}, R. and {Vincenzi}, M. and {Davis}, T.~M. and {Armstrong}, P. and {Brout}, D. and {Galbany}, L. and {Kelsey}, L. and {Lidman}, C. and {M{\"o}ller}, A. and {Rose}, B. and {Sako}, M. and {Sullivan}, M. and {Taylor}, G. and {Wiseman}, P. and {Asorey}, J. and {Carr}, A. and {Conselice}, C. and {Kuehn}, K. and {Lewis}, G.~F. and {Macaulay}, E. and {Rodriguez-Monroy}, M. and {Tucker}, B.~E. and {Abbott}, T.~M.~C. and {Aguena}, M. and {Allam}, S. and {Andrade-Oliveira}, F. and {Annis}, J. and {Bacon}, D. and {Bertin}, E. and {Bocquet}, S. and {Brooks}, D. and {Burke}, D.~L. and {Carnero Rosell}, A. and {Carrasco Kind}, M. and {Carretero}, J. and {Cawthon}, R. and {Costanzi}, M. and {da Costa}, L.~N. and {Pereira}, M.~E.~S. and {Desai}, S. and {Diehl}, H.~T. and {Doel}, P. and {Everett}, S. and {Ferrero}, I. and {Flaugher}, B. and {Friedel}, D. and {Frieman}, J. and {Garc{\'\i}a-Bellido}, J. and {Gatti}, M. and {Gaztanaga}, E. and {Gruen}, D. and {Hinton}, S.~R. and {Hollowood}, D.~L. and {Honscheid}, K. and {James}, D.~J. and {Lahav}, O. and {Lima}, M. and {March}, M. and {Menanteau}, F. and {Miquel}, R. and {Morgan}, R. and {Palmese}, A. and {Paz-Chinch{\'o}n}, F. and {Pieres}, A. and {Plazas Malag{\'o}n}, A.~A. and {Prat}, J. and {Romer}, A.~K. and {Roodman}, A. and {Sanchez}, E. and {Schubnell}, M. and {Serrano}, S. and {Sevilla-Noarbe}, I. and {Smith}, M. and {Soares-Santos}, M. and {Suchyta}, E. and {Tarle}, G. and {Thomas}, D. and {To}, C. and {Tucker}, D.~L. and {Varga}, T.~N.},
        title = "{Measuring Cosmological Parameters with Type Ia Supernovae in redMaGiC galaxies}",
      journal = {arXiv e-prints},
         year = 2022,
        month = feb,
          eid = {arXiv:2202.10480},
        pages = {arXiv:2202.10480},
archivePrefix = {arXiv},
       eprint = {2202.10480},
 primaryClass = {astro-ph.CO},
       adsurl = {https://ui.adsabs.harvard.edu/abs/2022arXiv220210480C}
}

@phdthesis{Qu:2024fja,
    author = "Qu, Helen",
    title = "{Towards Precision Photometric Type ia Supernova Cosmology With Machine Learning}",
    eprint = "2406.04529",
    archivePrefix = "arXiv",
    primaryClass = "astro-ph.CO",
    reportNumber = "FERMILAB-THESIS-2024-10",
    school = "UPenn, Philadelphia",
    year = "2024"
}

@article{Mitra:2022ykq,
    author = "Mitra, Ayan and Kessler, Richard and More, Surhud and Hlozek, Renee",
    collaboration = "LSST Dark Energy Science",
    title = "{Using Host Galaxy Photometric Redshifts to Improve Cosmological Constraints with Type Ia Supernovae in the LSST Era}",
    eprint = "2210.07560",
    archivePrefix = "arXiv",
    primaryClass = "astro-ph.CO",
    doi = "10.3847/1538-4357/acb057",
    journal = "Astrophys. J.",
    volume = "944",
    number = "2",
    pages = "212",
    year = "2023"
}

@ARTICLE{campbell,
       author = {{Campbell}, Heather and {D'Andrea}, Chris B. and {Nichol}, Robert C. and {Sako}, Masao and {Smith}, Mathew and {Lampeitl}, Hubert and {Olmstead}, Matthew D. and {Bassett}, Bruce and {Biswas}, Rahul and {Brown}, Peter and {Cinabro}, David and {Dawson}, Kyle S. and {Dilday}, Ben and {Foley}, Ryan J. and {Frieman}, Joshua A. and {Garnavich}, Peter and {Hlozek}, Renee and {Jha}, Saurabh W. and {Kuhlmann}, Steve and {Kunz}, Martin and {Marriner}, John and {Miquel}, Ramon and {Richmond}, Michael and {Riess}, Adam and {Schneider}, Donald P. and {Sollerman}, Jesper and {Taylor}, Matt and {Zhao}, Gong-Bo},
        title = "{Cosmology with Photometrically Classified Type Ia Supernovae from the SDSS-II Supernova Survey}",
      journal = {Astrophys. J.s},
         year = 2013,
        month = feb,
       volume = {763},
       number = {2},
          eid = {88},
        pages = {88},
          doi = {10.1088/0004-637X/763/2/88},
archivePrefix = {arXiv},
       eprint = {1211.4480},
 primaryClass = {astro-ph.CO},
       adsurl = {https://ui.adsabs.harvard.edu/abs/2013ApJ...763...88C}
}

@article{Gagliano:2020ucg,
    author = "Gagliano, Alex and Narayan, Gautham and Engel, Andrew and Carrasco Kind, Matias",
    collaboration = "LSST Dark Energy Science",
    title = "{GHOST: Using Only Host Galaxy Information to Accurately Associate and Distinguish Supernovae}",
    eprint = "2008.09630",
    archivePrefix = "arXiv",
    primaryClass = "astro-ph.GA",
    doi = "10.3847/1538-4357/abd02b",
    journal = "Astrophys. J.",
    volume = "908",
    number = "2",
    pages = "170",
    year = "2021"
}

@article{Foley:2013jrk,
    author = "Foley, Ryan J. and Mandel, Kaisey",
    title = "{Classifying Supernovae Using Only Galaxy Data}",
    eprint = "1309.2630",
    archivePrefix = "arXiv",
    primaryClass = "astro-ph.CO",
    doi = "10.1088/0004-637X/778/2/167",
    journal = "Astrophys. J.",
    volume = "778",
    pages = "167",
    year = "2013"
}

@article{Baldeschi:2020whr,
    author = "Baldeschi, A. and Miller, A. and Stroh, M. and Margutti, R. and Coppejans, D. L.",
    title = "{Star Formation and Morphological Properties of Galaxies in the Pan-STARRS 3$\pi$ Survey. I. A Machine-learning Approach to Galaxy and Supernova Classification}",
    eprint = "2005.00155",
    archivePrefix = "arXiv",
    primaryClass = "astro-ph.HE",
    doi = "10.3847/1538-4357/abb1c0",
    journal = "Astrophys. J.",
    volume = "902",
    number = "1",
    pages = "60",
    year = "2020"
}

@ARTICLE{sako,
       author = {{Sako}, Masao and {Bassett}, Bruce and {Connolly}, Brian and {Dilday}, Benjamin and {Cambell}, Heather and {Frieman}, Joshua A. and {Gladney}, Larry and {Kessler}, Richard and {Lampeitl}, Hubert and {Marriner}, John and {Miquel}, Ramon and {Nichol}, Robert C. and {Schneider}, Donald P. and {Smith}, Mathew and {Sollerman}, Jesper},
        title = "{Photometric Type Ia Supernova Candidates from the Three-year SDSS-II SN Survey Data}",
      journal = {Astrophys. J.s},
         year = 2011,
        month = sep,
       volume = {738},
       number = {2},
          eid = {162},
        pages = {162},
          doi = {10.1088/0004-637X/738/2/162},
archivePrefix = {arXiv},
       eprint = {1107.5106},
 primaryClass = {astro-ph.CO},
       adsurl = {https://ui.adsabs.harvard.edu/abs/2011ApJ...738..162S}
}

@ARTICLE{pantheon_new,
       author = {{Brout}, Dillon and {Scolnic}, Dan and {Popovic}, Brodie and {Riess}, Adam G. and {Zuntz}, Joe and {Kessler}, Rick and {Carr}, Anthony and {Davis}, Tamara M. and {Hinton}, Samuel and {Jones}, David and {Kenworthy}, W. D'Arcy and {Peterson}, Erik R. and {Said}, Khaled and {Taylor}, Georgie and {Ali}, Noor and {Armstrong}, Patrick and {Charvu}, Pranav and {Dwomoh}, Arianna and {Palmese}, Antonella and {Qu}, Helen and {Rose}, Benjamin M. and {Stubbs}, Christopher W. and {Vincenzi}, Maria and {Wood}, Charlotte M. and {Brown}, Peter J. and {Chen}, Rebecca and {Chambers}, Ken and {Coulter}, David A. and {Dai}, Mi and {Dimitriadis}, Georgios and {Filippenko}, Alexei V. and {Foley}, Ryan J. and {Jha}, Saurabh W. and {Kelsey}, Lisa and {Kirshner}, Robert P. and {M{\"o}ller}, Anais and {Muir}, Jessie and {Nadathur}, Seshadri and {Pan}, Yen-Chen and {Rest}, Armin and {Rojas-Bravo}, Cesar and {Sako}, Masao and {Siebert}, Matthew R. and {Smith}, Mat and {Stahl}, Benjamin E. and {Wiseman}, Phil},
        title = "{The Pantheon+ Analysis: Cosmological Constraints}",
      journal = {arXiv e-prints},
         year = 2022,
        month = feb,
          eid = {arXiv:2202.04077},
        pages = {arXiv:2202.04077},
archivePrefix = {arXiv},
       eprint = {2202.04077},
 primaryClass = {astro-ph.CO},
       adsurl = {https://ui.adsabs.harvard.edu/abs/2022arXiv220204077B}
}

@ARTICLE{DIA1,
       author = {{Alard}, C. and {Lupton}, Robert H.},
        title = "{A Method for Optimal Image Subtraction}",
      journal = {Astrophys. J.},
         year = 1998,
        month = aug,
       volume = {503},
       number = {1},
        pages = {325-331},
          doi = {10.1086/305984},
archivePrefix = {arXiv},
       eprint = {astro-ph/9712287},
 primaryClass = {astro-ph},
       adsurl = {https://ui.adsabs.harvard.edu/abs/1998ApJ...503..325A}
}

@article{ivezic,
    author = "Ivezi\'c, \v{Z}eljko and others",
    collaboration = "LSST",
    title = "{LSST: from Science Drivers to Reference Design and Anticipated Data Products}",
    eprint = "0805.2366",
    archivePrefix = "arXiv",
    primaryClass = "astro-ph",
    reportNumber = "SLAC-PUB-16076",
    doi = "10.3847/1538-4357/ab042c",
    journal = "Astrophys. J.",
    volume = "873",
    number = "2",
    pages = "111",
    year = "2019"
}

@ARTICLE{kessler2019,
       author = {{Kessler}, R. and {Brout}, D. and {D'Andrea}, C.~B. and {Davis}, T.~M. and {Hinton}, S.~R. and {Kim}, A.~G. and {Lasker}, J. and {Lidman}, C. and {Macaulay}, E. and {M{\"o}ller}, A. and {Sako}, M. and {Scolnic}, D. and {Smith}, M. and {Sullivan}, M. and {Zhang}, B. and {Andersen}, P. and {Asorey}, J. and {Avelino}, A. and {Calcino}, J. and {Carollo}, D. and {Challis}, P. and {Childress}, M. and {Clocchiatti}, A. and {Crawford}, S. and {Filippenko}, A.~V. and {Foley}, R.~J. and {Glazebrook}, K. and {Hoormann}, J.~K. and {Kasai}, E. and {Kirshner}, R.~P. and {Lewis}, G.~F. and {Mandel}, K.~S. and {March}, M. and {Morganson}, E. and {Muthukrishna}, D. and {Nugent}, P. and {Pan}, Y. -C. and {Sommer}, N.~E. and {Swann}, E. and {Thomas}, R.~C. and {Tucker}, B.~E. and {Uddin}, S.~A. and {Abbott}, T.~M.~C. and {Allam}, S. and {Annis}, J. and {Avila}, S. and {Banerji}, M. and {Bechtol}, K. and {Bertin}, E. and {Brooks}, D. and {Buckley-Geer}, E. and {Burke}, D.~L. and {Carnero Rosell}, A. and {Carrasco Kind}, M. and {Carretero}, J. and {Castander}, F.~J. and {Crocce}, M. and {da Costa}, L.~N. and {Davis}, C. and {De Vicente}, J. and {Desai}, S. and {Diehl}, H.~T. and {Doel}, P. and {Eifler}, T.~F. and {Flaugher}, B. and {Fosalba}, P. and {Frieman}, J. and {Garc{\'\i}a-Bellido}, J. and {Gaztanaga}, E. and {Gerdes}, D.~W. and {Gruen}, D. and {Gruendl}, R.~A. and {Gutierrez}, G. and {Hartley}, W.~G. and {Hollowood}, D.~L. and {Honscheid}, K. and {James}, D.~J. and {Johnson}, M.~W.~G. and {Johnson}, M.~D. and {Krause}, E. and {Kuehn}, K. and {Kuropatkin}, N. and {Lahav}, O. and {Li}, T.~S. and {Lima}, M. and {Marshall}, J.~L. and {Martini}, P. and {Menanteau}, F. and {Miller}, C.~J. and {Miquel}, R. and {Nord}, B. and {Plazas}, A.~A. and {Roodman}, A. and {Sanchez}, E. and {Scarpine}, V. and {Schindler}, R. and {Schubnell}, M. and {Serrano}, S. and {Sevilla-Noarbe}, I. and {Soares-Santos}, M. and {Sobreira}, F. and {Suchyta}, E. and {Tarle}, G. and {Thomas}, D. and {Walker}, A.~R. and {Zhang}, Y. and {DES Collaboration}},
        title = "{First cosmology results using Type Ia supernova from the Dark Energy Survey: simulations to correct supernova distance biases}",
      journal = {Mon. Not. Roy. Astron. Soc.},
         year = 2019,
        month = may,
       volume = {485},
       number = {1},
        pages = {1171-1187},
          doi = {10.1093/mnras/stz463},
archivePrefix = {arXiv},
       eprint = {1811.02379},
 primaryClass = {astro-ph.CO},
       adsurl = {https://ui.adsabs.harvard.edu/abs/2019MNRAS.485.1171K}
}

@ARTICLE{Kessler2010,
       author = "{Kessler}, Richard and {Cinabro}, David and {Bassett}, Bruce and {Dilday}, Benjamin and {Frieman}, Joshua A. and  others",
        title = "{Photometric Estimates of Redshifts and Distance Moduli for Type Ia Supernovae}",
      journal = {Astrophys. J.},
         year = 2010,
        month = jul,
       volume = {717},
       number = {1},
        pages = {40-57},
          doi = {10.1088/0004-637X/717/1/40},
archivePrefix = {arXiv},
       eprint = {1001.0738},
 primaryClass = {astro-ph.CO},
       adsurl = {https://ui.adsabs.harvard.edu/abs/2010ApJ...717...40K}
}

@ARTICLE{Palanque2010,
       author = {{Palanque-Delabrouille}, N. and {Ruhlmann-Kleider}, V. and {Pascal}, S. and {Rich}, J. and {Guy}, J. and {Bazin}, G. and {Astier}, P. and {Balland}, C. and {Basa}, S. and {Carlberg}, R.~G. and {Conley}, A. and {Fouchez}, D. and {Hardin}, D. and {Hook}, I.~M. and {Howell}, D.~A. and {Pain}, R. and {Perrett}, K. and {Pritchet}, C.~J. and {Regnault}, N. and {Sullivan}, M.},
        title = "{Photometric redshifts for type Ia supernovae in the supernova legacy survey}",
      journal = {Astron. Astrophys.},
         year = 2010,
        month = may,
       volume = {514},
          eid = {A63},
        pages = {A63},
          doi = {10.1051/0004-6361/200913283},
archivePrefix = {arXiv},
       eprint = {0911.1629},
 primaryClass = {astro-ph.CO},
       adsurl = {https://ui.adsabs.harvard.edu/abs/2010A&A...514A..63P}
}

@article{panstarrs,
    author = "Rest, A. and others",
    title = "{Cosmological Constraints from Measurements of Type Ia Supernovae discovered during the first 1.5 yr of the Pan-STARRS1 Survey}",
    eprint = "1310.3828",
    archivePrefix = "arXiv",
    primaryClass = "astro-ph.CO",
    doi = "10.1088/0004-637X/795/1/44",
    journal = "Astrophys. J.",
    volume = "795",
    number = "1",
    pages = "44",
    year = "2014"
}

@article{DES:2024jxu,
    author = "Abbott, T. M. C. and others",
    collaboration = "DES",
    title = "{The Dark Energy Survey: Cosmology Results with \ensuremath{\sim}1500 New High-redshift Type Ia Supernovae Using the Full 5 yr Data Set}",
    eprint = "2401.02929",
    archivePrefix = "arXiv",
    primaryClass = "astro-ph.CO",
    reportNumber = "FERMILAB-PUB-23-0821-PPD, DES-2023-805",
    doi = "10.3847/2041-8213/ad6f9f",
    journal = "Astrophys. J. Lett.",
    volume = "973",
    number = "1",
    pages = "L14",
    year = "2024"
}

@ARTICLE{adam,
       author = {{Riess}, Adam G. and {Filippenko}, Alexei V. and {Challis}, Peter and {Clocchiatti}, Alejandro and {Diercks}, Alan and others},
        title = "{Observational Evidence from Supernovae for an Accelerating Universe and a Cosmological Constant [arXiv:astro-ph/9805201]}",
      journal = {Astron. J},
         year = 1998,
        month = sep,
       volume = {116},
       number = {3},
        pages = {1009-1038},
          doi = {10.1086/300499},
archivePrefix = {arXiv},
       eprint = {astro-ph/9805201},
 primaryClass = {astro-ph},
       adsurl = {https://ui.adsabs.harvard.edu/abs/1998AJ....116.1009R}
}

@ARTICLE{kessler3,
       author = {{Kessler}, Richard and others},
        title = "{First-Year Sloan Digital Sky Survey-II Supernova Results: Hubble Diagram and Cosmological Parameters}",
      journal = {Astrophys. J. Lett. Suppl.},
         year = 2009,
        month = nov,
       volume = {185},
       number = {1},
        pages = {32-84},
          doi = {10.1088/0067-0049/185/1/32},
archivePrefix = {arXiv},
       eprint = {0908.4274},
 primaryClass = {astro-ph.CO},
       adsurl = {https://ui.adsabs.harvard.edu/abs/2009ApJS..185...32K}
}

@ARTICLE{2009ApJS..185...32K,
       author = {{Kessler}, Richard and {Becker}, Andrew C. and {Cinabro}, David and {Vanderplas}, Jake and {Frieman}, Joshua A. and {Marriner}, John and {Davis}, Tamara M. and {Dilday}, Benjamin and {Holtzman}, Jon and {Jha}, Saurabh W. and {Lampeitl}, Hubert and {Sako}, Masao and {Smith}, Mathew and {Zheng}, Chen and {Nichol}, Robert C. and {Bassett}, Bruce and {Bender}, Ralf and {Depoy}, Darren L. and {Doi}, Mamoru and {Elson}, Ed and {Filippenko}, Alexei V. and {Foley}, Ryan J. and {Garnavich}, Peter M. and {Hopp}, Ulrich and {Ihara}, Yutaka and {Ketzeback}, William and {Kollatschny}, W. and {Konishi}, Kohki and {Marshall}, Jennifer L. and {McMillan}, Russet J. and {Miknaitis}, Gajus and {Morokuma}, Tomoki and {M{\"o}rtsell}, Edvard and {Pan}, Kaike and {Prieto}, Jose Luis and {Richmond}, Michael W. and {Riess}, Adam G. and {Romani}, Roger and {Schneider}, Donald P. and {Sollerman}, Jesper and {Takanashi}, Naohiro and {Tokita}, Kouichi and {van der Heyden}, Kurt and {Wheeler}, J.~C. and {Yasuda}, Naoki and {York}, Donald},
        title = "{First-Year Sloan Digital Sky Survey-II Supernova Results: Hubble Diagram and Cosmological Parameters}",
      journal = {Astrophys. J. Lett. Suppl.},
         year = 2009,
        month = nov,
       volume = {185},
       number = {1},
        pages = {32-84},
          doi = {10.1088/0067-0049/185/1/32},
archivePrefix = {arXiv},
       eprint = {0908.4274},
 primaryClass = {astro-ph.CO},
       adsurl = {https://ui.adsabs.harvard.edu/abs/2009ApJS..185...32K}
}

@ARTICLE{2018ApJ...859..101S,
       author = {{Scolnic}, D.~M. and {Jones}, D.~O. and {Rest}, A. and {Pan}, Y.~C. and {Chornock}, R. and {Foley}, R.~J. and {Huber}, M.~E. and {Kessler}, R. and {Narayan}, G. and {Riess}, A.~G. and {Rodney}, S. and {Berger}, E. and {Brout}, D.~J. and {Challis}, P.~J. and {Drout}, M. and {Finkbeiner}, D. and {Lunnan}, R. and {Kirshner}, R.~P. and {Sanders}, N.~E. and {Schlafly}, E. and {Smartt}, S. and {Stubbs}, C.~W. and {Tonry}, J. and {Wood-Vasey}, W.~M. and {Foley}, M. and {Hand}, J. and {Johnson}, E. and {Burgett}, W.~S. and {Chambers}, K.~C. and {Draper}, P.~W. and {Hodapp}, K.~W. and {Kaiser}, N. and {Kudritzki}, R.~P. and {Magnier}, E.~A. and {Metcalfe}, N. and {Bresolin}, F. and {Gall}, E. and {Kotak}, R. and {McCrum}, M. and {Smith}, K.~W.},
        title = "{The Complete Light-curve Sample of Spectroscopically Confirmed SNe Ia from Pan-STARRS1 and Cosmological Constraints from the Combined Pantheon Sample}",
      journal = {Astrophys. J.s},
         year = 2018,
        month = jun,
       volume = {859},
       number = {2},
          eid = {101},
        pages = {101},
          doi = {10.3847/1538-4357/aab9bb},
archivePrefix = {arXiv},
       eprint = {1710.00845},
 primaryClass = {astro-ph.CO},
       adsurl = {https://ui.adsabs.harvard.edu/abs/2018ApJ...859..101S}
}

@ARTICLE{Betoule2014,
       author = {{Betoule}, M. and others},
        title = "{Improved cosmological constraints from a joint analysis of the SDSS-II and SNLS supernova samples}",
      journal = {Astron. Astrophys. },
         year = 2014,
        month = aug,
       volume = {568},
          eid = {A22},
        pages = {A22},
          doi = {10.1051/0004-6361/201423413},
archivePrefix = {arXiv},
       eprint = {1401.4064},
 primaryClass = {astro-ph.CO},
       adsurl = {https://ui.adsabs.harvard.edu/abs/2014A&A...568A..22B}
}

@ARTICLE{Trip1998,
       author = {{Tripp}, Robert},
        title = "{A two-parameter luminosity correction for Type IA supernovae}",
      journal = {Astron. Astrophys. },
     year = 1998,
        month = mar,
       volume = {331},
        pages = {815-820},
       adsurl = {https://ui.adsabs.harvard.edu/abs/1998A&A...331..815T}
}

@ARTICLE{DES3YR,
       author = "{Abbott}, T.~M.~C. and {Allam}, S. and {Andersen}, P. and {Angus}, C. and {Asorey}, J. and others",
       collaboration = "DES Collaboration",
      journal = {Astrophys. J. Lett. },
         year = 2019,
        month = feb,
       volume = {872},
       number = {2},
          eid = {L30},
        pages = {L30},
          doi = {10.3847/2041-8213/ab04fa},
archivePrefix = {arXiv},
       eprint = {1811.02374},
 primaryClass = {astro-ph.CO},
       adsurl = {https://ui.adsabs.harvard.edu/abs/2019ApJ...872L..30A}
}

@ARTICLE{Scolnic2018,
       author = {{Scolnic}, D.~M. and {Jones}, D.~O. and {Rest}, A. and {Pan}, Y.~C. and {Chornock}, R. and {Foley}, R.~J. and {Huber}, M.~E. and {Kessler}, R. and {Narayan}, G. and {Riess}, A.~G. and {Rodney}, S. and {Berger}, E. and {Brout}, D.~J. and {Challis}, P.~J. and {Drout}, M. and {Finkbeiner}, D. and {Lunnan}, R. and {Kirshner}, R.~P. and {Sanders}, N.~E. and {Schlafly}, E. and {Smartt}, S. and {Stubbs}, C.~W. and {Tonry}, J. and {Wood-Vasey}, W.~M. and {Foley}, M. and {Hand}, J. and {Johnson}, E. and {Burgett}, W.~S. and {Chambers}, K.~C. and {Draper}, P.~W. and {Hodapp}, K.~W. and {Kaiser}, N. and {Kudritzki}, R.~P. and {Magnier}, E.~A. and {Metcalfe}, N. and {Bresolin}, F. and {Gall}, E. and {Kotak}, R. and {McCrum}, M. and {Smith}, K.~W.},
        title = "{The Complete Light-curve Sample of Spectroscopically Confirmed SNe Ia from Pan-STARRS1 and Cosmological Constraints from the Combined Pantheon Sample}",
      journal = {Astrophys. J.},
         year = 2018,
        month = jun,
       volume = {859},
       number = {2},
          eid = {101},
        pages = {101},
          doi = {10.3847/1538-4357/aab9bb},
archivePrefix = {arXiv},
       eprint = {1710.00845},
 primaryClass = {astro-ph.CO},
       adsurl = {https://ui.adsabs.harvard.edu/abs/2018ApJ...859..101S}
}

@ARTICLE{pantheon+,
       author = {{Scolnic}, Dan and {Brout}, Dillon and {Carr}, Anthony and {Riess}, Adam G. and {Davis}, Tamara M. and {Dwomoh}, Arianna and {Jones}, David O. and {Ali}, Noor and {Charvu}, Pranav and {Chen}, Rebecca and {Peterson}, Erik R. and {Popovic}, Brodie and {Rose}, Benjamin M. and {Wood}, Charlotte M. and {Brown}, Peter J. and {Chambers}, Ken and {Coulter}, David A. and {Dettman}, Kyle G. and {Dimitriadis}, Georgios and {Filippenko}, Alexei V. and {Foley}, Ryan J. and {Jha}, Saurabh W. and {Kilpatrick}, Charles D. and {Kirshner}, Robert P. and {Pan}, Yen-Chen and {Rest}, Armin and {Rojas-Bravo}, Cesar and {Siebert}, Matthew R. and {Stahl}, Benjamin E. and {Zheng}, WeiKang},
        title = "{The Pantheon+ Analysis: The Full Data Set and Light-curve Release}",
      journal = {Astrophys. J.s},
         year = 2022,
        month = oct,
       volume = {938},
       number = {2},
          eid = {113},
        pages = {113},
          doi = {10.3847/1538-4357/ac8b7a},
archivePrefix = {arXiv},
       eprint = {2112.03863},
 primaryClass = {astro-ph.CO},
       adsurl = {https://ui.adsabs.harvard.edu/abs/2022ApJ...938..113S}
}

@article{4most_tides,
    author = "Frohmaier, C. and others",
    collaboration = "LSST Dark Energy Science",
    title = "{TiDES: The 4MOST Time Domain Extragalactic Survey}",
    journal = "Astrophys. J.",
    eprint = "2501.16311",
    archivePrefix = "arXiv",
    primaryClass = "astro-ph.HE",
    doi = "10.3847/1538-4357/adff4e",
    month = "1",
    year = "2025"
}

@article{Popovic:2021cwq,
    author = "Popovic, Brodie and Brout, Dillon and Kessler, Richard and Scolnic, Dan and Lu, Lisa",
    title = "{Improved Treatment of Host-Galaxy Correlations in Cosmological Analyses With Type Ia Supernovae}",
    eprint = "2102.01776",
    archivePrefix = "arXiv",
    primaryClass = "astro-ph.CO",
    doi = "10.3847/1538-4357/abf14f",
    journal = "Astrophys. J.",
    volume = "913",
    number = "1",
    pages = "49",
    year = "2021"
}

@article{roman1,
    author = "Dor\'e, Olivier and others",
    title = "{WFIRST: The Essential Cosmology Space Observatory for the Coming Decade}",
    eprint = "1904.01174",
    journal = "Bull. Am. Astron. Soc.",
    archivePrefix = "arXiv",
    primaryClass = "astro-ph.CO",
    month = "4",
    year = "2019"
}

@article{Mitra2021,
    author = "Mitra, Ayan and Linder, Eric V.",
    title = "{Cosmology requirements on supernova photometric redshift systematics for the Rubin LSST and Roman Space Telescope}",
    eprint = "2011.08206",
    archivePrefix = "arXiv",
    primaryClass = "astro-ph.CO",
    doi = "10.1103/PhysRevD.103.023524",
    journal = "Phys. Rev. D",
    volume = "103",
    number = "2",
    pages = "023524",
    year = "2021"
}

@ARTICLE{Linder2019,
       author = {{Linder}, Eric V. and {Mitra}, Ayan},
        title = "{Photometric supernovae redshift systematics requirements [arXiv:1907.00985]}",
      journal = {Phys. Rev. D},
         year = 2019,
        month = aug,
       volume = {100},
       number = {4},
          eid = {043542},
        pages = {043542},
          doi = {10.1103/PhysRevD.100.043542},
archivePrefix = {arXiv},
       eprint = {1907.00985},
 primaryClass = {astro-ph.CO},
       adsurl = {https://ui.adsabs.harvard.edu/abs/2019PhRvD.100d3542L}
}

@ARTICLE{perl,
       author = {{Perlmutter}, S. and {Aldering}, G. and {Goldhaber}, G. and {Knop}, R.~A. and {Nugent}, P and others},
        title = "{Measurements of {\ensuremath{\Omega}} and {\ensuremath{\Lambda}} from 42 High-Redshift Supernovae [astro-ph/9812133]}",
      journal = {Astrophys. J.},
         year = 1999,
        month = jun,
       volume = {517},
       number = {2},
        pages = {565-586},
          doi = {10.1086/307221},
archivePrefix = {arXiv},
       eprint = {astro-ph/9812133},
 primaryClass = {astro-ph},
       adsurl = {https://ui.adsabs.harvard.edu/abs/1999ApJ...517..565P}
}

@article{Popovic:2021yuo,
    author = "Popovic, Brodie and Brout, Dillon and Kessler, Richard and Scolnic, Daniel",
    title = "{The Pantheon+ Analysis: Forward Modeling the Dust and Intrinsic Color Distributions of Type Ia Supernovae, and Quantifying Their Impact on Cosmological Inferences}",
    eprint = "2112.04456",
    archivePrefix = "arXiv",
    primaryClass = "astro-ph.CO",
    doi = "10.3847/1538-4357/aca273",
    journal = "Astrophys. J.",
    volume = "945",
    number = "1",
    pages = "84",
    year = "2023"
}

@ARTICLE{w2,
       author = {{Zhai}, Zhongxu and {Wang}, Yun and {Scolnic}, Dan},
        title = "{Forecasting Cosmological Constraints from the Weak Lensing Magnification of Type Ia Supernovae Measured by the Nancy Grace Roman Space Telescope [arXiv:2008.06804]}",
      journal = {arXiv e-prints},
         year = 2020,
        month = aug,
          eid = {arXiv:2008.06804},
        pages = {arXiv:2008.06804},
archivePrefix = {arXiv},
       eprint = {2008.06804},
 primaryClass = {astro-ph.CO},
       adsurl = {https://ui.adsabs.harvard.edu/abs/2020arXiv200806804Z}
}

@ARTICLE{snls,
       author = {{Guy}, J. and others},
        title = "{SALT2: using distant supernovae to improve the use of type Ia supernovae as distance indicators [arXiv:astro-ph/0701828]}",
      journal = {Astron. Astrophys.},
         year = 2007,
        month = apr,
       volume = {466},
       number = {1},
        pages = {11-21},
          doi = {10.1051/0004-6361:20066930},
archivePrefix = {arXiv},
       eprint = {astro-ph/0701828},
 primaryClass = {astro-ph},
       adsurl = {https://ui.adsabs.harvard.edu/abs/2007A&A...466...11G}
}

@article{DESI:2024mwx,
    author = "Adame, A. G. and others",
    collaboration = "DESI",
    title = "{DESI 2024 VI: cosmological constraints from the measurements of baryon acoustic oscillations}",
    eprint = "2404.03002",
    archivePrefix = "arXiv",
    primaryClass = "astro-ph.CO",
    reportNumber = "FERMILAB-PUB-24-0154-PPD",
    doi = "10.1088/1475-7516/2025/02/021",
    journal = "JCAP",
    volume = "02",
    pages = "021",
    year = "2025"
}

@ARTICLE{komatsu,
       author = {{Komatsu}, E. and others},
        title = "{Five-Year Wilkinson Microwave Anisotropy Probe Observations: Cosmological Interpretation}",
      journal = {Astrophys. J.s},
         year = 2009,
        month = feb,
       volume = {180},
       number = {2},
        pages = {330-376},
          doi = {10.1088/0067-0049/180/2/330},
archivePrefix = {arXiv},
       eprint = {0803.0547},
 primaryClass = {astro-ph},
       adsurl = {https://ui.adsabs.harvard.edu/abs/2009ApJS..180..330K}
}

@article{Albrecht2006_FoM,
    author = "Albrecht, Andreas and others",
    title = "{Report of the Dark Energy Task Force}",
    journal = {arXiv e-prints},
    eprint = "astro-ph/0609591",
    archivePrefix = "arXiv",
    reportNumber = "FERMILAB-FN-0793-A",
    month = "9",
    year = "2006"
}

@ARTICLE{Wang2008_FoM,
       author = {{Wang}, Yun},
        title = "{Figure of merit for dark energy constraints from current observational data}",
      journal = {\prd},
         year = 2008,
        month = jun,
       volume = {77},
       number = {12},
          eid = {123525},
        pages = {123525},
          doi = {10.1103/PhysRevD.77.123525},
archivePrefix = {arXiv},
       eprint = {0803.4295},
 primaryClass = {astro-ph},
       adsurl = {https://ui.adsabs.harvard.edu/abs/2008PhRvD..77l3525W}
}

@ARTICLE{Sanchez,
       author = {{S{\'a}nchez}, B. and {Kessler}, R. and {Scolnic}, D. and {Armstrong}, B. and {Biswas}, R. and {Bogart}, J. and {Chiang}, J. and {Cohen-Tanugi}, J. and {Fouchez}, D. and {Gris}, Ph. and {Heitmann}, K. and {Hlo{\v{z}}ek}, R. and {Jha}, S. and {Kelly}, H. and {Liu}, S. and {Narayan}, G. and {Racine}, B. and {Rykoff}, E. and {Sullivan}, M. and {Walter}, C. and {Wood-Vasey}, M. and {The LSST Dark Energy Science Collaboration}},
        title = "{SNIa-Cosmology Analysis Results from Simulated LSST Images: from Difference Imaging to Constraints on Dark Energy}",
      journal = {arXiv e-prints},
         year = 2021,
        month = nov,
          eid = {arXiv:2111.06858},
        pages = {arXiv:2111.06858},
archivePrefix = {arXiv},
       eprint = {2111.06858},
 primaryClass = {astro-ph.CO},
       adsurl = {https://ui.adsabs.harvard.edu/abs/2021arXiv211106858S}
}

@ARTICLE{wfit2,
       author = {{James}, F. and {Roos}, M.},
        title = "{Minuit - a system for function minimization and analysis of the parameter errors and correlations}",
      journal = {Computer Physics Communications},
         year = 1975,
        month = dec,
       volume = {10},
       number = {6},
        pages = {343-367},
          doi = {10.1016/0010-4655(75)90039-9},
       adsurl = {https://ui.adsabs.harvard.edu/abs/1975CoPhC..10..343J}
}

@ARTICLE{conley11,
       author = {{Conley}, A. and {Guy}, J. and {Sullivan}, M. and {Regnault}, N. and {Astier}, P. and {Balland}, C. and {Basa}, S. and {Carlberg}, R.~G. and {Fouchez}, D. and {Hardin}, D. and {Hook}, I.~M. and {Howell}, D.~A. and {Pain}, R. and {Palanque-Delabrouille}, N. and {Perrett}, K.~M. and {Pritchet}, C.~J. and {Rich}, J. and {Ruhlmann-Kleider}, V. and {Balam}, D. and {Baumont}, S. and {Ellis}, R.~S. and {Fabbro}, S. and {Fakhouri}, H.~K. and {Fourmanoit}, N. and {Gonz{\'a}lez-Gait{\'a}n}, S. and {Graham}, M.~L. and {Hudson}, M.~J. and {Hsiao}, E. and {Kronborg}, T. and {Lidman}, C. and {Mourao}, A.~M. and {Neill}, J.~D. and {Perlmutter}, S. and {Ripoche}, P. and {Suzuki}, N. and {Walker}, E.~S.},
        title = "{Supernova Constraints and Systematic Uncertainties from the First Three Years of the Supernova Legacy Survey}",
      journal = {Astrophys. J. Lett. Suppl.},
         year = 2011,
        month = jan,
       volume = {192},
       number = {1},
          eid = {1},
        pages = {1},
          doi = {10.1088/0067-0049/192/1/1},
archivePrefix = {arXiv},
       eprint = {1104.1443},
 primaryClass = {astro-ph.CO},
       adsurl = {https://ui.adsabs.harvard.edu/abs/2011ApJS..192....1C}
}

@article{DES:2023tfm,
    author = "Qu, H. and others",
    collaboration = "DES",
    title = "{The Dark Energy Survey Supernova Program: Cosmological Biases from Host Galaxy Mismatch of Type Ia Supernovae}",
    eprint = "2307.13696",
    archivePrefix = "arXiv",
    primaryClass = "astro-ph.CO",
    reportNumber = "FERMILAB-PUB-23-400-PPD",
    doi = "10.3847/1538-4357/ad251d",
    journal = "Astrophys. J.",
    volume = "964",
    number = "2",
    pages = "134",
    year = "2024"
}

@ARTICLE{LSSTDarkEnergyScience:2025lah,
       author = {{OpenUniverse} and {The LSST Dark Energy Science Collaboration} and {The Roman HLIS Project Infrastructure Team} and {The Roman RAPID Project Infrastructure Team} and {The Roman Supernova Cosmology Project Infrastructure Team} and {Alarcon}, A. and {Aldoroty}, L. and {Beltz-Mohrmann}, G. and {Bera}, A. and {Blazek}, J. and {Bogart}, J. and {Braeunlich}, G. and {Broughton}, A. and {Cao}, K. and {Chiang}, J. and {Chisari}, N.~E. and {Desai}, V. and {Fang}, Y. and {Galbany}, L. and {Hearin}, A. and {Heitmann}, K. and {Hirata}, C. and {Hounsell}, R. and {Jain}, B. and {Jarvis}, M. and {Jencson}, J. and {Kannawadi}, A. and {Kasliwal}, M.~K. and {Kessler}, R. and {Kiessling}, A. and {Knop}, R. and {Kovacs}, E. and {Laher}, R. and {Laliotis}, K. and {Lin}, C. and {Lopes}, I. and {Mahabal}, A. and {Mandelbaum}, R. and {Masiero}, J. and {Mau}, S. and {Meehan}, C. and {Meyers}, J. and {Moraes}, B. and {Paladini}, R. and {Pearl}, A. and {Plazas Malagon}, A. and {Rose}, B. and {Rubin}, D. and {Rusholme}, B. and {Santos}, A. and {{\v{S}}ar{\v{c}}evi{\'c}}, N. and {Singhal}, J. and {Scolnic}, D. and {Troxel}, M.~A. and {Van Alfen}, N. and {Van Dyke}, S. and {Walter}, C.~W. and {Wu}, T. and {Yamamoto}, M. and {Yan}, Y. and {Zhang}, T.},
        title = "{OpenUniverse2024: A shared, simulated view of the sky for the next generation of cosmological surveys}",
      journal = {arXiv e-prints},
         year = 2025,
        month = jan,
          eid = {arXiv:2501.05632},
        pages = {arXiv:2501.05632},
          doi = {10.48550/arXiv.2501.05632},
archivePrefix = {arXiv},
       eprint = {2501.05632},
 primaryClass = {astro-ph.CO},
       adsurl = {https://ui.adsabs.harvard.edu/abs/2025arXiv250105632O}
}

@ARTICLE{Rubin:2023ovl,
       author = {{Rubin}, David and {Aldering}, Greg and {Betoule}, Marc and {Fruchter}, Andy and {Huang}, Xiaosheng and {Kim}, Alex G. and {Lidman}, Chris and {Linder}, Eric and {Perlmutter}, Saul and {Ruiz-Lapuente}, Pilar and {Suzuki}, Nao},
        title = "{Union through UNITY: Cosmology with 2000 SNe Using a Unified Bayesian Framework}",
      journal = {\apj},
         year = 2025,
        month = jun,
       volume = {986},
       number = {2},
          eid = {231},
        pages = {231},
          doi = {10.3847/1538-4357/adc0a5},
archivePrefix = {arXiv},
       eprint = {2311.12098},
 primaryClass = {astro-ph.CO},
       adsurl = {https://ui.adsabs.harvard.edu/abs/2025ApJ...986..231R}
}

@ARTICLE{2012ApJ...746...85S,
       author = {{Suzuki}, N. and {Rubin}, D. and {Lidman}, C. and {Aldering}, G. and {Amanullah}, R. and {Barbary}, K. and {Barrientos}, L.~F. and {Botyanszki}, J. and {Brodwin}, M. and {Connolly}, N. and {Dawson}, K.~S. and {Dey}, A. and {Doi}, M. and {Donahue}, M. and {Deustua}, S. and {Eisenhardt}, P. and {Ellingson}, E. and {Faccioli}, L. and {Fadeyev}, V. and {Fakhouri}, H.~K. and {Fruchter}, A.~S. and {Gilbank}, D.~G. and {Gladders}, M.~D. and {Goldhaber}, G. and {Gonzalez}, A.~H. and {Goobar}, A. and {Gude}, A. and {Hattori}, T. and {Hoekstra}, H. and {Hsiao}, E. and {Huang}, X. and {Ihara}, Y. and {Jee}, M.~J. and {Johnston}, D. and {Kashikawa}, N. and {Koester}, B. and {Konishi}, K. and {Kowalski}, M. and {Linder}, E.~V. and {Lubin}, L. and {Melbourne}, J. and {Meyers}, J. and {Morokuma}, T. and {Munshi}, F. and {Mullis}, C. and {Oda}, T. and {Panagia}, N. and {Perlmutter}, S. and {Postman}, M. and {Pritchard}, T. and {Rhodes}, J. and {Ripoche}, P. and {Rosati}, P. and {Schlegel}, D.~J. and {Spadafora}, A. and {Stanford}, S.~A. and {Stanishev}, V. and {Stern}, D. and {Strovink}, M. and {Takanashi}, N. and {Tokita}, K. and {Wagner}, M. and {Wang}, L. and {Yasuda}, N. and {Yee}, H.~K.~C. and {Supernova Cosmology Project}, The},
        title = "{The Hubble Space Telescope Cluster Supernova Survey. V. Improving the Dark-energy Constraints above z > 1 and Building an Early-type-hosted Supernova Sample}",
      journal = {Astrophys. J.s},
         year = 2012,
        month = feb,
       volume = {746},
       number = {1},
          eid = {85},
        pages = {85},
          doi = {10.1088/0004-637X/746/1/85},
archivePrefix = {arXiv},
       eprint = {1105.3470},
 primaryClass = {astro-ph.CO},
       adsurl = {https://ui.adsabs.harvard.edu/abs/2012ApJ...746...85S}
}

@article{SupernovaSearchTeam:2004lze,
    author = "Riess, Adam G. and others",
    collaboration = "Supernova Search Team",
    title = "{Type Ia supernova discoveries at z \ensuremath{>} 1 from the Hubble Space Telescope: Evidence for past deceleration and constraints on dark energy evolution}",
    eprint = "astro-ph/0402512",
    archivePrefix = "arXiv",
    doi = "10.1086/383612",
    journal = "Astrophys. J.",
    volume = "607",
    pages = "665--687",
    year = "2004"
}

@ARTICLE{2013PhR...530...87W,
       author = {{Weinberg}, David H. and {Mortonson}, Michael J. and {Eisenstein}, Daniel J. and {Hirata}, Christopher and {Riess}, Adam G. and {Rozo}, Eduardo},
        title = "{Observational probes of cosmic acceleration}",
      journal = {Phys. Rep.},
         year = 2013,
        month = sep,
       volume = {530},
       number = {2},
        pages = {87-255},
          doi = {10.1016/j.physrep.2013.05.001},
archivePrefix = {arXiv},
       eprint = {1201.2434},
 primaryClass = {astro-ph.CO},
       adsurl = {https://ui.adsabs.harvard.edu/abs/2013PhR...530...87W}
}

@article{Frieman:2008sn,
    author = "Frieman, Joshua and Turner, Michael and Huterer, Dragan",
    title = "{Dark Energy and the Accelerating Universe}",
    eprint = "0803.0982",
    archivePrefix = "arXiv",
    primaryClass = "astro-ph",
    reportNumber = "FERMILAB-PUB-08-613-A",
    doi = "10.1146/annurev.astro.46.060407.145243",
    journal = "Ann. Rev. Astron. Astrophys.",
    volume = "46",
    pages = "385--432",
    year = "2008"
}

@ARTICLE{kunz,
       author = {{Kunz}, Martin and {Bassett}, Bruce A. and {Hlozek}, Ren{\'e}e A.},
        title = "{Bayesian estimation applied to multiple species}",
      journal = {\prd},
         year = 2007,
        month = may,
       volume = {75},
       number = {10},
          eid = {103508},
        pages = {103508},
          doi = {10.1103/PhysRevD.75.103508},
archivePrefix = {arXiv},
       eprint = {astro-ph/0611004},
 primaryClass = {astro-ph},
       adsurl = {https://ui.adsabs.harvard.edu/abs/2007PhRvD..75j3508K}
}

@article{2012MNRAS.421..913N,
    author = "Newling, James and Bassett, Bruce. A. and Hlozek, Renee and Kunz, Martin and Smith, Mathew and Varughese, Melvin",
    title = "{Parameter Estimation with BEAMS in the presence of biases and correlations}",
    eprint = "1110.6178",
    archivePrefix = "arXiv",
    primaryClass = "astro-ph.IM",
    doi = "10.1111/j.1365-2966.2011.20147.x",
    journal = "Mon. Not. Roy. Astron. Soc.",
    volume = "421",
    pages = "913",
    year = "2012"
}

@ARTICLE{2020MNRAS.497..210S,
       author = {{S{\'a}nchez}, J. and {Walter}, C.~W. and {Awan}, H. and {Chiang}, J. and {Daniel}, S.~F. and {Gawiser}, E. and {Glanzman}, T. and {Kirkby}, D. and {Mandelbaum}, R. and {Slosar}, A. and {Wood-Vasey}, W.~M. and {AlSayyad}, Y. and {Burke}, C.~J. and {Digel}, S.~W. and {Jarvis}, M. and {Johnson}, T. and {Kelly}, H. and {Krughoff}, S. and {Lupton}, R.~H. and {Marshall}, P.~J. and {Peterson}, J.~R. and {Price}, P.~A. and {Sembroski}, G. and {Van Klaveren}, B. and {Wiesner}, M.~P. and {Xin}, B. and {LSST Dark Energy Science Collaboration}},
        title = "{The LSST DESC data challenge 1: generation and analysis of synthetic images for next-generation surveys}",
      journal = {Mon. Not. Roy. Astron. Soc.},
         year = 2020,
        month = sep,
       volume = {497},
       number = {1},
        pages = {210-228},
          doi = {10.1093/mnras/staa1957},
archivePrefix = {arXiv},
       eprint = {2001.00941},
 primaryClass = {astro-ph.IM},
       adsurl = {https://ui.adsabs.harvard.edu/abs/2020MNRAS.497..210S}
}

@ARTICLE{snn,
       author = {{M{\"o}ller}, A. and {de Boissi{\`e}re}, T.},
        title = "{SuperNNova: an open-source framework for Bayesian, neural network-based supernova classification}",
      journal = {Mon. Not. Roy. Astron. Soc.},
         year = 2020,
        month = jan,
       volume = {491},
       number = {3},
        pages = {4277-4293},
          doi = {10.1093/mnras/stz3312},
archivePrefix = {arXiv},
       eprint = {1901.06384},
 primaryClass = {astro-ph.IM},
       adsurl = {https://ui.adsabs.harvard.edu/abs/2020MNRAS.491.4277M}
}

@ARTICLE{Sullivan_2010,
       author = {{Sullivan}, M. and {Conley}, A. and {Howell}, D.~A. and {Neill}, J.~D. and {Astier}, P. and {Balland}, C. and {Basa}, S. and {Carlberg}, R.~G. and {Fouchez}, D. and {Guy}, J. and {Hardin}, D. and {Hook}, I.~M. and {Pain}, R. and {Palanque-Delabrouille}, N. and {Perrett}, K.~M. and {Pritchet}, C.~J. and {Regnault}, N. and {Rich}, J. and {Ruhlmann-Kleider}, V. and {Baumont}, S. and {Hsiao}, E. and {Kronborg}, T. and {Lidman}, C. and {Perlmutter}, S. and {Walker}, E.~S.},
        title = "{The dependence of Type Ia Supernovae luminosities on their host galaxies}",
      journal = {Mon. Not. Roy. Astron. Soc.},
         year = 2010,
        month = aug,
       volume = {406},
       number = {2},
        pages = {782-802},
          doi = {10.1111/j.1365-2966.2010.16731.x},
archivePrefix = {arXiv},
       eprint = {1003.5119},
 primaryClass = {astro-ph.CO},
       adsurl = {https://ui.adsabs.harvard.edu/abs/2010MNRAS.406..782S}
}

@ARTICLE{cpl2,
       author = {{Chevallier}, Michel and {Polarski}, David},
        title = "{Accelerating Universes with Scaling Dark Matter [arXiv:gr-qc/0009008]}",
      journal = {International Journal of Modern Physics D},
         year = 2001,
        month = jan,
       volume = {10},
       number = {2},
        pages = {213-223},
          doi = {10.1142/S0218271801000822},
archivePrefix = {arXiv},
       eprint = {gr-qc/0009008},
 primaryClass = {gr-qc},
       adsurl = {https://ui.adsabs.harvard.edu/abs/2001IJMPD..10..213C}
}

@ARTICLE{2025arXiv250815899K,
       author = {{Karchev}, Konstantin and {Trotta}, Roberto and {Jimenez}, Raul},
        title = "{CIGaRS I: Combined simulation-based inference from SNae Ia and host photometry}",
      journal = {arXiv e-prints},
         year = 2025,
        month = aug,
          eid = {arXiv:2508.15899},
        pages = {arXiv:2508.15899},
          doi = {10.48550/arXiv.2508.15899},
archivePrefix = {arXiv},
       eprint = {2508.15899},
 primaryClass = {astro-ph.CO},
       adsurl = {https://ui.adsabs.harvard.edu/abs/2025arXiv250815899K}
}

@INPROCEEDINGS{2023AAS...24111701N,
       author = {{Narayan}, Gautham and {ELAsTiCC Team}},
        title = "{The Extended LSST Astronomical Time-series Classification Challenge (ELAsTiCC)}",
    booktitle = {American Astronomical Society Meeting Abstracts \#241},
         year = 2023,
       series = {American Astronomical Society Meeting Abstracts},
       volume = {241},
        month = jan,
          eid = {117.01},
        pages = {117.01},
       adsurl = {https://ui.adsabs.harvard.edu/abs/2023AAS...24111701N}
}

@article{Knights:2012if,
    author = "Knights, Michelle and Bassett, Bruce A. and Varughese, Melvin and Hlozek, Ren{\'e}e and Kunz, Martin and Smith, Mat and Newling, James",
    title = "{Extending BEAMS to incorporate correlated systematic uncertainties}",
    eprint = "1205.3493",
    archivePrefix = "arXiv",
    primaryClass = "astro-ph.CO",
    doi = "10.1088/1475-7516/2013/01/039",
    journal = "JCAP",
    volume = "01",
    pages = "039",
    year = "2013"
}

@ARTICLE{2024AJ....168...80C,
       author = {{Crenshaw}, John Franklin and {Kalmbach}, J. Bryce and {Gagliano}, Alexander and {Yan}, Ziang and {Connolly}, Andrew J. and {Malz}, Alex I. and {Schmidt}, Samuel J. and {The LSST Dark Energy Science Collaboration}},
        title = "{Probabilistic Forward Modeling of Galaxy Catalogs with Normalizing Flows}",
      journal = {Astron. J.},
         year = 2024,
        month = aug,
       volume = {168},
       number = {2},
          eid = {80},
        pages = {80},
          doi = {10.3847/1538-3881/ad54bf},
archivePrefix = {arXiv},
       eprint = {2405.04740},
 primaryClass = {astro-ph.IM},
       adsurl = {https://ui.adsabs.harvard.edu/abs/2024AJ....168...80C}
}
\bibliographystyle{mnras}
\end{document}